\documentclass[aps,prb,10pt, twocolumn]{revtex4}

\usepackage[T1]{fontenc}
\usepackage[latin1]{inputenc}
\usepackage{amsmath}
\usepackage[dvipdfmx]{graphicx} 
\usepackage{amssymb}

\usepackage{color} 
\usepackage{bm}
\usepackage{amsmath}
\usepackage{threeparttable}
\makeatletter

\begin{document}
\newcommand{\kv}{{\bm k}}
\newcommand{\qv}{{\bm q}}
\newcommand{\pv}{{\bm p}}
\newcommand{\rv}{{\bm r}}
\newcommand{\gkv}{{\bm \gamma}_{\bm k}}
\newcommand{\hgkv}{\hat{{\bm \gamma}}_{\bm k}}
\newcommand{\hav}{\hat{{\bm \alpha}}}
\newcommand{\hkv}{\hat{{\bm k}}}
\newcommand{\gqv}{{\bm \gamma}_{\bm q}}
\newcommand{\tgkv}{\tilde{{\bm \gamma}}_{\bm k}}
\newcommand{\vare}{\varepsilon}
\newcommand{\beqa}{\begin{eqnarray}}
\newcommand{\eeqa}{\end{eqnarray}}
\newcommand{\beq}{\begin{equation}}
\newcommand{\eeq}{\end{equation}}
\newcommand{\bars}{\bar{\sigma}}
\newcommand{\upa}{\uparrow}
\newcommand{\doa}{\downarrow}
\newcommand{\Gt}{\tilde\Gamma_2}
\newcommand{\Gi}{\tilde\Gamma_1}
\newcommand{\bR}{\bar{R}}
\newcommand{\bA}{\bar{A}}
\newcommand{\varex}{\xi}
\newcommand{\mv}{{\bm m}}

\title{Theory of electromagnetic wave propagation in ferromagnetic Rashba conductor}
\author{Junya Shibata}
\email{j_shibata@toyo.jp}
\affiliation{
Department of Electrical, Electronic and Communications Engineering, 
Toyo University, Kawagoe, Saitama, 350-8585, Japan}
\author{Akihito Takeuchi}
\email{akihito@phys.aoyama.ac.jp}
\affiliation{Department of Physics and Mathematics, Aoyama Gakuin University, Sagamihara, Kanagawa, 252-5258, Japan}
\author{Hiroshi Kohno}
\email{kohno@s.phys.nagoya-u.ac.jp}
\affiliation{
Department of Physics, Nagoya University, Furo-cho, Chiku
sa-ku, Nagoya, 464-8602, Japan}
\author{Gen Tatara}
\email{gen.tatara@riken.jp}
\affiliation{RIKEN Center for Emergent Matter Science (CEMS),
2-1 Hirosawa, Wako, Saitama, 351-0198, Japan}
\date{\today}

\begin{abstract}
We present a comprehensive study of various electromagnetic wave propagation 
phenomena in a ferromagnetic bulk Rashba conductor 
from the perspective of quantum mechanical transport. 
In this system, both the space inversion and time reversal symmetries are broken, 
as characterized by the Rashba field ${\bm \alpha}$ and magnetization ${\bm M}$, respectively. 
First, 
we present a general phenomenological analysis of electromagnetic wave propagation 
in media with broken space inversion and time reversal symmetries 
based on 
the dielectric tensor. The dependence of the dielectric tensor on the wave vector $\qv$ and ${\bm M}$ 
are retained to first order. 
Then, we calculate the microscopic electromagnetic response of the current and spin of conduction electrons subjected to ${\bm \alpha}$ and ${\bm M}$, 
based on linear response theory and the Green's function method; the results are used to study the system optical properties. 
Firstly, it is found that a large ${\bm \alpha}$ enhances the anisotropic properties of the system and enlarges the frequency range in which the 
electromagnetic waves have hyperbolic dispersion surfaces and  
exhibit unusual propagations known as negative refraction and backward waves. 
Secondly, we consider the electromagnetic cross-correlation effects (direct and inverse Edelstein effects) on the wave propagation. 
These effects stem from the lack of space inversion symmetry and yield 
$\qv$-linear off-diagonal components in the dielectric tensor. This induces a Rashba-induced birefringence,  
in which the polarization vector rotates around the vector $({\bm \alpha}\times\qv)$. 
In the presence of ${\bm M}$, which breaks time reversal symmetry, there arises an anomalous Hall effect 
and the dielectric tensor acquires off-diagonal components linear in ${\bm M}$. 
 For ${\bm \alpha}\parallel{\bm M}$, these components yield the Faraday effect for the Faraday configuration 
$\qv \parallel {\bm M}$, and the Cotton-Mouton effect for the Voigt configuration ($\qv \perp {\bm M}$). 
 When ${\bm \alpha}$ and ${\bm M}$ are noncollinear,  
${\bm M}$- and ${\bm q}$-induced optical phenomena are possible,  
which include nonreciprocal directional dichroism in the Voigt configuration. 
 In these nonreciprocal optical phenomena, 
a ``toroidal moment,'' ${\bm \alpha}\times {\bm M}$,  
and a ``quadrupole moment,'' $\alpha_{i}M_{j} + \alpha_{j}M_{i}$,  play central roles. 
These phenomena are strongly enhanced at the spin-split transition edge in the electron band. 
\end{abstract}  

\maketitle
\section{Introduction}

\begin{figure*}[t]
\includegraphics[width=16.8cm]{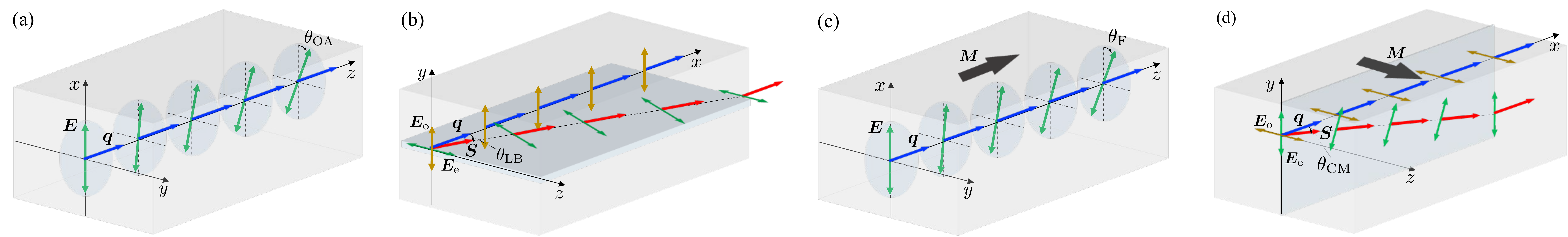}
\caption{Schematic illustration of (a) optical activity, 
(b) linear birefringence, (c) Faraday effect, and (d) Cotton-Mouton effect. 
 In the figures, ${\bm E}$ is the polarization vector,  ${\bm q}$ is the wave vector, and ${\bm M}$ is the magnetization vector. 
(a) The optical activity causes rotation of ${\bm E}$ around the propagation direction ${\bm q}$. 
(b) The linear birefringence, also known as double refraction, induces a rotation of one particular linear polarization, 
called the extraordinary wave ${\bm E}_{\rm e}$, while the other, i.e., the ordinary wave ${\bm E}_{\rm o}$, is unaffected. 
 As a result, the Poynting vector ${\bm S}$ is not parallel to ${\bm q}$.  
(c) The Faraday effect is a magneto-optical effect that occurs in the presence of ${\bm M}$ and induces rotation of ${\bm E}$ for linearly polarized waves. 
This effect is maximal when ${\bm M}$ and $\qv$ are parallel or anti-parallel (Faraday configuration). 
 (d) The Cotton-Mouton effect is a magnetization-induced birefringence that occurs under the Voigt configuration (${\bm M}\perp \qv$). 
}
\label{Sec1-OP}
\end{figure*}

Momentum-dependent spin-orbit coupled systems 
such as topological insulators, Weyl semimetals, and Rashba conductors 
having broken symmetries have recently attracted widespread attention.  
These media exhibit electromagnetic cross-correlation effects 
due to the momentum-dependent spin-orbit interaction 
and, thus, a rich variety of optical phenomena can be expected 
\cite{Tse-MacDonald10,Ohnoutek16,Grushin12,Burkov12,Tewari13,Franz13,Kargarian15,Ma15,Zhong16,STKT16}.  

Among them,  
optical phenomena induced by the spatial dispersion and/or magnetization 
in media lacking space inversion and/or time reversal symmetries  
are important for elucidating the electromagnetic responses of electrons. Such phenomena have been the subject of numerous studies performed in gasses, 
liquids, and solids \cite{{Barron82},{Agranovich84}}. 
These optical phenomena are well described by the Maxwell equations and governed by the symmetry properties of the dielectric tensor $\vare_{ij}(\qv,\omega,{\bm M})$ of the medium 
\cite{{Agranovich84},{Landau}}. 
 Here, $\qv$ and $\omega$ are the wave vector and (angular) frequency 
of the electromagnetic wave, respectively, and ${\bm M}$ is either the spontaneous magnetization, 
which exists in a ferromagnetic medium, or an applied static magnetic field. 
Microscopically, $\vare_{ij}$ is related to the quantum mechanical transport coefficients and satisfies the Onsager reciprocity relation 
\begin{align}
\label{or-epsilon}
\vare_{ji}(-\qv,\omega,-{\bm M}) = \vare_{ij}(\qv,\omega,{\bm M}) , 
\end{align}
as a consequence of microscopic reversibility. 
 Various optical phenomena caused by the absence of spatial inversion and/or time reversal symmetries 
(or the presence of ${\bm M}$) can be deduced from the expansion of $\vare_{ij}$ 
with respect to $\qv$ and ${\bm M}$ \cite{{Portigal71}, {Rikken98}}:  
\begin{align}
\label{expand-epsilon}
\vare_{ij}(\qv,\omega,{\bm M}) &= \vare^{(0)}_{ij}
+ \alpha_{ijk}q_{k} + \beta_{ijk}M_{k} + 
\gamma_{ijkl}q_{k}M_{l}+\cdots,   
\end{align}
where $\vare^{(0)}_{ij}(\omega)$, $\alpha_{ijk}(\omega)$, $\beta_{ijk}(\omega)$, 
and $\gamma_{ijkl}(\omega)$ are the tensor coefficients and functions of $\omega$. 
 Note that, in the above expression and hereafter, repeated indices in a single term imply summation. 
 The Onsager relation (\ref{or-epsilon}) requires that the tensors 
$\vare^{(0)}_{ij}$ and $\gamma_{ijkl}$ are symmetric with respect to $i$ and $j$, 
and that $\alpha_{ijk}$ and $\beta_{ijk}$ are antisymmetric. That is, $\vare^{(0)}_{ji}=\vare^{(0)}_{ij}$, $\alpha_{jik} = -\alpha_{ijk}$,  
$\beta_{jik}=-\beta_{ijk}$, and $\gamma_{jikl}=\gamma_{ijkl}$.

\begin{figure}
\includegraphics[width=8.4cm]{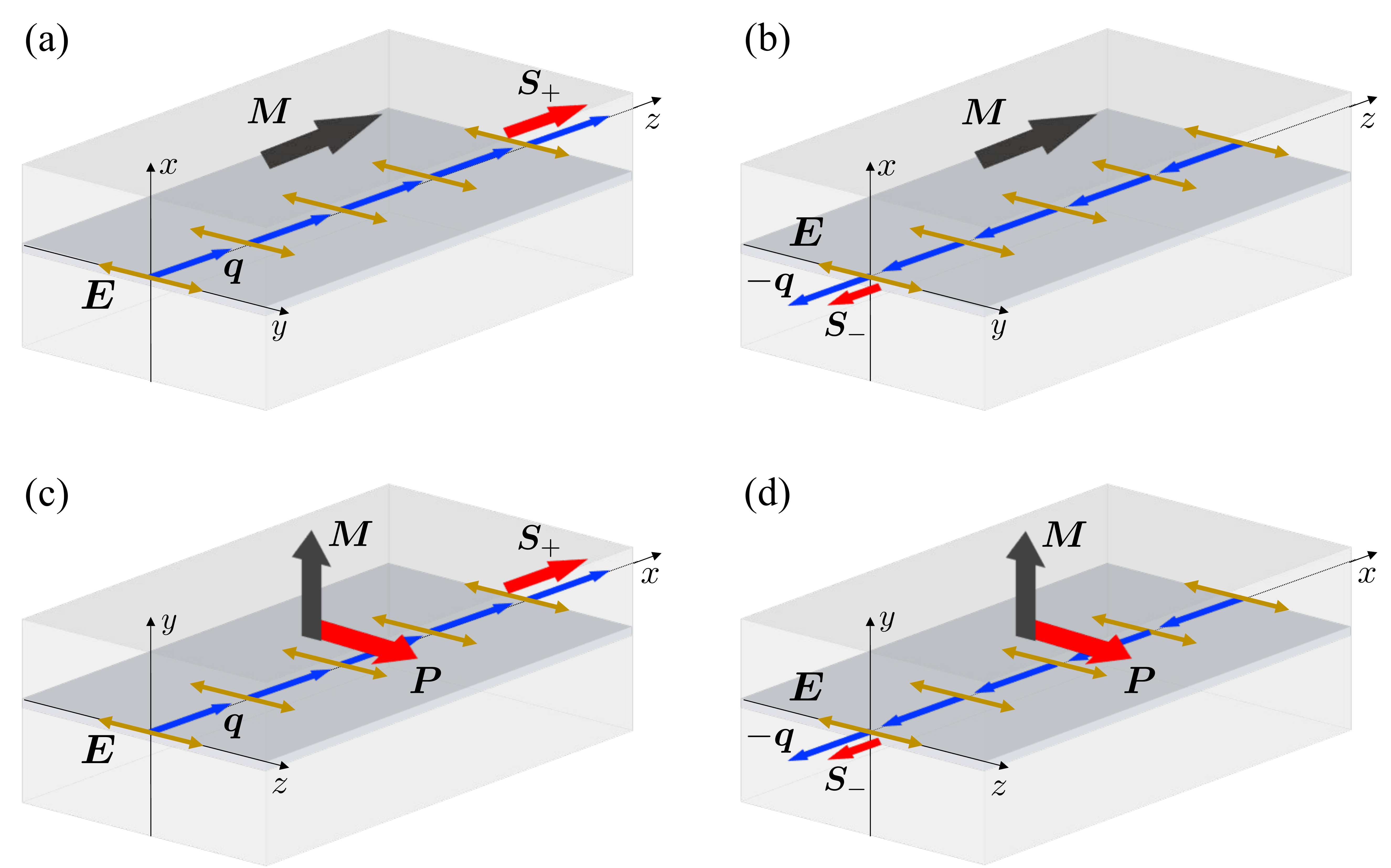}
\caption{Schematic illustration of two types of directional dichroism. 
(a, b) Magneto-chiral dichroism in Faraday configuration. 
(c, d) Nonreciprocal directional dichroism in Voigt configuration. 
The Poynting vectors of the forward ($\qv$) and backward ($-\qv$) waves 
are represented by ${\bm S}_{+}$ and ${\bm S}_{-}$, respectively. 
There exists an absorption difference for counter-propagating waves.}
\label{Sec1-MOP}
\end{figure}

 Each term of the right-hand side of Eq.~(\ref{expand-epsilon}) has the following significance with regard to optical phenomena. 
The first term, $\vare_{ij}^{(0)}$,  
determines the fundamental properties of the electromagnetic wave propagation,
such as the dispersion relation. 
 The second term, $\alpha_{ijk}q_{k}$, can only exist when the medium has no spatial inversion symmetry as, otherwise, $\vare_{ij}$ would be an even function of ${\bm q}$.  
 Generally, the ${\bm q}$-dependence of $\vare_{ij}$ is termed the ``spatial dispersion''; however, 
the $\alpha_{ijk}q_{k}$ term is more specialized in that it is odd in ${\bm q}$, 
generating two types of rotation of the polarization vector ${\bm E}$ of linearly polarized light. These two behaviors are known as natural optical activity (Fig.~\ref{Sec1-OP}(a)) 
and linear birefringence (Fig.~\ref{Sec1-OP}(b)) \cite{{Barron82},{Raab05}}. 
 In this paper, we refer to these optical phenomena as spatial-dispersion-induced phenomena, or simply, 
 ${\bm q}$-induced phenomena.  
 The third term, $\beta_{ijk}M_{k}$, can exist only when the system breaks the time reversal symmetry, 
or more specifically, in the presence of magnetization or an applied (static) magnetic field (i.e., ${\bm M}$). 
This term yields magneto-optical phenomena known as the Faraday effect (Fig.~\ref{Sec1-OP}(c)) 
and the Cotton-Mouton effect (Fig.~\ref{Sec1-OP}(d)) \cite{{Faraday},{Cotton-Mouton}}, 
in which ${\bm E}$ is rotated around ${\bm M}$.  
We call these behaviors ${\bm M}$-induced phenomena.
For the ${\bm M}$- and $\qv$-induced phenomena  
such as magneto-chiral birefringence and dichroism (Fig.~\ref{Sec1-MOP}(a)), 
and nonreciprocal directional dichroism (Fig.~\ref{Sec1-MOP}(b)) 
\cite{{HS68},{Barron84},{Rikken97},{Wagniere98},{Wagniere99},{Krichevtsov00},{Vallet01},{Train08},{Tokura11},{Mochizuki13},{Furukawa14},{Tomita14}},  
the system is required to simultaneously break the space inversion and time reversal symmetries. 
These phenomena are described by the fourth term of Eq.~(\ref{expand-epsilon}),  
$\gamma_{ijkl}q_{k}M_{l}$, 
which is bilinear in $\qv$ and ${\bm M}$ (i.e., linear in both $\qv$ and ${\bm M}$).

 The ${\bm q}$- and ${\bm M}$-induced phenomena described above 
have a common aspect of ${\bm E}$ rotation; however, 
there is an essential difference in the reciprocity.  
 That is, the wave propagations in the ${\bm q}$-induced phenomena are reciprocal, 
i.e., the polarization vectors (${\bm E}$) of the linearly polarized forward (${\bm q}$) and backward ($-{\bm q}$) waves 
rotate in mutually opposite directions. 
In contrast, 
the wave propagations in the magnetization-induced phenomena are nonreciprocal,  
i.e., ${\bm E}$ rotates in the same direction. 

In this paper, we study ${\bm q}$- and/or ${\bm M}$-induced optical phenomena in a medium with low symmetry 
based on each term in Eq.~(\ref{expand-epsilon}) for $\vare_{ij}$. 
  To do so, we evaluate each term microscopically and investigate the electromagnetic wave propagations 
by solving the Maxwell (wave) equation. 
As a concrete microscopic model, we focus on a free electron model with Rashba-type spin-orbit coupling and ${\bm M}$
(or a static external magnetic field), e.g., a ferromagnetic bulk Rashba conductor \cite{STKT16}. 

The remainder of this paper is organized as follows. In the first half of the paper (Secs.~II--IV), general aspects of electromagnetic wave propagation are described 
based on a phenomenological symmetry argument. In detail, in Sec.~II, a brief overview of the ferromagnetic Rashba conductor is provided.  
In Sec.~III, we derive the wave equation for an electric field propagating in a ferromagnetic and electromagnetically cross-correlated material.  
The relations between $\vare_{ij}$, Eq.~(\ref{expand-epsilon}), 
and various transport coefficients are also given. 
In Sec.~IV, we solve the wave equation by considering each term in Eq.~(\ref{expand-epsilon}) 
for $\vare_{ij}$, and discuss possible optical phenomena.

In the second half of the paper (Secs.~V--IX), we present microscopic analyses by considering 
a ferromagnetic Rashba conductor specifically. 
In Sec.~V, after defining the model, 
we formulate current and spin responses to an electromagnetic field based on linear response theory.  
Calculated results for the transport properties for a nonmagnetic Rashba conductor and ferromagnetic bulk Rashba conductor are presented in Secs.~VI 
and VII, respectively. 
 These results are then used in Sec.~VIII to demonstrate various wave propagations in a nonmagnetic bulk Rashba conductor, 
which include negative refraction, backward waves, and Rashba-induced birefringence. 
In Sec.~IX, optical phenomena in the ferromagnetic Rashba conductor are studied, including 
the Faraday and Kerr effects and the nonreciprocal directional dichroism. 
Finally, the findings of this study are summarized in Sec.~X.   
Various calculation details are given in the Appendices.

\section{Ferromagnetic Rashba conductor}

Recently, momentum-dependent giant spin splitting was found in the electron band 
of the BiTeI polar semiconductor \cite{{Ishizaka11},{Lee-Tokura11},{Tokura12}}.  
This behavior, called the ``Rashba effect,'' is ascribed to the Rashba spin-orbit interaction (RSOI),  
which is expected in systems without space inversion symmetry \cite{Rashba60}. 
The BiTeI polar semiconductor has a layered structure stacked along the $c$ axis with a trigonal crystal symmetry. 
The spin splitting is $\sim 200~{\rm meV}$, which corresponds to $|{\bm \alpha}|=3.05~{\rm eV}$\AA, where ${\bm \alpha}$ is the Rashba spin-orbit field specifying the strength 
and direction of the RSOI:
\begin{align}
{\cal H}_{\rm R} = \frac{1}{\hbar}{\bm \alpha}\cdot ({\bm \sigma}\times \hat{\bm p})
=-\frac{1}{\hbar}({\bm \alpha}\times\hat{\bm p})\cdot{\bm \sigma}, 
\label{H-R}
\end{align}
where $\hat{\bm p}$ is a momentum operator and 
${\bm \sigma}=(\sigma_{x},\sigma_{y},\sigma_{z})$ 
is a vector of Pauli spin matrices.  
Furthermore, it was proposed that ${\bm \alpha}$   
provides a highly anisotropic property to the medium   
and that
the bulk Rashba conductor can be regarded as a kind of hyperbolic material \cite{STKT16}. 

 To date, hyperbolic media have been realized through artificial engineering in metal-dielectric multilayer systems  
having metallic in-plane and insulating inter-plane properties 
\cite{{Podolskiy05},{Hoffman07},{Liu08},{Hoffman09},{Harish12},{JOpt12},{APN12},{Poddubny13}, {Esslinger14},{Narimanov15},{PQE15}}. 
 Such materials have hyperbolic dispersion surfaces for electromagnetic waves in a certain frequency range 
and, thus, exhibit unusual electromagnetic responses  
such as negative refraction and backward waves 
\cite{{Veselago}, {Pendry96},{Pendry00}, {Smith00},{Lindell01},{Smith03},{Belov03}}. 
This hyperbolic frequency region has also been found in natural materials, e.g., 
tetradymites (${\rm Bi_{2}Te_{2}S}$, ${\rm Bi_{2}Se_{3}}$, and ${\rm Bi_{2}Te_{3}}$)
\cite{{Poddubny13},{Esslinger14},{Narimanov15}}. 
Interestingly, these behaviors are already described by the first term 
$\vare_{ij}^{(0)}$ of Eq.~(\ref{expand-epsilon}) (see Secs. IV-A and VIII-A).

 The Rashba effect has also attracted attention in the field of spintronics,    
because of interesting electromagnetic cross-correlation effects. 
 One is the Edelstein effect \cite{Edelstein}, in which a nonequilibrium spin accumulation 
\begin{align}
{\bm \sigma}_{\rm E}(\qv,\omega)=\kappa_{\rm E}(\omega)\hat{\bm \alpha}\times{\bm E}(\qv,\omega) 
\label{s-Edelstein}
\end{align}
 is induced by 
an external electric field ${\bm E}$ 
\cite{{Obata08},{Manchon2008},{Matos2009},{Manchon12},{Kim12a},{MacDonald12},{Duine12},{Titov15}}. 
Here, $\kappa_{\rm E}(\omega)$ is a frequency-dependent coefficient, 
$\hat{\bm \alpha}={\bm \alpha}/|{\bm \alpha}|$ is a unit vector, 
and ${\bm E}(\qv,\omega)$ is the Fourier amplitude of the external electric field. 
Such a spin accumulation is observable as a torque on ${\bm M}$, 
which is called the spin-orbit torque \cite{Brataas14},  
through the exchange interaction between the electron spin and magnetization, such that 
\begin{align}
{\cal H}_{\rm ex} = -{\bm M}\cdot {\bm \sigma}. 
\label{H-ex}
\end{align}
Indeed, 
magnetization reversal and magnetic domain wall motion on the surface of a ferromagnetic 
ultrathin metal sandwiched between a heavy metal layer and an oxide layer 
have been proposed and performed experimentally 
\cite{{Miron08},{Miron10},{Miron11a},{Miron11b}}.

As a reciprocal cross-correlation effect, the inverse Edelstein effect has also been studied intensively 
\cite{{Fert-Nat-Comm-13},{Nomura15},{Sangiao15},{Zhang15},{Isasa16}}. 
 In this effect, a non-equilibrium spin current is pumped into a heavy metal layer (such as Bi/Ag bilayer) 
from a ferromagnetic layer (such as NiFe) by the ferromagnetic resonance, 
and is converted to a charge Hall current through the RSOI 
\cite{Fert-Nat-Comm-13}. 
The generation of motive force by the dynamics of magnetization in a ferromagnetic Rashba metal
has also been studied theoretically in Refs \onlinecite{Kim12b} and \onlinecite{Nakabayashi-Tatara13}. 
Theoretically, the polarization current ${\bm j}_{\rm IE}$ induced by the inverse Edelstein effect is given by 
\cite{{Raimondi14},{STKT16}} 
\begin{align}
{\bm j}_{\rm IE}(\qv,\omega)=\kappa_{\rm IE}(\omega)\hat{\bm \alpha}\times{\bm B}(\qv,\omega),  
\label{j-IEdelstein}
\end{align}
where 
${\bm B}(\qv,\omega)$ is a time-dependent external magnetic field 
and $\kappa_{\rm IE}(\omega)$ is the transport coefficient, 
which is related to $\kappa_{\rm E}(\omega)$ 
through the Onsager relation \cite{STKT16} 
\begin{align}
\kappa_{\rm IE}(\omega) = i\hbar\gamma\omega \kappa_{\rm E}(\omega), 
\end{align}
where $\gamma$ is the gyromagnetic ratio (see Sec.~VI). 
Thus, the total current induced by the electromagnetic field 
is given by 
\begin{align}
{\bm j}_{\rm E-IE} &= {\bm j}_{\rm E} + {\bm j}_{\rm IE} , 
\end{align}
where ${\bm j}_{\rm E} = -i\hbar\gamma\qv \times {\bm \sigma}_{\rm E}$ is  
the magnetization current density due to the Edelstein effect.  
Through combination with Faraday's law, 
$ -i\omega {\bm B}(\qv,\omega) +i\qv\times{\bm E}(\qv,\omega)=0$ [Eq.~(\ref{F-law})], the following expression is obtained\cite{STKT16}:
\begin{align}
{\bm j}_{\rm E-IE}(\qv,\omega) = i\hbar\kappa_{\rm E}(\omega){\bm B}_{\rm eff}(\qv)\times {\bm E}(\qv,\omega), 
\label{j-Edelstein-IEdelstein}
\end{align}
where ${\bm B}_{\rm eff}(\qv) = \hat{\bm \alpha}\times \qv$ is a 
$\qv$-induced effective magnetic field (see Sec.~VI). 
Thus, a momentum-dependent spin-orbit coupling yields an electromagnetic response 
that involves a Hall effect due to ${\bm B}_{\rm eff}$,  
which induces a rotation of ${\bm E}$ around ${\bm B}_{\rm eff}$. 

The corresponding optical conductivity $\sigma_{ij}^{\rm E-IE}$, defined by 
$({\bm j}_{\rm E-IE})_{i} =\sigma_{ij}^{\rm E-IE} E_{j}$, 
is read from Eq.~(\ref{j-Edelstein-IEdelstein}), such that
\begin{align}
\sigma_{ij}^{\rm E-IE}(\qv,\omega) = -i\hbar\kappa_{\rm E}(\omega)\vare_{ijk}(\hat{\bm \alpha}\times\qv)_{k} , 
\end{align}
where $\vare_{ijk}$ is the completely antisymmetric tensor with $\vare_{xyz}=1$. 
Thus, in a Rashba conductor, 
the electromagnetic cross-correlation effects 
contribute to the antisymmetric (hence, off-diagonal) part of $\vare_{ij}$, 
\begin{align}
\alpha_{ijk}q_{k} &= \frac{1}{\vare_{0}}\frac{i}{\omega}\sigma^{\rm E-IE}_{ij}(\qv,\omega),
\nonumber\\
&=\frac{\hbar}{\vare_{0}}\frac{\kappa_{\rm E}(\omega)}{\omega} 
\left(\hat{\alpha}_{i}q_{j}-q_{i}\hat{\alpha}_{j}\right), 
\end{align}
where $\vare_{0}$ is the vacuum permittivity. 
 This is linear in $\qv$ and induces linear birefringence (see Sec.~IV-B).

In the presence of ${\bm M}$, 
a Rashba conductor may exhibit an anomalous Hall effect 
\cite{{Inoue},{Kovalev},{BRV11},{Titov16}}. 
This effect is maximal when ${\bm \alpha}$ and ${\bm M}$ are parallel or antiparallel, 
and the induced current has the form 
\begin{align}
\label{Sec1-AHc}
{\bm j}_{\rm AH}=\sigma_{\rm AH}(\omega)(\hat{\bm \alpha}\cdot{\bm M}) \, \hat{\bm \alpha}\times{\bm E}(\qv,\omega),  
\end{align}
where 
$\sigma_{\rm AH}(\omega)$ is the anomalous Hall conductivity (see Sec.~VII).    
This effect exists even at $\qv={\bm 0}$ and 
contributes to the third term of Eq.~(\ref{expand-epsilon}) (see Sec.~VII-C): 
\begin{align}
\beta_{ijk}M_{k} =-\frac{i}{\vare_{0}}\frac{\sigma_{\rm AH}(\omega)}{\omega}
(\hat{\bm \alpha}\cdot {\bm M}) \, \vare_{ijk} \hat{\alpha}_k  . 
\end{align}
These off-diagonal components, which are linear in ${\bm M}$,  
yield the Faraday and Kerr effects for the Faraday configuration (${\bm M}\parallel \qv $), 
and the Cotton-Mouton effect for the Voigt configuration (${\bm M}\perp\qv$); 
see Sec.~IV-C. 
 Recently, a large Kerr effect was observed in BiTeI under a static magnetic field \cite{Tokura12}.

Finally, the bilinear  term in Eq.~(\ref{expand-epsilon}), $\gamma_{ijkl}q_{k}M_{l}$, 
which describes the ${\bm M}$-induced spatial dispersion,  
can be deduced from the magnetoresistance effect \cite{Potter75} 
due to ${\bm B}_{\rm eff}(\qv)$ and ${\bm M}$. 
The relevant current density may be written as 
\begin{align}
j_{i}(\qv,\omega) = \sigma_{ijkl}(\omega)B_{{\rm eff},k}(\qv)M_{l}E_{j}(\qv,\omega), 
\end{align}
where $\sigma_{ijkl}(\omega)$ is a tensor symmetric under $i \leftrightarrow j$.  
If $\sigma_{ijkl}= \sigma_{\rm TM}(\omega) \, \delta_{ij} \delta_{kl}$ (neglecting other possible terms),  
where $\sigma_{\rm TM}(\omega)$ is a frequency-dependent coefficient, 
a ``Doppler shift'' term \cite{Kawaguchi-Tatara16} ${\cal {\bm T}}\cdot \qv$ (with 
${\cal {\bm T}}={\bm \alpha}\times {\bm M}$) appears in the diagonal components of $\vare_{ij}$, with 
\begin{align}
\gamma_{ijkl}q_{k}M_{l} = \frac{i}{\vare_{0}}\frac{\sigma_{\rm TM}(\omega)}{\omega}
( {\cal {\bm T}}\cdot \qv) \, \delta_{ij}. 
\label{eq:Tq}
\end{align}
As ${\bm \alpha}$ reflects the broken inversion symmetry, 
it is a polar vector similar to an electric polarization vector ${\bm P}$. 
Thus, the vector ${\cal {\bm T}}$ is an analog of the toroidal moment 
${\bm P}\times{\bm M}$ discussed in the context of multiferroics \cite{{Spaldin08},{TSN14}}. 
Recently, a giant nonreciprocal directional dichroism induced by the toroidal moment 
was observed in ${\rm CuB_{2}O_{4}}$ \cite{Arima08-1}. 
Such optical phenomena 
are described by a term linear in both $\qv$ and ${\bm M}$ 
and in the {\it diagonal component} of $\vare_{ij}$. 
These points are pursued further in Sec.~VII.

\section{Wave equation}

In this section, we derive the wave equation for the electric field that 
propagates in electromagnetically cross-correlated materials with broken 
space inversion and time reversal symmetries.  
This derivation clarifies the connections of the various correlation functions to $\vare_{ij}$. 

In general, 
the Fourier components of the electric and magnetic fields are expressed as 
\begin{align}
\label{DEP}
{\bm D}(\qv,\omega) &= \vare_{0}{\bm E}(\qv,\omega) 
+ {\bm P}_{\rm e}(\qv,\omega), 
\\
\label{BHM}
{\bm B}(\qv,\omega) &= \mu_{0}{\bm H}(\qv,\omega)
+{\bm M}_{\rm e}(\qv,\omega), 
\end{align}
where $\vare_{0}$ and $\mu_{0}$ are the permittivity 
and magnetic permeability of free space, respectively.
${\bm D}$ and ${\bm H}$ 
are the electric displacement and magnetic field intensity, respectively, 
which are related to ${\bm E}$ and ${\bm B}$ 
through the electric polarization ${\bm P}_{\rm e}$ 
and magnetization ${\bm M}_{\rm e}$ 
of the medium that includes the conduction electrons. 
The Fourier representation of 
Faraday's law and the Maxwell-Amp\'ere equation 
in the absence of external electric currents are respectively expressed as   
\begin{align}
\label{F-law}
-i\omega {\bm B}(\qv,\omega) + i\qv \times {\bm E}(\qv,\omega) &= {\bm 0}, \\
\label{M-A}
 i\qv\times {\bm H}(\qv,\omega)
 +i\omega
{\bm D}(\qv,\omega)&= {\bm 0}. 
\end{align}
Using Eqs.~(\ref{DEP}) and (\ref{BHM}) and 
substituting Eq.~(\ref{F-law}) into Eq.~(\ref{M-A}), 
we have 
\begin{align}
\label{Feq1}
&c^2i\qv\times(i\qv \times {\bm E})
+(i\omega)^2{\bm E}
=\frac{i\omega}{\vare_{0}} {\bm j}, 
\end{align}
where  ${\bm j}(\qv,\omega)$ is an induced current density 
that consists of the polarization current density 
${\bm j}_{P}(\qv,\omega)=-i\omega {\bm P}_{\rm e}(\qv,\omega)$ 
and the magnetization current density 
${\bm j}_{M}(\qv,\omega)=i\qv\times{\bm M}_{\rm e}(\qv,\omega)/\mu_{0}$. That is, 
\begin{align}
{\bm j}= 
{\bm j}_{P}+{\bm j}_{M}=-i\omega {\bm P}_{\rm e}+\frac{1}{\mu_{0}}i\qv\times{\bm M}_{\rm e}. 
\end{align}
Microscopically, 
the ${\bm j}_{P}(\qv,\omega)$ and ${\bm M}_{\rm e}$ induced by electromagnetic fields 
are evaluated on the basis of linear response theory, as 
\begin{align}
\label{jp-1}
{\bm j}_{P}(\qv,\omega) &= {\cal K}_{ij}^{jj}(\qv,\omega,{\bm M})E_{j}-i\hbar\gamma\omega{\cal K}_{ij}^{js}(\qv,\omega,{\bm M})B_{j}, \\
\label{s-1}
-\frac{{\bm M}_{\rm e}(\qv,\omega)}{\mu_{0}\hbar\gamma}
&={\cal K}_{ij}^{sj}(\qv,\omega,{\bm M})E_{j} - i\hbar\gamma\omega
{\cal K}^{ss}_{ij}(\qv,\omega,{\bm M})B_{j} , 
\end{align}
where 
${\cal K}^{jj}_{ij}$, ${\cal K}^{js}_{ij}$, ${\cal K}^{sj}_{ij}$, and ${\cal K}^{ss}_{ij}$ 
are the current-current, current-spin, spin-current, and spin-spin correlation (or 
response) functions, respectively. 
The second and first terms on the right-hand sides of Eqs.~(\ref{jp-1}) and  (\ref{s-1}) 
represent the electromagnetic cross-correlation effects, 
which are mutually related through the Onsager reciprocity relation  
\begin{align}
\label{or-sj-js}
{\cal K}^{sj}_{ji}(-\qv,\omega,-{\bm M}) = {\cal K}^{js}_{ij}(\qv,\omega,{\bm M}). 
\end{align}
Similarly, the other two response functions satisfy 
\begin{align}
\label{or-jj}
&{\cal K}^{jj}_{ji}(-\qv,\omega,-{\bm M}) = {\cal K}^{jj}_{ij}(\qv,\omega,{\bm M}), \\
\label{or-ss}
&{\cal K}^{ss}_{ji}(-\qv,\omega,-{\bm M}) = {\cal K}^{ss}_{ij}(\qv,\omega,{\bm M}). 
\end{align}
Proof of these relations under certain conditions is given in Appendix A.

From Eqs.~(\ref{jp-1}) and (\ref{s-1}), 
the induced total current density is expressed as 
\begin{align}
j_{i}&={\cal K}^{jj}_{ij}E_{j} -i\hbar \gamma\{
{\cal K}^{js}_{ij}\omega B_{j} + \vare_{ijk}q_{k}{\cal K}^{sj}_{kl}E_{l}
\}\nonumber\\
&-(\hbar\gamma)^2\vare_{ijk}q_{j}{\cal K}^{ss}_{kl}\omega B_{l}. 
\label{j-2}
\end{align}
Using Faraday's law [Eq.~(\ref{F-law})] to eliminate the magnetic field, we obtain 
\begin{align}
\label{j-ocE}
j_{i}(\qv,\omega) = \sigma_{ij}(\qv,\omega,{\bm M}) E_{j}(\qv,\omega), 
\end{align}
where $\sigma_{ij}$ is the optical conductivity, with
\begin{align}
\label{oc-1}
\sigma_{ij} &= {\cal K}_{ij}^{jj}
-i\hbar\gamma
\left\{
{\cal K}^{js}_{il}\vare_{lkj}+\vare_{ikl}{\cal K}^{sj}_{lj}
\right\}q_{k}\nonumber\\
&
-(\hbar\gamma)^2\vare_{ikm}\vare_{nlj}{\cal K}^{ss}_{mn}q_{k}q_{l} . 
\end{align}
From Eqs.~(\ref{or-sj-js})--(\ref{or-ss}), 
$\sigma_{ij}$ also satisfies the Onsager relation:
\begin{align} 
\sigma_{ji}(-\qv,\omega,-{\bm M}) = \sigma_{ij}(\qv,\omega,{\bm M}). 
\label{or-oc}
\end{align}
Substituting Eq.~(\ref{j-ocE}) into Eq.~(\ref{Feq1}),  
we obtain the wave equation for ${\bm E}$, 
\begin{align}
\label{weq-0}
\left[
c^2(q^2\delta_{ij}-q_{i}q_{j})-\omega^2\vare_{ij}(\qv,\omega,{\bm M})
\right]E_{j}(\qv,\omega)=0, 
\end{align}
where 
\begin{align}
\label{dielectric}
\vare_{ij}(\qv,\omega,{\bm M}) &= 
\delta_{ij} + \frac{1}{\vare_{0}}\frac{i}{\omega}
\sigma_{ij}(\qv,\omega,{\bm M})
\end{align}
is the dielectric tensor. 
It is apparent that the optical properties of cross-correlated materials   
are governed by all types of correlation functions 
through the $\sigma_{ij}$ given in Eq.~(\ref{oc-1}). 
From Eqs.~(\ref{oc-1}) and (\ref{dielectric}),  
each term of $\vare_{ij}$  in Eq.~(\ref{expand-epsilon}) is 
expressed in a microscopic sense as    
\begin{align}
&\vare^{(0)}_{ij}(\omega) = \delta_{ij} + \frac{1}{\vare_{0}}\frac{i}{\omega} {\cal K}^{jj}_{ij}({\bm 0},\omega,{\bm 0}),\label{vare-0}\\
&\alpha_{ijk}(\omega)= \frac{\hbar\gamma}{\vare_{0}}\left\{
{\cal K}^{js}_{il}({\bm 0},\omega,{\bm 0})\vare_{lkj}
+\vare_{ikl}{\cal K}^{sj}_{lj}({\bm 0},\omega,{\bm 0})
\right\},\label{alpha-ijk}\\
&\beta_{ijk}(\omega) = \frac{1}{\vare_{0}}\frac{i}{\omega} \left.
\frac{\partial}{\partial M_{k}}{\cal K}^{jj}_{ij}({\bm 0},\omega,{\bm M})
\right|_{{\bm M}={\bm 0}},\label{beta-ijk}\\
&\gamma_{ijkl}(\omega) =  \frac{1}{\vare_{0}}\frac{i}{\omega}\left.
\frac{\partial^2}{\partial q_{k}\partial M_{l}}{\cal K}^{jj}_{ij}(\qv,\omega,{\bm M})
\right|_{\qv={\bm 0},{\bm M}={\bm 0}}\nonumber\\
&+\frac{\hbar\gamma}{\vare_{0}}\frac{\partial}{\partial M_{l}} \left.\left\{
{\cal K}^{js}_{im}({\bm 0},\omega,{\bm M})\vare_{mkj}
+\vare_{ikm}{\cal K}^{sj}_{mj}({\bm 0},\omega,{\bm M})
\right\}\right|_{{\bm M}={\bm 0}}. 
\label{gamma-ijkl}
\end{align}
Note that explicit evaluation of these expressions is performed in Secs.~V and VI. 
Here, it is sufficient to calculate 
${\cal K}^{jj}$ to first order in $\qv$ and ${\bm M}$, and
${\cal K}^{js}$ and ${\cal K}^{sj}$ to first order in ${\bm M}$ 
at $\qv={\bm 0}$. 
As for ${\cal K}^{ss}$, 
the last term of Eq.~(\ref{oc-1}) can be dropped as it is already second-order in $\qv$. However, 
we note that all correlation functions for materials  with momentum-dependent spin-orbit coupling involve ${\cal K}^{ss}$,  
because of the anomalous velocity that contains a spin operator.  

\section{Phenomenological analysis of wave propagation}  

In this section, we study electromagnetic wave propagations in low-symmetry media  
based on the wave equation (\ref{weq-0}), 
by considering each term of Eq.~(\ref{expand-epsilon}) for $\vare_{ij}$  phenomenologically. 
Various optical phenomena, as listed in Table I, 
are classified according to the form of $\vare_{ij}$. 
By choosing the wave propagation direction to be in the $x$-$z$ plane, we have $\qv = q_{x}{\bm e}_{x} + q_{z}{\bm e}_{z}$, 
where ${\bm e}_{i}~(i=x,y,z)$ is a unit vector in the $i$-direction. Then, 
we can write Eq.~(\ref{weq-0}) as 
\begin{widetext}
\begin{align}
\label{weq-a}
\begin{pmatrix}
c^2q_{z}^2-\omega^2 \vare_{xx}& -\omega^2\vare_{xy}& -c^2q_{x}q_{z}-\omega^2\vare_{xz}\\
-\omega \vare_{yx}& c^2q^2-\omega^2\vare_{yy}& -\omega^2\vare_{yz}\\
-c^2q_{x}q_{z}-\omega \vare_{zx}& -\omega^2\vare_{zy}& c^2q_{x}^2-\omega^2\vare_{zz}
\end{pmatrix}
\begin{pmatrix}
E_{x}\\
E_{y}\\
E_{z}
\end{pmatrix}
=0, 
\end{align}
\end{widetext}
where $q^2=q_{x}^2+q_{z}^2$. 
Various optical phenomena expected in electromagnetically cross-correlated media 
are contained within this wave equation. 

\begin{table*}[t]
\caption{Dielectric tensor $\vare_{ij}$ and possible optical phenomena. 
The presence or absence of symmetry (I: space inversion, T: time reversal) 
is indicated by $+$ or $-$, respectively. 
The last column shows the presence ($\bigcirc$) or absence ($-$) of the given effect in a  
ferromagnetic Rashba conductor. }
\label{Table1}
\begin{threeparttable}
\begin{tabular}{lcccc}
\hline\hline
Dielectric tensor &I & T &Optical phenomenon &Ferromagnetic Rashba conductor\\
\hline
$\vare^{(0)}_{ij}=\vare_{x}(\delta_{ix}\delta_{jx}+\delta_{iy}\delta_{jy})+\vare_{\parallel}\delta_{iz}\delta_{jz}~~~~$ &$+$ & $+$&~~~~Negative refraction and backward wave&$\bigcirc$\\\hline
$\alpha_{ijk}q_{k}\sim \vare_{ijk}q_{k}$& $-$ &$+$ & Optical activity &$-$\\
$\alpha_{ijk}q_{k}\sim q_{i}\delta_{jz}-\delta_{iz}q_{j}$&$-$& $+$ 
& Linear birefringence & $\bigcirc$\\
\hline
$\beta_{ijk}M_{k}\sim\vare_{ijk}M_{k}$~(Faraday: $\qv\parallel{\bm M}$)& $+$ & $-$ &Faraday and Kerr effects&$\bigcirc$\\
$\beta_{ijk}M_{k}\sim\vare_{ijk}M_{k}$~(Voigt: $\qv\perp{\bm M}$)& $+$ & $-$ &Cotton-Mouton effect&$\bigcirc$\\
\hline
$\gamma_{ijk}q_{k}M_{l} \sim \delta_{ij} \qv\cdot{\bm M}$ (Faraday) 
& $-$& $-$ & MChD$^{c}$ and MChB$^{d}$& $-$\\
$\gamma_{ijk}q_{k}M_{l} \sim\delta_{ij} \qv\cdot ({\bm P}\times{\bm M}) $ (Voigt)&$-$&$-$& NB$^{e}$and NDD$^{f}$
&$\bigcirc$\\
\hline\hline
\end{tabular}
\footnotesize
\begin{flushleft}
$^{a}$Magneto-chiral dichroism, $^{b}$Magneto-chiral birefringence, 
$^{c}$Nonreciprocal birefringence, $^{d}$Nonreciprocal directional dichroism 
\end{flushleft}
\end{threeparttable}
\end{table*}
\subsection{Effects of anisotropy in $\vare^{(0)}_{ij}$}

\begin{figure}[t]
\includegraphics[width=8.4cm]{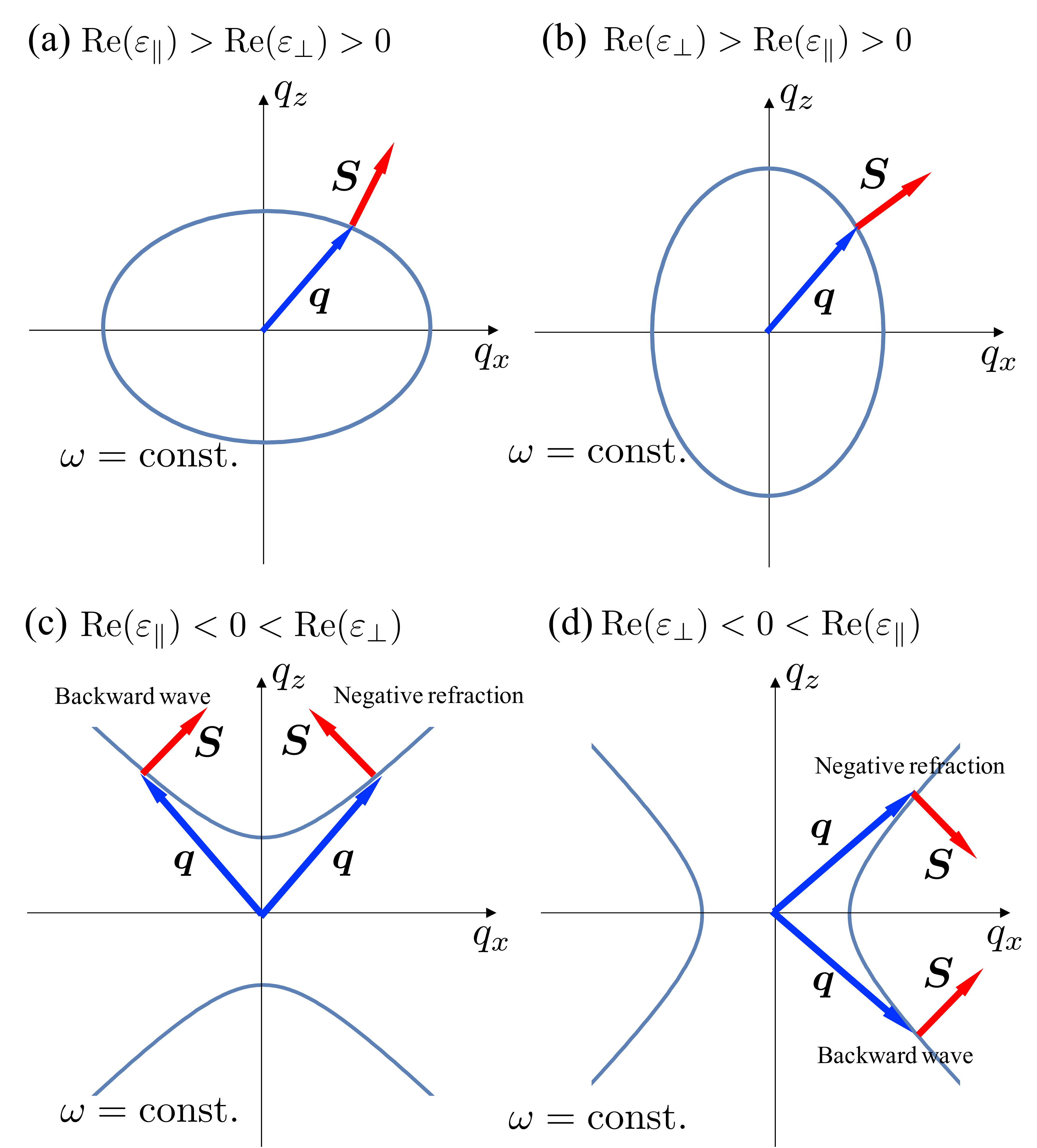}
\caption{Schematic illustration of equifrequency dispersion curve for 
(a) ${\rm Re}(\vare_{\parallel})>{\rm Re}(\vare_{\perp})>0$, 
(b) ${\rm Re}(\vare_{\perp})>{\rm Re}(\vare_{\parallel})>0$, 
(c) ${\rm Re}(\vare_{\parallel})<0<{\rm Re}(\vare_{\perp})$, 
and 
(d) ${\rm Re}(\vare_{\perp})<0<{\rm Re}(\vare_{\parallel})0$.  
In cases (a) and (b), an extraordinary wave can propagate in the medium, 
where the wave vector $\qv$ and the Poynting vector ${\bm S}$  
are refracted to the positive side but are not parallel to each other. 
Negative refraction occurs 
when the medium interface involves the (c) $x$- or (d) $z$-axis. 
A backward wave occurs when the medium interface involves the (c) $z$- or (d) $x$-axis. 
}
\label{Sec3-EW-NR-BW}
\end{figure}

For an isotropic metal or semiconductor, 
the first term $\vare^{(0)}_{ij}(\omega)$ in Eq.~(\ref{expand-epsilon})
describes the conventional symmetric tensor, 
which takes the diagonal form of $\vare_{ij}^{(0)}(\omega)=\vare(\omega)\delta_{ij}$, 
with $\vare(\omega)$ and $\delta_{ij}$ being a complex function of $\omega$ and the Kronecker delta in three dimensions, respectively. 
In this case, the wave equation (\ref{weq-a}) becomes 
\begin{align}
\begin{pmatrix}
c^2q_{z}^2-\omega^2\vare& 0& -c^2q_{x}q_{z}\\
0& c^2q^2-\omega^2\vare & 0\\
-c^2q_{x}q_{z} & 0& c^2q_{x}^2-\omega^2\vare
\end{pmatrix}
\begin{pmatrix}
E_{x}\\
E_{y}\\
E_{z}
\end{pmatrix}
=0. 
\end{align}
The plane wave solution exists if and only if $\qv$ and $\omega$ satisfy the 
characteristic equation 
\begin{align}
(c^2q^2-\omega^2\vare)^2\omega^2=0, 
\end{align}
which yields the dispersion relations $q= q_{\rm TM}(\omega)$ and $q=q_{\rm TE}(\omega)$, 
where  
\begin{align}
q_{\rm TM}(\omega )=q_{\rm TE}(\omega) = \frac{\omega}{c}\sqrt{\vare(\omega)}. 
\end{align}
For $q_{\rm TM}(\omega)$, the eigen vector is given by 
${\bm E}_{\rm TM} = E_{0}{\bm e}_{y}$ with $E_{0}$ being a scalar. 
This solution represents the linearly polarized wave. 
On the other hand, for $q_{\rm TE}(\omega)$, 
the eigen vector is given by 
${\bm E}_{\rm TE} = E_{0x}{\bm e}_{x} + E_{0z}{\bm e}_{z}$, 
where $E_{0x}$ and $E_{0z}$ are components 
of the electric field vector satisfying 
the orthogonality condition $\qv \cdot {\bm E}_{\rm TE}=q_{x}E_{0x}+q_{z}E_{0z}=0$.  
This wave is linearly polarized in the $x$-$z$ plane. 
Thus, in the case of an isotropic medium, it is apparent that conventional wave propagation is obtained 
in the frequency region satisfying ${\rm Re}(\vare(\omega))>0$. 
However, if the medium is anisotropic in nature,   
the property of the wave propagation changes dramatically. 
In the case of a uniaxially anisotropic medium for which the optic axis is the $z$-axis, 
the dielectric tensor has the form 
\begin{align}
\vare^{(0)}_{ij} &= 
\begin{pmatrix}
\vare_{\perp}&0&0\\
0&\vare_{\perp}&0\\
0&0&\vare_{\parallel}
\end{pmatrix},  
\label{dielectric-0-1}
\end{align}
where $\vare_{\perp}$ and $\vare_{\parallel}$ 
are the dielectric constants in the perpendicular and parallel directions, 
respectively, with respect to the anisotropy (optic) axis. 
Substituting this tensor into the wave equation (\ref{weq-a}), we have 
\begin{align}
\label{weq-2}
\begin{pmatrix}
c^2q_{z}^2-\omega^2 \vare_{\perp}& 0& -c^2q_{x}q_{z}\\
0& c^2q^2-\omega^2\vare_{\perp}&0\\
-c^2q_{x}q_{z}& 0& c^2q_{x}^2-\omega^2\vare_{\parallel}
\end{pmatrix}
\begin{pmatrix}
E_{x}\\
E_{y}\\
E_{z}
\end{pmatrix}
=0. 
\end{align}
The characteristic equation is given by 
\begin{align}
(c^2q^2-\omega^2\vare_{\perp})(
\vare_{\perp}c^2q^2_{x}+\vare_{\parallel}c^2q_{z}^2-\omega^2\vare_{\perp}\vare_{\parallel}
)\omega^2=0,
\end{align}
which yields two types of wave propagation, i.e., ordinary and extraordinary waves, respectively. 
For the ordinary wave,  
the dispersion relation is $cq=cq_{\rm o}(\omega) 
= \omega\sqrt{\vare_{\perp}(\omega)}$ and the eigen vector is 
${\bm E}_{\rm o} = E_{0}{\bm e}_{y}$. 
This wave is linearly polarized in the $y$-direction 
and can propagate in a conducting medium for ${\rm Re}(\vare_{\perp})>0$. 
On the other hand, for an extraordinary wave, 
the dispersion relation is given by
\begin{align}
\frac{q_{x}^2}{\vare_{\parallel}}+\frac{q_{z}^2}{\vare_{\perp}} = \frac{\omega^2}{c^2}, 
\label{i-disp1}
\end{align}
and the eigen vector is given by 
${\bm E}_{\rm e} = E_{0x}{\bm e}_{x} + E_{0z}{\bm e}_{z}$, which satisfies 
\begin{align}
(c^2q_{z}^2-\omega\vare_{\perp})E_{0x} = c^2q_{x}q_{z}E_{0z}. 
\end{align}
Thus, the orthogonality condition is not satisfied, i.e., 
${\bm E}_{\rm e} \cdot \qv \ne 0$. This means that 
the wavefront propagation direction $\qv$ and the Poynting vector 
${\bm S}\sim {\bm E}\times{\bm B}$ are not parallel. 
As noted above, such a wave is called an extraordinary wave. 
When ${\rm Re}(\vare_{\perp})>0$ and ${\rm Re}(\vare_{\parallel})>0$, 
the equifrequency contour of Eq.~(\ref{i-disp1}) is elliptical in the 
$(q_{x},q_{z})$-plane, the wave propagation of which is conventional 
\cite{Landau} (Fig.~\ref{Sec3-EW-NR-BW}(a,b)), 
where $\qv$ and ${\bm S}$ are refracted to the positive side. 
On the other hand, when ${\rm Re}(\vare_{\perp})$ and ${\rm Re}(\vare_{\parallel})$ have opposite sign, 
the equifrequency dispersion curve becomes hyperbolic, as illustrated in 
Fig.~\ref{Sec3-EW-NR-BW}(c, d). 
Both figures indicate that the transverse component of ${\bm S}$ 
can have the opposite sign to that of $\qv$. 
This indicates that the energy flow of an obliquely incident wave 
is refracted to the negative side with respect to the interface normal of the medium. 
This unusual optical phenomenon is called ``negative refraction'' 
\cite{{Lindell01},{Smith03},{Belov03}}. 
On the other hand, although the energy flow direction should be positive, 
it is possible for the vertical component of the wave vector to be negative.  
Such an optical phenomenon is called a ``backward wave'' \cite{{Lindell01},{Smith03},{Belov03}}. 

Recently, materials possessing hyperbolic dispersion and known as ``hyperbolic materials'' have become the focus of research attention 
\cite{{JOpt12},{Poddubny13},{PQE15}}. 
These materials consist of a metal-dielectric multilayer 
\cite{{Hoffman07},{Liu08},{Hoffman09},{Harish12}} 
and natural materials \cite{{Poddubny13},{Esslinger14},{Narimanov15}}. 
In the case of a Rashba conductor, 
the existence of ${\bm \alpha}$ implies 
that the system has, at least, a uniaxial anisotropy. The dielectric tensor $\vare^{(0)}_{ij}$ takes the uniaxial form, 
$\vare^{(0)}_{ij}= \vare_{\perp}(\delta_{ij}-\hat{\alpha}_{i}\hat{\alpha}_{j})+ \vare_{\parallel}\hat{\alpha}_{i}\hat{\alpha}_{j}$.  
In Sec. VIII, we demonstrate that ${\bm \alpha}$ enhances the anisotropy,   
suggesting that a medium with large Rashba spin-split bands becomes a kind of hyperbolic material. Such a medium is expected to exhibit unusual electromagnetic wave propagation behavior, as demonstrated in Sec.~VIII.

\subsection{Effects of $\alpha_{ijk}q_{k}$ caused by broken space inversion symmetry}

When the system breaks the space inversion symmetry, 
the second term in Eq.~(\ref{expand-epsilon}), $\alpha_{ijk}q_{k}$, appears in the expression for $\vare_{ij}$. Recall that $\alpha_{ijk}$ is antisymmetric under $i \leftrightarrow j$. 
Such $\qv$-linear off-diagonal components  
originate from the electromagnetic cross-correlation effects 
and generate two types of ${\bm q}$-induced optical phenomena, 
known as ``natural optical activity'' \cite{{Barron82},{Agranovich84}} 
and ``linear birefringence'' \cite{{Raab05}}.

\subsubsection{Optical activity}

The term ``optical activity'' refers to the phenomenon in which the ${\bm E}$  of the linearly polarized light 
is rotated around $\qv$ (Fig.~\ref{Sec1-OP}(a)) \cite{Barron82}.  
This indicates that the refractive indexes of the left- and right-handed circularly polarized waves differ. 
This difference occurs when the antisymmetric part of the $\vare_{ij}$ expression takes the form 
\begin{align}
\alpha_{ijk}(\omega)q_{k} = -i\frac{c}{\omega}\alpha_{\rm OA}(\omega)\vare_{ijk}q_{k}, 
\end{align}
where $\alpha_{\rm OA}(\omega)$ is a complex function of $\omega$.  
To examine this behavior more closely, we consider the case of  
$\qv = q_{z}{\bm e}_{z}$ and $\vare_{ij}^{(0)}(\omega)= 
\vare_{\perp}(\omega)(\delta_{ix}\delta_{jx}+\delta_{iy}\delta_{jy})+\vare_{\parallel}(\omega)\delta_{iz}\delta_{jz}$. 
Thus, the dielectric tensor is given by 
\begin{align}
\vare^{(0)}_{ij} + \alpha_{ijk}q_{k} = 
\begin{pmatrix}
\vare_{\perp}&-ic\alpha_{\rm OA}q_{z}/\omega&0\\
ic\alpha_{\rm OA}q_{z}/\omega&\vare_{\perp}&0\\
0&0&\vare_{\parallel} 
\end{pmatrix}  . 
\end{align}
This setup yields the wave equation 
\begin{align}
\begin{pmatrix}
c^2q_{z}^2-\omega^2\vare_{\perp}&ic\omega\alpha_{\rm OA}q_{z} & 0\\
-ic\omega\alpha_{\rm OA} q_{z}& c^2q^2_{z}-\omega^2\vare_{\perp}&0\\
0& 0& -\omega^2\vare_{\parallel}
\end{pmatrix}
\begin{pmatrix}
E_{x}\\
E_{y}\\
E_{z}
\end{pmatrix}
=0, 
\label{weq-oa}
\end{align}
which in turn yields 
\begin{align}
\left\{(c^2q_{z}^2-\omega^2\vare_{\perp})^2-c^2\omega^2\alpha_{\rm OA}^2q_{z}^2\right\}
\omega^2\vare_{\parallel}=0. 
\end{align}
As $E_{z}=0$, the plane wave solution exhibits a transverse wave propagating 
in the $z$-direction. 
Solving for $q_{z}$, we obtain the dispersion relations 
\begin{align}
q_{\rm \pm} &=  
\frac{\omega}{c} \left\{
\sqrt{\vare_{\perp} + (\alpha_{\rm OA}/2)^2 } \pm (\alpha_{\rm OA}/2) 
\right\} . 
\label{dispersion-oa}
\end{align}
Substituting this expression into Eq.~(\ref{weq-oa}), 
we then find the eigen vector 
\begin{align}
{\bm E}_{\pm} = \frac{1}{\sqrt{2}}({\bm e}_{x} \pm i {\bm e}_{y})E_{0}, 
\end{align}
which represents the left- ($+$) and right-handed ($-$) 
circularly polarized waves, respectively. 
In this study, we state that a circularly polarized plane wave is left-handed (right-handed), 
${\bm E}_{\pm}(z,t)={\bm E}_{\pm}e^{i(q_{\pm}z-\omega t)}$, 
if the electric field vector rotates counter-clockwise (clockwise) 
at a fixed point when viewed from the wave propagation direction\cite{Jackson}.

Let us consider the optical rotation of the ${\bm E}$ of a linearly polarized wave and evaluate the rotation angle. 
If an incident wave is initially polarized in the $x$-direction in a vacuum, with 
${\bm E}(z,0) = E_{0}{\bm e}_{x}e^{i\omega z/c}$, 
the wave passing through the conductor is given by 
\begin{align}
{\bm E}(z,t) &= \frac{1}{\sqrt{2}}
\left({\bm E}_{+}e^{i(q_{+}z-\omega t)}+{\bm E}_{-}e^{i(q_{-}z-\omega t)}\right),\nonumber\\
&= E_{0}\left\{
{\bm e}_{x}\cos\psi(z)+{\bm e}_{y}
\sin\psi(z)
\right\}
e^{i\{(q_{+}+q_{-})z/2 -\omega t\}}, 
\end{align}
where 
\begin{align}
\psi(z) = \theta_{\rm OA}(z)+i\kappa_{\rm OA}z, 
\end{align}
with 
\begin{align}
\label{theta-OA}
&\theta_{\rm OA}(z) = \frac{{\rm Re}(q_{+}-q_{-})z}{2}
={\rm Re}(\alpha_{\rm OA})\frac{\omega}{2c}z,\\
\label{kappa-OA}
&\kappa_{\rm OA}=\frac{{\rm Im}(q_{+}-q_{-})}{2}
={\rm Im}(\alpha_{\rm OA})\frac{\omega}{2c}. 
\end{align}
Thus, the rotation angle following passage through a conductor of thickness $d$ is given by $\theta_{\rm OA}(d)$. 
Note that this optical rotation is reciprocal, 
i.e., the ${\bm E}$ of the wave propagating backward ($q_z<0$) in the sample 
rotates in the opposite direction ($ \theta_{\rm OA} (-d) = -\theta_{\rm OA}(d)$) to that of the forward wave ($q_z>0$).
This reciprocity originates from the fact that the present effect comes from the $\qv$-linear term 
in the {\it off-diagonal component} of the dielectric tensor. 
Indeed, the electric current induced by this $\qv$-linear term 
has the form ${\bm j}_{\rm OA} \sim \alpha_{\rm OA}\qv\times{\bm E}$, 
indicating that the electric field vector rotates around the $\qv$ vector. 

The imaginary part of $\psi$ represents 
the difference in absorption between left- and right-handed circularly polarized waves, 
which is called ``(optical) circular dichroism.'' 
Thus, the absorption rate per unit length is given by $\kappa_{\rm OA}$ in Eq.~(\ref{kappa-OA}), 
which is proportional to the imaginary part of $\alpha_{\rm OA}$.

\subsubsection{Linear birefringence}

Linear birefringence is a phenomenon known as 
``double refraction,'' which is due to two types of linearly polarized waves, i.e., an ordinary and extraordinary wave. 
The polarization plane of the latter, which is defined by ${\bm E}$ and ${\bm B}$ vectors, 
is rotated by the $\qv$-linear term in the off-diagonal component of 
$\vare_{ij}$ (Fig.~\ref{Sec1-OP}(b))\cite{Raab05}.  
This phenomenon occurs when the antisymmetric part of $\vare_{ij}$ takes the form 
\begin{align}
\alpha_{ijk}(\omega)q_{k} &= -i\frac{c}{\omega}\alpha_{\rm LB}(\omega)
\vare_{ijk}(\hat{\bm c}\times\qv)_{k},\nonumber\\
&=-i\frac{c}{\omega}\alpha_{\rm LB}(\omega)
(q_{i}\hat{c}_{j} - \hat{c}_{i}q_{j}), 
\label{dielectric-lb}
\end{align} 
where $\alpha_{\rm LB}(\omega)$ is a complex function of real $\omega$ 
and $\hat{\bm c}$ is a unit vector pointing to the optic axis of the medium. 
Setting $\hat{\bm c} = {\bm e}_{z}$ and $\qv = q_{x}{\bm e}_{x}$, 
we obtain 
\begin{align}
\vare^{(0)}_{ij} + \alpha_{ijk}q_{k} 
= 
\begin{pmatrix}
\vare_{\perp}&0&-ic\alpha_{\rm LB}q_{x}/\omega\\
0&\vare_{\perp}&0\\
ic\alpha_{\rm LB}q_{x}/\omega&0&\vare_{\parallel}
\end{pmatrix}.
\end{align}
Thus, the wave equation is given by 
\begin{align}
\begin{pmatrix}
-\omega^2\vare_{\perp}&0&ic\omega\alpha_{\rm LB} q_{x} \\
0& c^2q^2_{x}-\omega^2\vare_{\perp}&0\\
-ic\omega\alpha_{\rm LB} q_{x}& 0& c^2q^2_{x}-\omega^2\vare_{\parallel}
\end{pmatrix}
\begin{pmatrix}
E_{x}\\
E_{y}\\
E_{z}
\end{pmatrix}
=0, 
\label{weq-lb}
\end{align}
which yields
\begin{align}
(c^2q_{x}^2-\omega^2\vare_{\perp})\left\{
c^2(\vare_{\perp}+\alpha_{\rm LB}^2)q_{x}^2-\omega^2\vare_{\perp}\vare_{\parallel}
\right\}\omega^2=0. 
\end{align}
There are two types of solution, i.e., $q_x=q_{\rm o}$ and $q_{\rm e}$, where 
\begin{align}
&q_{\rm o}= \frac{\omega}{c}\sqrt{\vare_{\perp}},\\
&q_{\rm e} = 
\frac{\omega}{c}\sqrt{\frac{\vare_{\perp}\vare_{\parallel}}
{\vare_{\perp}+\alpha^{2}_{\rm LB}}}. 
\end{align}
The former represents the ordinary wave, the eigen vector of which is 
${\bm E}_{\rm o} = E_{0}{\bm e}_{y}$, and 
the latter represents the extraordinary wave, 
${\bm E}_{\rm e} = E_{0x}{\bm e}_{x} + E_{0z}{\bm e}_{z}$, with
\begin{align}
E_{0x} = i \frac{\omega}{c}q_{\rm e}\frac{\alpha_{\rm LB}}{\vare_{\perp}}E_{0z}, 
\label{eigen-lb}
\end{align}
which yields $\qv\cdot{\bm E}_{\rm e}\neq 0$. 
Thus, the off-diagonal components of $\vare_{ij}$
induce a longitudinal component of the electric field. 
Hence, 
the direction of ${\bm S}$ (the energy flow) of the extraordinary 
wave is not parallel to $\qv$.

The degree of birefringence, $n_{\rm LB}$,  
is defined by the difference in the refractive indexes of the ordinary and extraordinary waves 
\cite{Raab05}, such that 
\begin{align}
n_{\rm LB}=\frac{\omega}{c}\left[{\rm Re}(q_{\rm o})-{\rm Re}(q_{\rm e})\right] 
\simeq {\rm Re}(\sqrt{\vare_{\perp}})-{\rm Re}(\sqrt{\vare_{\parallel}}) . 
\end{align}
The second equality follows when $|\alpha_{\rm LB}|^2\ll |\vare_{\perp}|$.  
On the other hand, the tilt angle of the polarization plane (spanned by ${\bm E}$ and ${\bm B}$) 
for the extraordinary wave, 
which is denoted by $\theta_{\rm LB}$, is first order in $|\alpha_{\rm LB}|$, with
\begin{align}
\theta_{\rm LB} = 
\tan^{-1}\frac{{\rm Re}(E_{0x})}{{\rm Re}(E_{0z})}
\simeq {\rm Re}\left(i\frac{\sqrt{\vare_{\parallel}}}{\vare_{\perp}}\alpha_{\rm LB}\right).  
\end{align} 

In the case of a Rashba conductor, 
the $\alpha_{ijk}q_{k}$ term originates from the combination of the direct and inverse Edelstein effects. The induced current has the form ${\bm j}_{\rm E-IE}\sim ({\bm \alpha}\times\qv)\times {\bm E}$, 
meaning that the electric field vector rotates around the vector ${\bm \alpha}\times\qv$. 
Thus, linear birefringence is expected to occur in the bulk Rashba conductor. 
This is indeed the case, as shown in Sec.~VIII-B.

\subsection{Effects of $\beta_{ijk}M_{k}$ caused by broken time reversal symmetry}

When the system breaks the time reversal symmetry, 
there appears an ${\bm M}$-linear term, $\beta_{ijk}M_{k}$, in the electric tensor, which originates from the anomalous Hall effect. The dielectric tensor takes the form 
\begin{align}
\beta_{ijk}(\omega)M_{k} = \beta_{\rm AH}(\omega)\vare_{ijk}M_{k}, 
\label{tensor-Faraday}
\end{align}
where $\beta_{\rm AH}$ is a complex function of $\omega$.  
Similar to the case of the previous subsection, 
the off-diagonal components of the dielectric tensor yield two types of ${\bm M}$-induced optical 
phenomena, which are known as the Faraday effect and the Cotton-Mouton effect. 
The former corresponds to the optical activity and the latter to the linear birefringence.  

\subsubsection{Faraday effect}

The Faraday effect refers to rotation of the ${\bm E}$ of the linearly polarized wave 
due to ${\bm M}$
(or an external dc magnetic field); this is called Faraday rotation\cite{Faraday}. 
This effect occurs when ${\bm M}$ and $\qv$ 
are parallel (or anti-parallel), i.e., they are in the Faraday configuration 
(Fig.~\ref{Sec1-OP}(c)).  
Here, setting ${\bm M} = M{\bm e}_{z}$, ${\bm q} = q_{z}{\bm e}_{z}$, 
$\vare_{ij}^{(0)}(\omega) = \vare(\omega)\delta_{ij}$, and $\alpha_{ijk}(\omega)q_{k}=0$, 
we express the wave equation (\ref{weq-a}) as  
\begin{align}
\begin{pmatrix}
c^2q_{z}^2-\omega^2\vare&-\omega^2M\beta_{\rm AH}& 0\\
\omega^2M\beta_{\rm AH}& c^2q^2_{z}-\omega^2\vare&0\\
0& 0& -\omega^2\vare
\end{pmatrix}
\begin{pmatrix}
E_{x}\\
E_{y}\\
E_{z}
\end{pmatrix}
=0, 
\label{weq-FK}
\end{align}
which yields 
\begin{align}
\left\{
(c^2q_{z}^2-\omega^2\vare)^2+\omega^4M^2\beta_{\rm AH}^2
\right\}\omega^2\vare=0. 
\end{align}
Hence, we obtain the dispersion relation $q_{z} = q^{M}_{\pm}(\omega)$, where
\begin{align}
q^{M}_{\pm}(\omega) = \frac{\omega}{c}\sqrt{\vare_{\pm}(\omega)}, 
\end{align}
with
\begin{align}
\vare_{\pm}(\omega) = \vare(\omega)\pm i\beta_{\rm AH}(\omega)M. 
\end{align}
We also obtain the eigen vector 
\begin{align}
{\bm E}_{\pm} = \frac{1}{\sqrt{2}}({\bm e}_{x} \pm i {\bm e}_{y})E_{0}, 
\end{align}
The above expressions are for the left- ($+$) and right-handed ($-$) 
circularly polarized waves.  
Thus, similar to the optical activity case, 
${\bm E}$ rotates in the $x$-$y$ plane. 
The rotation angle after passage through a medium of thickness $d$, 
$\theta^{M}_{\rm F}(d)$, is given by 
\begin{align}
\label{Sec3-Faraday-angle}
\theta^{M}_{\rm F}(d) = 
\frac{{\rm Re}(q^{M}_{+}-q^{M}_{-})d}{2}
\simeq {\rm Re}\left(
\frac{i\beta_{F}}{\sqrt{\vare}}
\right)M\frac{\omega}{2c}d. 
\end{align}
Note that, contrary to the case of optical rotation,  
this ${\bm M}$-induced rotation does not depend on the propagating direction, 
where the polarization vectors of the forward and backward waves 
rotate in the same direction ($\theta^{M}_{\rm F}(d)= \theta^{-M}_{\rm F}(-d)$).  
Indeed, the ${\bm M}$-induced current ${\bm j}_{\rm F} \sim \beta_{\rm M}{\bm M}\times{\bm E}$, 
which does not depend on $\qv$. 

The imaginary part of $q^{M}_{+}-q^{M}_{-}$ 
represents the difference in absorption between the left- and right-handed circularly 
polarized waves, which causes magnetic circular dichroism (MCD). 
The MCD per unit length $\kappa_{\rm MCD}$ is given by 
\begin{align}
\label{sec-3-MCD}
\kappa_{\rm MCD}
=\frac{{\rm Im}(q^{M}_{+}-q^{M}_{-})}{2}
 \simeq {\rm Im}\left(
\frac{i\beta_{\rm AH}}{\sqrt{\vare}}
\right)M\frac{\omega}{2c}. 
\end{align}

In the case of a ferromagnetic Rashba conductor,  
this term ($\sim \beta_{ijk}M_{k}$) is derived from the anomalous Hall effect 
due to the RSOI and the exchange field (${\bm M}$), 
in which the induced current takes the form  
${\bm j}_{\rm AH} \sim ({\bm \alpha}\cdot{\bm M}){\bm \alpha}\times {\bm E}$. 
Thus, the ${\bm M}$ parallel to the Rashba field contributes to 
the Faraday rotation and the MCD. 
The relevant details are given in Sec.~IX-A.

\subsubsection{Cotton-Mouton effect}

The ``Voigt configuration'' is obtained when ${\bm M}$ is perpendicular to $\qv$ (Fig.~\ref{Sec1-OP}(d)). In that case, a similar phenomenon to the linear birefringence occurs, 
which is called the ``Cotton-Mouton effect.'' \cite{Cotton-Mouton}
Here, we set ${\bm M} = M{\bm e}_{z}$, ${\bm q} = q_{x}{\bm e}_{x}$, 
$\vare_{ij}^{(0)}(\omega) = \vare(\omega)\delta_{ij}$, and $\alpha_{ijk}(\omega)q_{k}=0$; then, 
\begin{align}
\begin{pmatrix}
-\omega^2\vare&-\omega^2\beta_{\rm AH}M& 0\\
 \omega^2\beta_{\rm AH}M& c^2q^2_{x}-\omega^2\vare&0\\
0&0& c^2q_{x}^2-\omega^2\vare
\end{pmatrix}
\begin{pmatrix}
E_{x}\\
E_{y}\\
E_{z}
\end{pmatrix}
=0. 
\label{weq-CM}
\end{align}
The characteristic equation has two types of solution. 
One is an ordinary wave with dispersion relation $q_{\rm o}(\omega) = \omega
\sqrt{\vare(\omega)}$ and eigen vector 
${\bm E}_{\rm o} = E_{0}{\bm e}_{z}$. 
The other is an extraordinary wave, 
with dispersion relation 
\begin{align}
q_{\rm CM} = \frac{\omega}{c}\sqrt{\vare + 
\frac{\beta^2_{\rm AH}}{\vare}M^2}.
\end{align}
The eigen vector is ${\bm E}_{\rm CM} = E_{0x}{\bm e}_{x} + E_{0y}{\bm e}_{y}$ 
with 
\begin{align}
E_{0x}  = -\frac{\beta_{\rm AH}}{\vare}ME_{0y}.  
\end{align}
Thus, similar to the linear birefringence, 
${\bm E}$ rotates around ${\bm M}$ and the electric field acquires a longitudinal component. 
The ${\bm M}$-induced birefringence, $n_{\rm CM}$,  is thus given by 
\begin{align}
n_{\rm CM} = \frac{\omega}{c}{\rm Re}(q_{\rm e}-q_{\rm o})
\simeq \frac{\omega}{2c}{\rm Re}\left(
\frac{\beta_{\rm AH}^2}{\sqrt{\vare}\vare} . 
\right) M^2 . 
\end{align}
The second equality holds when $|\vare_{\rm CM}|M \ll |\vare|$. 
Thus, the ${\bm M}$-induced birefringence is the second-order effect in $M$. 
On the other hand, the tilt angle, 
\begin{align}
\theta_{\rm CM} \simeq -{\rm Re}\left(\frac{\beta_{\rm AH}}{\vare}\right)M, 
\end{align} 
is first-order in $M$.

\subsection{Effects of $\gamma_{ijkl}q_{k}M_{l}$ caused by 
broken space inversion and time reversal symmetries}

When the system simultaneously breaks the space inversion and the time reversal symmetries, 
the fourth term in Eq.~(\ref{expand-epsilon}), $\gamma_{ijkl}q_{k}M_{l}$, 
appears as a leading term that expresses ${\bm M}$-induced spatial dispersion. 
Such media exhibit magneto-chiral birefringence and dichroism 
for the Faraday configuration (Fig.~\ref{Sec1-MOP}(a, b)), 
or 
nonreciprocal directional birefringence and dichroism 
for the Voigt configuration (Fig.~\ref{Sec1-MOP}(c, d)).    
Here, the term ``nonreciprocal'' refers to the directional dependence of the wave  propagation 
between the two counter-propagating waves ($\qv \leftrightarrow -\qv$) (Fig.~\ref{Sec1-MOP}).  
These phenomena are described by the
diagonal components of the dielectric tensor 
bilinear in $\qv$ and ${\bm M}$\cite{{Rikken01},{Rikken05}}: 
\begin{align}
\gamma_{ijkl}(\omega)q_{k}M_{l} = 
\begin{cases}
\gamma^{\rm F}_{ij}(\omega)\qv\cdot{\bm M}~~{\rm (Faraday)},\\
\gamma^{\rm V}_{ij}(\omega)
\qv\cdot({\bm P} \times {\bm M})~~{\rm (Voigt)},
\end{cases}
\label{dielectric4-1}
\end{align}
where $\gamma^{\rm F}_{ij}(\omega)$ and $\gamma^{\rm V}_{ij}(\omega)$ 
are symmetric tensors  
and ${\bm P}$ is a polar vector representing 
a spontaneous polarization or an external static electric field. 
The former (Faraday configuration) yields the magneto-chiral dichroism (MChD) 
and birefringence (MChB), 
which have been observed in the chiral molecule \cite{{Portigal71}, {Barron84}}, 
chiral ferromagnets \cite{{Rikken97}, {Wagniere98}, {Rikken98}, {Wagniere99}, {Vallet01}, {Train08}}, 
and artificial media \cite{Tomita14}. 
The latter (Voigt configuration) generates 
nonreciprocal linear birefringence \cite{{HS68},{Krichevtsov00},{Raab05}} 
and 
nonreciprocal directional dichroism (NDD) \cite{{Tokura11},{Mochizuki13},{Furukawa14}}.  

To demonstrate these phenomena, we consider the following diagonal components: 
\begin{align}
\gamma_{ijkl}(\omega)q_{k}M_{l} = -2\gamma_{\rm N}(\omega)\omega c M q_{z} \delta_{iy}\delta_{jy},  
\end{align}
where $\gamma_{\rm N}(\omega)$ is a complex function of $\omega$, 
and we set ${\bm q}=q_z {\bm e}_z$.  
Assuming $\vare^{(0)}_{ij} = \vare\delta_{ij}$, 
$\alpha_{ijk}\sim 0$, and $\beta_{ijk}\sim 0$ for simplicity, 
we obtain the wave equation 
\begin{align}
\begin{pmatrix}
c^2q_{z}^2-\omega^2\vare&0& 0\\
0& c^2q^2_{z}-\omega^2\vare-2\gamma_{\rm N}c\omega Mq_{z} &0\\
0& 0& -\omega^2\vare
\end{pmatrix}
\begin{pmatrix}
E_{x}\\
E_{y}\\
E_{z}
\end{pmatrix}
=0, 
\label{weq-ndd}
\end{align}
which yields
\begin{align}
(c^2q_{z}^2-\omega^2\vare)(c^2q_{z}^2-\omega^2\vare-\omega^2\vare-\gamma_{\rm N}c\omega Mq_{z} )\omega^2\vare=0. 
\label{chara-eq}
\end{align}
For $E_{y}\neq 0$, we have 
\begin{align}
c^2q^2_{z} - 2\omega\gamma_{\rm N}Mcq_{z}-\omega^2\vare=0. 
\label{chara-eq_Ey}
\end{align} 
Note that the second term ($\sim Mq_z$) changes sign 
depending on the sign of $\qv\cdot{\bm M}$ or $\qv\cdot({\bm P}\times {\bm M})$ in the case of the Faraday or Voigt configurations, respectively. 
The corresponding dispersion relation is given by 
\begin{align}
&q_{\pm}(\omega) = \frac{\omega}{c}\left(
\sqrt{\vare+\gamma^2_{\rm N}M^2} \pm \gamma_{\rm N} M
\right). 
\end{align}

The magnitude of the nonreciprocal directional birefringence is expressed as 
\begin{align}
n_{\rm N} = \frac{c}{\omega}{\rm Re}(q_{+}-q_{-}) = 2{\rm Re}(\gamma_{\rm N})M , 
\end{align}
and 
that of the nonreciprocal directional dichroism as
\begin{align}
\kappa_{\rm N} = \frac{c}{\omega}{\rm Im}(q_{+}-q_{-}) = 2{\rm Im}(\gamma_{\rm N})M. 
\end{align}

In the case of a ferromagnetic Rashba conductor, ${\bm \alpha}$ is a polar vector similar to ${\bm P}$. 
Thus, it is possible that the $\vare_{ij}$ expression contains the term $\gamma_{ijkl}q_{k}M_{l} 
= \gamma^{\rm R}_{ij}\qv\cdot ({\bm \alpha}\times {\bm M})$, 
and a nonreciprocal wave propagation can be expected.  
The details are presented in Sec.~IX.

\section{Microscopic Model and Formulation}

In this section, we consider a ferromagnetic bulk Rashba conductor 
as a concrete microscopic model  
and formulate the calculation of the current and spin responses induced by 
a space- and time-varying electromagnetic field 
on the basis of linear response theory 
and using path-ordered Green's functions. 

\subsection{Hamiltonian and Green's function}

\begin{figure}
\begin{center}
\resizebox{8.4cm}{!}{\includegraphics[angle=0]{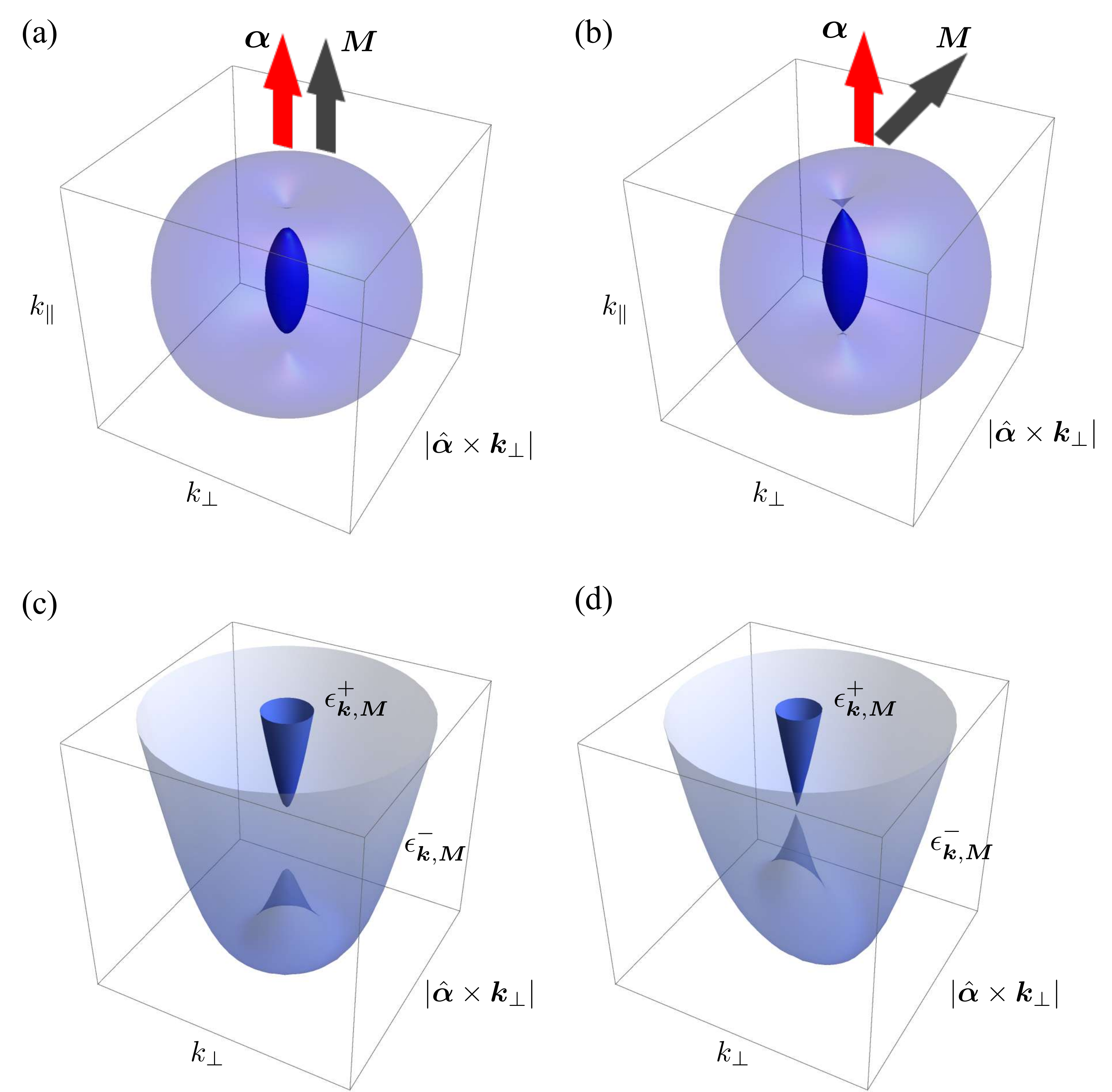}}
\end{center}
\caption{(a, b) Fermi surfaces in $\kv$-space and (c, d) energy dispersions in $\kv_{\perp}$-plane 
for ${\bm \alpha}\times{\bm M}={\bm 0}$ (a, c) and 
${\bm \alpha}\times {\bm M}\neq{\bm 0}$ (b, d).} 
\label{FE-ED}
\end{figure}

We consider a ferromagnetic bulk Rashba conductor, 
in which the conduction electrons experience a momentum-dependent spin orbit interaction 
characterized by a certain Rashba field, ${\bm \alpha}$, 
and  
an exchange field due to some magnetization ${\bm M}$. 
The former and latter break the space inversion and time reversal symmetries, respectively. 
We assume that the system  
has uniaxial anisotropy, with the anisotropy axis set by the Rashba field such that $\hat{\bm \alpha}={\bm \alpha}/|{\bm \alpha}|$\cite{{Tsutsui-Murakami12},{Ye-Tokura15},{Maiti15}}. 
Then, the Hamiltonian for the conduction electrons is given by 
\begin{subequations}
\label{Hamiltonian}
\begin{align}
H &= \sum_{\kv}c^{\dagger}_{\kv}({\cal H}_{\kv}-\epsilon_{\rm F})c_{\kv}, 
\label{H}\\
{\cal H}_{\kv} &= \epsilon_{\kv}
+ {\bm \alpha}\cdot ({\bm \sigma}\times \kv)
-{\bm M}\cdot{\bm \sigma}, \label{Hk}\\
\epsilon_{\kv} &= \dfrac{\hbar^2 \kv^2_{\perp}}{2m_{\perp}} + \dfrac{\hbar^2\kv^2_{\parallel}}{2m_{\parallel}}, 
\end{align}
\end{subequations}
where $c^{\dagger}_{\kv} = 
(c^{\dagger}_{\kv,\uparrow},c^{\dagger}_{\kv,\downarrow})$ 
is the electron creation operator with wave vector $\kv$ and 
spin projection ($\uparrow$ or $\downarrow$) along the $z$-axis. Further, 
$\kv_{\parallel}=(\hat{\bm \alpha}\cdot \kv)\hat{\bm \alpha}$ and 
${\bm k}_{\perp}= \kv - \kv_{\parallel}$ are parallel and perpendicular components of ${\bm k}$, respectively, 
with respect to $\hat{\bm \alpha}$. In addition, $m_{\parallel}$ and $m_{\perp}$ are the effective masses in the respective directions \cite{perp}, 
$\epsilon_{\rm F}$ is the Fermi energy, and 
${\bm \sigma}=(\sigma_{x},\sigma_{y},\sigma_{z})$ 
is a vector of Pauli spin matrices. 
The direction of ${\bm M}$ is arbitrary and its magnitude, 
$M=|{\bm M}|$, is taken to have the unit of energy in this study. 
The eigenenergies of ${\cal H}_{\kv}$ [Eq.~(\ref{Hk})] are given by 
\begin{align}
\epsilon_{\kv, {\bm M}}^{s}  &= \epsilon_{\kv}
+s \sqrt{\alpha^2\kv_{\perp}^2-2
({\bm \alpha}\times  {\bm M})\cdot \kv_{\perp} + {\bm M}^2}, 
\label{ek}
\end{align} 
where 
$s = \pm 1$ specifies the 
spin-split upper ($s = 1$) and lower ($s = -1$) bands.  
The Fermi surfaces in $\kv$-space and
 the energy dispersions in the $\kv_{\perp}$-plane  
for ${\bm \alpha}\times{\bm M}={\bm 0}$ and 
$\neq{\bm 0}$ are illustrated in Fig.~\ref{FE-ED}. 
For ${\bm \alpha}\times{\bm M}={\bm 0}$, 
the Fermi surface (Fig.~\ref{FE-ED}(a)) 
has rotation invariance around the $\hat{\bm \alpha}$-axis and, 
thus, the energy dispersion is symmetric with respect to 
the $\kv_{\perp}$-plane in $\kv$-space (Fig.~\ref{FE-ED}(b)).  
The exchange interaction removes the band degeneracy at 
$\kv_{\perp}={\bm 0}$. 
For ${\bm \alpha}\times {\bm M}\neq{\bm 0}$, 
the Fermi surface (Fig.~\ref{FE-ED}(c)) and 
energy dispersion become asymmetric 
and the energy degeneracy is recovered at the Dirac point 
satisfying  $|\kv_{\perp}\times{\bm \alpha} -{\bm M}|=0$. 
This asymmetry implies that 
the light absorption due to electronic interband transitions 
depends on whether $\kv_{\perp}$ is parallel or antiparallel 
to ${\bm \alpha}\times  {\bm M}$, 
which yields a difference in absorption 
for counterpropagating light beams relative to $\kv_{\perp}$ 
(see Sec. IX-B). 

The Green's function of electrons 
corresponding to the Hamiltonian $H$ [Eq.~(\ref{H})]  
is given by 
\begin{align}
\label{G1}
G_{\kv,{\bm M}}(z) = 
\frac{1}{z-({\cal H}_{\kv}-\epsilon_{\rm F})}
=\frac{1}{2}\sum_{s=\pm 1}
\frac{1+s\hat{\bm \gamma}_{\kv,{\bm M}}\cdot{\bm \sigma}}{z-\epsilon_{\kv,{\bm M}}^{s}+\epsilon_{\rm F}}, 
\end{align}
where 
$\hat{\bm \gamma}_{\kv,{\bm M}} = {\bm \gamma}_{\kv,{\bm M}}/|{\bm 
\gamma_{\kv,{\bm M}}}|$ 
with ${\bm \gamma}_{\kv,{\bm M}}=\kv\times{\bm \alpha}-{\bm M}$. 
The retarded and advanced Green's functions are given by 
$G^{\rm R}_{\kv,{\bm M}}(\epsilon) = G_{\kv,{\bm M}}(\epsilon+i0)$ and 
$G^{\rm A}_{\kv,{\bm M}}(\epsilon) = G_{\kv,{\bm M}}(\epsilon-i0)$, respectively, where 
$0$ is a positive infinitesimal. 
In this study, 
we consider the clean limit and use the relation 
\begin{align}
\label{R-A}
&G^{\rm A}_{\kv,{\bm M}}(\epsilon) - G^{\rm R}_{\kv,{\bm M}}(\epsilon)
\nonumber\\&=
i\pi \sum_{s = \pm 1}
\delta(\epsilon-\epsilon^{s}_{\kv,{\bm M}}+\epsilon_{\rm F})(1 + s\hat{\bm \gamma}_{\kv,{\bm M}}
\cdot{\bm \sigma}), 
\end{align}
where $\delta(x)$ is the delta function. 

\subsection{Response functions for general $\qv$ and ${\bm M}$}

Let us consider current and spin responses to 
a space- and time-varying electromagnetic field
based on linear response theory\cite{Kubo57}. 
The current and spin density operators $\hat{\bm j}(\rv)$ and
$\hat{\bm \sigma}(\rv)$, respectively, are given by 
\begin{align}
\hat{\bm j}(\rv)&=\hat{\bm j}_{0}(\rv)+\hat{\bm j}_{\rm R}(\rv)\nonumber\\
&-\frac{e^2}{m_{\perp}}\hat{n}(\rv){\bm A}_{\perp}(\rv,t)
-\frac{e^2}{m_{\parallel}}\hat{n}(\rv){\bm A}_{\parallel}(\rv,t), \\
\hat{\bm j}_{0}(\rv)&=
-e\dfrac{\hbar}{2m_{\perp}i}\{
c^{\dagger}(\rv)\nabla_{\perp}c(\rv)
-(\nabla_{\perp}c^{\dagger}(\rv))c(\rv)
\}
\nonumber\\
&-e\dfrac{\hbar}{2m_{\parallel}i}
\{
c^{\dagger}(\rv)\nabla_{\parallel}c(\rv)
-(\nabla_{\parallel}c^{\dagger}(\rv))c(\rv)
\},\\
\hat{\bm j}_{\rm R}(\rv)&=-\frac{e}{\hbar}{\bm \alpha}\times \hat{\bm \sigma}(\rv),\\
\hat{n}(\rv) &= c^{\dagger}(\rv)c(\rv),\\
\hat{\bm \sigma}(\rv) &= c^{\dagger}(\rv){\bm \sigma}c(\rv), 
\end{align}
where $c(\rv) = \sum_{\kv}e^{i\kv\cdot\rv}c_{\kv}$. Further, 
${\bm A}_{\parallel} = \hat{\bm \alpha}(\hat{\bm \alpha}\cdot{\bm A})$ 
and ${\bm A}_{\perp} = {\bm A}-{\bm A}_{\parallel}$ 
are the parallel and perpendicular components of the vector potential ${\bm A}(\rv,t)$, respectively, which yields 
the electric field ${\bm E}(\rv,t) = -\dfrac{\partial}{\partial t}{\bm A}(\rv,t)$ 
and the magnetic field ${\bm B}(\rv,t ) = \nabla \times {\bm A}(\rv, t)$. In addition, $\nabla_{\parallel} = \hat{\bm \alpha}(\hat{\bm \alpha}\cdot \nabla)$ 
and 
$\nabla_{\perp} = \nabla-\nabla_{\parallel}$. 
The electromagnetic perturbation to the linear order is described by 
\begin{align}
H_{\rm int}(t) 
&= -\int d\rv~\bigg\{
\{\hat{\bm j}_{0}(\rv)+\hat{\bm j}_{\rm R}(\rv)\}\cdot {\bm A}(\rv,t)
\nonumber\\
&-\hbar\gamma\hat{\bm \sigma}(\rv)\cdot(\nabla \times {\bm A}(\rv,t))
\bigg\}. 
\end{align}
In what follows, 
we evaluate the expectation values of 
$\hat{\bm j}(\qv)=\int d\rv e^{-i\qv\cdot\rv}\hat{\bm j}(\rv)$ 
and 
$\hat{\bm \sigma}(\qv)=\int d\rv e^{-i\qv\cdot\rv}\hat{\bm \sigma}(\rv)$ in the Fourier space. 
The Fourier components of the electromagnetic field are given by 
${\bm E}(\qv,\omega) = i\omega {\bm A}(\qv,\omega)$
and 
${\bm B}(\qv,\omega) = i\qv \times {\bm A}(\qv,\omega)$.  
The expectation values of  
$\hat{\bm j}(\qv)$ and $\hat{\bm \sigma}(\qv)$ are expressed as 
\begin{align}
\langle \hat{\bm j}(\qv) \rangle_{\omega} 
&= -e\int_{-\infty}^{\infty}\frac{d\vare}{2\pi i}{\rm Tr}\left[
\sum_{\kv}\tilde{\bm v}G^{<}_{\kv_{+},\kv_{-}}(\vare_{+},\vare_{-})
\right]\nonumber\\
&-\frac{e^2n_{\rm e}}{m_{\perp}}{\bm A}_{\perp}(\qv,\omega)
-\frac{e^2n_{\rm e}}{m_{\parallel}}{\bm A}_{\parallel}(\qv,\omega),\\
\langle \hat{\bm \sigma}(\qv) \rangle_{\omega} 
&= -e\int_{-\infty}^{\infty}\frac{d\vare}{2\pi i}{\rm Tr}\left[
\sum_{\kv}{\bm \sigma}G^{<}_{\kv_{+},\kv_{-}}(\vare_{+},\vare_{-})
\right], 
\end{align}
\begin{align}
\tilde{\bm v} = {\bm v} + {\bm \alpha}\times{\bm \sigma}/\hbar, 
\end{align}
with ${\bm v} =\hbar\kv_{\perp}/ m_{\perp} + \hbar \kv_{\parallel}/m_{\parallel}$, 
$\kv_{\pm} = \kv\pm\qv/2$, and $\vare_{\pm} = \vare \pm \omega/2$. Here, 
$n_{\rm e}$ is the equilibrium electron density and $G^{<}_{\kv_{+},\kv_{-}}(\vare_{+},\vare_{-})$ is 
the lesser component of the path-ordered Green's function \cite{Rammer86} 
\begin{align}
G_{\kv,\kv'}(\vare,\vare') &= \int_{-\infty}^{\infty}dt  \int_{-\infty}^{\infty}dt'e^{-i\vare t+i\vare't'}G_{\kv,\kv'}(t,t'),\\ 
G_{\kv,\kv'}(t,t') &= -i\langle T_{\rm C}\left[
c_{\kv}(t)c_{\kv'}(t')
\right]\rangle_{H+H_{\rm int}(t)}. 
\end{align}
In the above expression, $T_{\rm C}$ is a path-ordering operator in the complex time plane \cite{Rammer86} and the bracket 
$\langle \cdots \rangle_{H+H_{\rm int}(t)}$ represents the expectation value 
in the nonequilibrium state. 
Performing a perturbative expansion of the path-ordered Green's function with respect to 
$H_{\rm int}(t)$, we evaluate the linear response of $\hat{\bm j}(\qv)$ and $\hat{\bm \sigma}(\qv)$ 
to ${\bm E}(\qv,\omega)$ and ${\bm B}(\qv,\omega)$, such that
\begin{align}
\langle \hat{j}_{i}(\qv) \rangle_{\omega} = {\cal K}^{jj}_{ij}(\qv,\omega,{\bm M})E_{j}
-i\hbar\gamma\omega {\cal K}^{js}_{ij}(\qv,\omega,{\bm M})B_{j},\\
\langle \hat{\sigma}_{i}(\qv) \rangle_{\omega} = {\cal K}^{sj}_{ij}(\qv,\omega,{\bm M})E_{j}
-i\hbar\gamma\omega {\cal K}^{ss}_{ij}(\qv,\omega,{\bm M})B_{j},
\end{align}
where the response functions are given by 
\begin{subequations}
\label{Kij}
\begin{align}
{\cal K}^{jj}_{ij}(\qv,\omega,{\bm M}) &= \chi^{jj}_{ij}(\qv,\omega,{\bm M})\nonumber\\
&-\frac{e^{2}}{\omega+i0^{{+}}}\frac{n_{\rm e}}{m_{\perp}}\delta^{\perp}_{ij}
-\frac{e^{2}}{\omega+i0^{{+}}}\frac{n_{\rm e}}{m_{\parallel}}\hat{\alpha}_{i}\hat{\alpha}_{j},\\
{\cal K}^{js}_{ij}(\qv,\omega,{\bm M}) &= \frac{ie}{\omega+i0}\chi^{js}_{ij}(\qv,\omega,{\bm M}),\\
{\cal K}^{sj}_{ij}(\qv,\omega,{\bm M}) &= \frac{ie}{\omega+i0}\chi^{sj}_{ij}(\qv,\omega,{\bm M}),\\
{\cal K}^{ss}_{ij}(\qv,\omega,{\bm M})  &= \frac{-i}{\omega +i0}\chi^{ss}_{ij}(\qv,\omega,{\bm M}) , 
\end{align}
\end{subequations}
with 
\begin{subequations}
\begin{align}
&\chi^{{j}{j}}_{ij}=-\sum_{\kv}\int_{-\infty}^{\infty}
\frac{d\epsilon}{2\pi i}
{\rm tr}\left[
\tilde{v}_{ i }G_{\kv_{+},{\bm M}}(\epsilon_{+})\tilde{v}_{ j }G_{\kv_{-},{\bm M}}(\epsilon_{-})
\right]^{<}, 
\label{cjj}
\\
&\chi^{{j}s}_{ij}=
-\sum_{\kv}\int_{-\infty}^{\infty}
\frac{d\epsilon}{2\pi i}
{\rm tr}\left[
\tilde{v}_{ i }G_{\kv_{+},{\bm M}}(\epsilon_{+}){\sigma}_{ j }G_{\kv_{-},{\bm M}}(\epsilon_{-})
\right]^{<}, 
\label{cjs}
\\
&\chi^{s{j}}_{ij} = -\sum_{\kv}\int_{-\infty}^{\infty}
\frac{d\epsilon}{2\pi i}
{\rm tr}\left[
{\sigma}_{ i }G_{\kv_{+},{\bm M}}(\epsilon_{+})
\tilde{v}_{ i }G_{\kv_{-},{\bm M}}(\epsilon_{-})
\right]^{<}, 
\label{csj}
\\
&\chi^{ss}_{ij} = 
-\sum_{\kv}\int_{-\infty}^{\infty}
\frac{d\epsilon}{2\pi i}
{\rm tr}\left[
{\sigma}_{ i }G_{\kv_{+},{\bm M}}(\epsilon_{+})
{\sigma}_{ j }G_{\kv_{-},{\bm M}}(\epsilon_{-})
\right]^{<}. 
\label{css}
\end{align}
\label{cfs}
\end{subequations}Here, $G_{\kv,{\bm M}}(\vare)$ is the one-particle path-ordered Green's function, 
the lesser component of which is given by \cite{{Langreth76},{Haug98}} 
\begin{align}
\label{lesser}
G^{<}_{\kv,{\bm M}}(\epsilon) = f(\epsilon)\left(
G^{\rm A}_{\kv,{\bm M}}(\epsilon)-G^{\rm R}_{\kv,{\bm M}}(\epsilon) 
\right),
\end{align}
with 
$f(\epsilon)=(1+e^{\epsilon/(k_{\rm B}T)})^{-1}$ being the Fermi distribution function 
($T$ is the temperature). 
The correlation functions [Eqs.~(\ref{cfs})] 
satisfy the Onsager reciprocity relations 
(see Appendix A):  
\begin{subequations}
\begin{align}
&\chi^{{j}{j}}_{ji}(-\qv,\omega,-{\bm  M}) = 
\chi^{{j}{j}}_{ij}(\qv,\omega,{\bm  M}) , 
\label{O-jj}\\
\label{O-js-sj}
&\chi^{{j}s}_{ j  i }(-\qv,\omega,-{\bm  M}) = 
\chi^{s{j}}_{ij}(\qv,\omega,{\bm  M}), \\
&\chi^{ss}_{ji}(-\qv,\omega,-{\bm  M}) = 
\chi^{ss}_{ij}(\qv,\omega,{\bm  M}) . 
\label{O-ss}
\end{align}
\end{subequations}
Note that these four correlation functions 
contain the spin-spin correlation function $\chi^{ss}_{ij}$, 
because of the anomalous velocity ${\bm \alpha}\times{\bm \sigma}/\hbar$ due to the RSOI. 
This implies that there exists a current to which the spin polarization of the electrons contributes, 
even at $\qv={\bm 0}$ (see the calculation results reported in Eq. (\ref{cfs0})).  

\section{Microscopic calculation: Nonmagnetic Rashba conductor}
\subsection{Response functions at $\qv={\bm 0}$ and ${\bm M}={\bm 0}$}
\begin{figure*}
\resizebox{16.8cm}{!}{\includegraphics[angle=0]{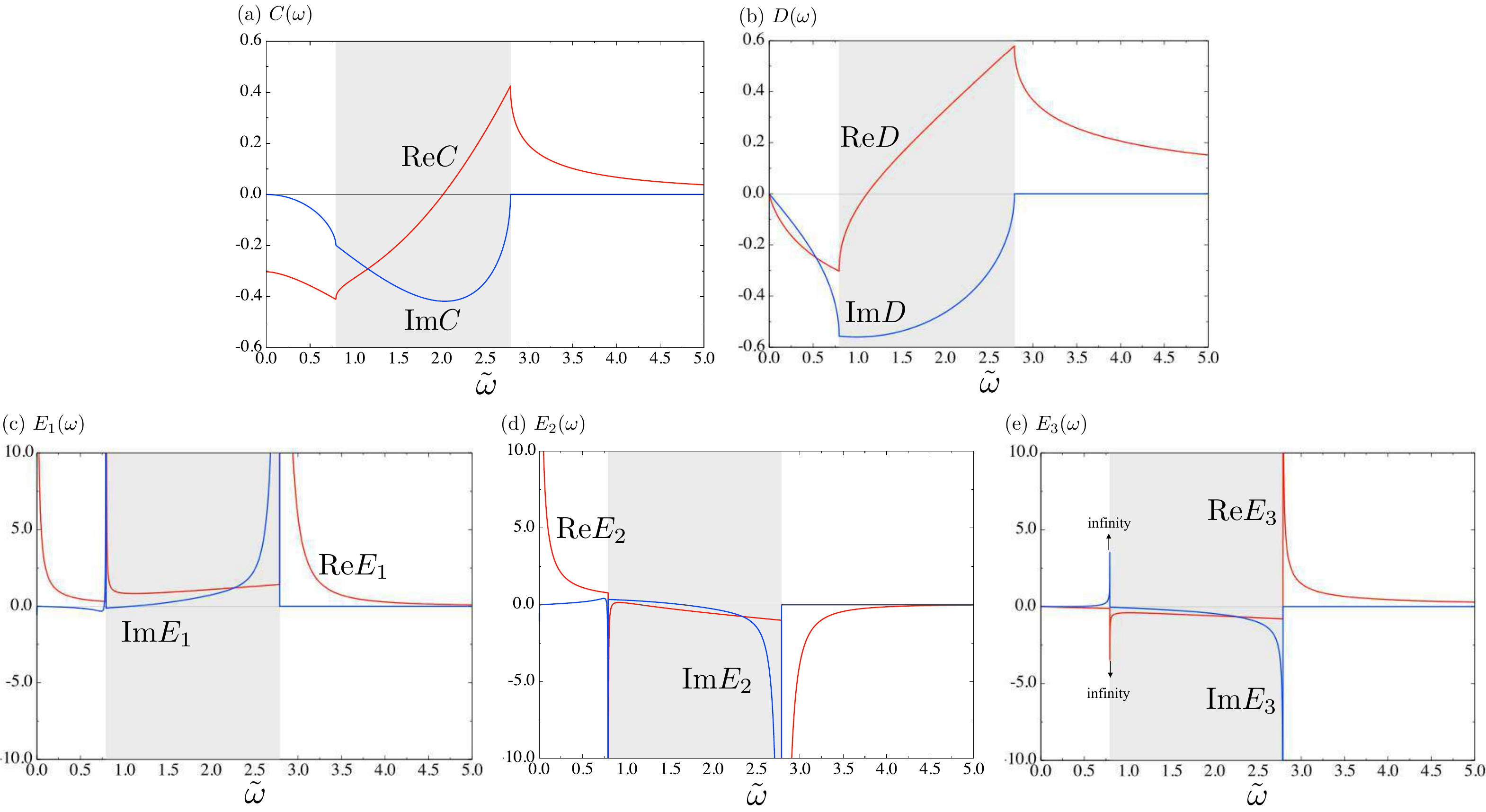}}
\caption{Real and imaginary parts of (a--e) $C(\omega)$, $D(\omega)$, 
$E_{1}(\omega)$, $E_{2}(\omega)$, and $E_{3}(\omega)$, respectively, plotted 
as functions of $\tilde{\omega} = \omega/\omega^{\parallel}_{\rm p}$ 
for the dimensionless Rashba parameter $\tilde{\alpha} = 0.67$. 
In this study, frequencies are measured in units of 
plasma frequency $\omega^{\parallel}_{\rm p} = 
\sqrt{\dfrac{e^2n_{\rm e}}{\vare_{0}m_{\parallel}}}$. 
For $C(\omega)$ and $D(\omega)$, 
cusps exist at the interband transition edges
$\tilde{\omega}_{\pm} = \omega_{\pm} /\omega^{\parallel}_{\rm p}$ 
($\tilde{\omega}_{-} \simeq 0.8$ and $\tilde{\omega}_{+}\simeq 2.8$), 
the region of which is shaded. 
For $E_{1}(\omega)$, $E_{2}(\omega)$, and $E_{3}(\omega)$, 
singularities exist at the $\tilde{\omega}_{\pm}$. }
\label{CDE}
\end{figure*}

We first calculate the response functions at $\qv={\bm 0}$, 
${\bm M}={\bm 0}$, and at zero temperature. 
The results are obtained from the following expressions: 
\cite{note1}
\begin{subequations}
\label{cfs0}
\begin{align}
\chi^{{j}{j}}_{ij}({\bm 0},\omega,{\bm 0}) &=-
\frac{n_{\rm e}}{m_{\perp}}C(\omega)
\delta^{\perp}_{ij}, 
\label{jj-00}
\\
\chi^{{j}s}_{ij}({\bm 0},\omega,{\bm 0})
&=\chi^{s{j}}_{ji}({\bm 0},\omega,{\bm 0})=
\frac{\hbar}{\alpha}\frac{n_{\rm e}}{m_{\perp}}C(\omega)
\vare_{ijk}\hat{\alpha}_{k}, 
\label{js-sj-00}
\\
\chi^{ss}_{ij}({\bm 0},\omega,{\bm 0})&=
\frac{\hbar^2}{\alpha^2}\frac{n_{\rm e}}{m_{\perp}}
C(\omega)(\delta^{\perp}_{ij}-2\delta_{ij}), 
\end{align}
\end{subequations}
where  
\begin{align}
C(\omega) = -\frac{4\tilde{\alpha}^2}{n_{\rm e}}\epsilon_{\rm F}
\sum_{\kv}
\frac{\gamma_{\kv}\sum_{s=\pm1}sf_{\kv}^{s}}{(\hbar\omega+i0)^2-4\gamma_{\kv}^2},
\label{C}
\end{align}
is a complex function of $\omega$, with 
$\tilde \alpha = \dfrac{m_{\perp}\alpha}{\hbar^2k^{\perp}_{\rm F}}$ being a dimensionless  
Rashba parameter and $f_{\kv}^{s} = f(\epsilon^{s}_{\kv,{\bm M=0}}-\epsilon_{\rm F})$ 
being the Fermi distribution function. 
Note that all correlation functions contain $C(\omega)$, 
which is related to the interband transitions between the spin-split bands.  
The explicit form of $C(\omega)$ is given by Eqs. (\ref{V-RC}) and (\ref{V-IC}) 
and plotted in Fig. \ref{CDE}(a). 
In what follows, we use the typical material parameters of 
BiTeI \cite{{Ishizaka11},{Lee-Tokura11},{Tokura12}} listed in Table II.  
The real and imaginary parts of $C(\omega)$ are finite in a wide frequency range 
between the lower and upper interband transition edges $\omega_{\pm}$, 
where 
\begin{align}
\omega_{\pm} = \dfrac{4\epsilon_{\rm F}}{\hbar}
\tilde{\alpha}(\sqrt{1+\tilde{\alpha}^2}\pm \tilde{\alpha}).
\end{align}
\begin{table*}
 \caption{List of material parameters of BiTeI and their typical values. 
 }
 \begin{tabular}{lcc}
\hline\hline
Quantity&Symbol&Value\\
\hline
Rashba interaction strength& $\alpha$ & 3.8~${\rm eV }$\AA\\
Non-dimensional Rashba interaction strength &$\tilde{\alpha}= m_{\perp}\alpha /(\hbar^2k^{\perp}_{\rm F})$&0.67\\
Effective mass (parallel to ${\bm \alpha}$) &$m_{\parallel}=5m_{\perp}$ &$4.3\times10^{-31} {\rm kg}$\\
Effective mass (perpendicular to ${\bm \alpha}$)&$m_{\perp}$ & $0.86\times10^{-31}{\rm kg}$\\
Fermi energy &$\epsilon_{\rm F}$ &0.2~{\rm eV}\\
Electron density&$n_{\rm e}$ & $4.5\times 10^{25}~{\rm m^{-3}}$\\
Plasma frequency (parallel to ${\bm \alpha}$)&
$\dfrac{\omega_{\rm p}^{\parallel}}{2\pi} =\dfrac{1}{2\pi}\sqrt{\dfrac{e^2n_{\rm e}}{\vare_{0}m_{\parallel}}} $ 
& $8.8\times10^{13}~{\rm Hz}$\\
&&($\hbar\omega_{\rm p}^{\parallel}\simeq 0.36~{\rm eV},~
2\pi c/\omega^{\parallel}_{\rm p}\simeq 3.4~{\rm \mu m}$)\\
&\\
Plasma frequency (perpendicular to ${\bm \alpha}$)&$\dfrac{\omega_{\rm p}^{\perp}}{2\pi}
=\sqrt{\dfrac{e^2n_{\rm e}}{\vare_{0}m_{\perp}}} $ & $1.9\times10^{14}~{\rm Hz}$\\
&&
($\hbar\omega_{\rm p}^{\perp}\simeq 0.77~{\rm eV},
~
2\pi c/\omega^{\perp}_{\rm p} \simeq 1.5~{\rm \mu m}$)\\
&\\
Interband transition edge (lower)&$\dfrac{\omega_{-}}{2\pi} = \dfrac{2\epsilon_{\rm F}}{h}
\left(\tilde{\alpha}
\sqrt{1+\tilde{\alpha}^2}-\tilde{\alpha}\right)
$&$7.0\times10^{13}~{\rm Hz}$\\
&\\
Interband transition edge (higher)&$\dfrac{\omega_{+}}{2\pi} = \dfrac{2\epsilon_{\rm F}}{h}
\left(\tilde{\alpha}
\sqrt{1+\tilde{\alpha}^2}+\tilde{\alpha}\right)$&$ 2.4\times10^{14}~{\rm Hz}$\\
&\\
\hline\hline
 \end{tabular}
 \end{table*}
\subsection{Induced current and spin}

\begin{figure*}
\includegraphics[width=16.8cm]{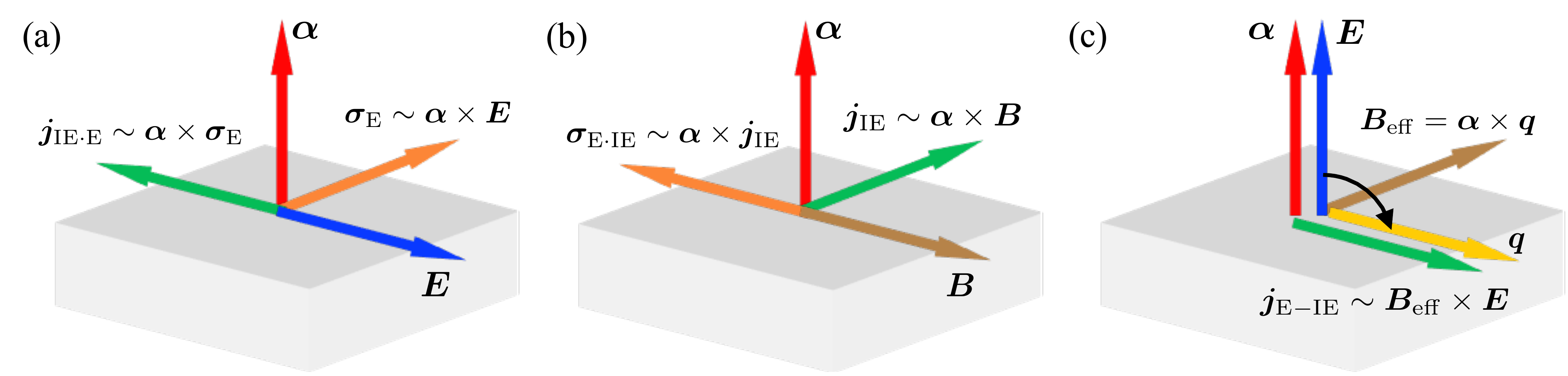}
\caption{Schematic illustration showing cross-correlation effects 
in (nonmagnetic) Rashba conductor: 
(a) Edelstein effect; 
(b) inverse Edelstein effect; and (c) combined direct and inverse Edelstein effects 
with spatial dispersion $\qv$. 
(a) A spin accumulation ${\bm \sigma}_{\rm E} \sim {\bm \alpha}\times{\bm E}$ 
is induced by an external electric field ${\bm E}$, and a charge current ${\bm j}_{\rm IE\cdot E}\sim {\bm \alpha}\times {\bm \sigma}_{\rm E}$ 
is then induced by this ${\bm \sigma}_{\rm E}$ because of the inverse Edelstein effect 
(the direct and inverse Edelstein effects), 
contributing to the plasma frequency modification. 
(b) A charge current ${\bm j}_{\rm IE} \sim {\bm \alpha}\times {\bm B}$ 
is induced by a time-varying external magnetic field, and 
a spin accumulation ${\bm \sigma}_{\rm IE\cdot E}\sim{\bm \alpha}\times {\bm j}_{\rm IE}$ is then induced by this ${\bm j}_{\rm IE}$ because of the Edelstein effect 
(the direct and inverse Edelstein effects), 
contributing to the modification of the magnetic susceptibility. 
(c) A ``Hall'' current ${\bm j}_{\rm E-IE}\sim {\bm B}_{\rm eff}(\qv)\times {\bm E}$ 
is induced by a $\qv$-dependent effective magnetic field ${\bm B}(\qv) 
=\hat{\bm \alpha}\times \qv$ and an ${\bm E} $
resulting from the combination of the direct and inverse Edelstein effects. }
\label{EE-IEE}
\end{figure*}

Substituting Eqs.~(\ref{Kij}) and (\ref{cfs0}) into Eqs.~(\ref{jp-1}) and (\ref{s-1}), 
we obtain the induced current and spin densities 
${\bm j}^{(0)}(\qv,\omega)$ 
and ${\bm \sigma}^{(0)}(\qv,\omega)$, respectively, as 
\begin{align}
{\bm j}^{(0)}
&= \sigma_{\rm D}^{\parallel}(\hat{\bm \alpha}\cdot{\bm E})\hat{\bm \alpha}
-\sigma_{\rm D}^{\perp}\hat{\bm \alpha}\times
(\hat{\bm \alpha}\times{\bm E})+\kappa_{\rm IE}\hat{\bm \alpha}\times {\bm B}, 
\label{j0}\\
{\bm \sigma}^{(0)} &= \kappa_{\rm E}
\hat{\bm \alpha} \times {\bm E}
+
\chi_{\rm B}
\left\{{\bm B}+\frac{1}{2}\hat{\bm \alpha}\times(\hat{\bm \alpha}\times{\bm B})\right\}, 
\label{s0}
\end{align}
where 
\begin{subequations}
\begin{align}
\label{sigma-D-para}
\sigma_{\rm D}^{\parallel}(\omega) 
&= \frac{ie^2}{\omega+i0}\frac{n_{\rm e}}{m_{\parallel}}
= \frac{i\vare_{0}(\omega_{\rm p}^{\parallel})^2}{\omega+i0} ,\\
\label{sigma-D-perp}
\sigma_{\rm D}^{\perp}(\omega) &= 
\frac{ie^2}{\omega+i0}\frac{n_{\rm e}}
{m_{\perp}}(1+C(\omega))\nonumber\\
&= \frac{i\vare_{0}(\omega_{\rm p}^{\perp})^2}{\omega+i0}(1+C(\omega)),\\
\label{kappa-E}
\kappa_{\rm E}(\omega)&= \dfrac{ie}{\omega+i0}\dfrac{\hbar}{\alpha}\dfrac{n_{\rm e}}{m_{\perp}}C(\omega),\\
\label{kappa-IE}
\kappa_{\rm IE}(\omega) &= i\hbar\gamma \omega\kappa_{\rm E}(\omega),\\
\label{chi-B}
\chi_{\rm B}(\omega)&=2\hbar\gamma \dfrac{n_{\rm e}}{m_{\perp}}\dfrac{\hbar^2}{\alpha^2}C(\omega), 
\end{align}
\end{subequations}
with $\omega^{\parallel}_{\rm p} = \sqrt{\dfrac{e^2n_{\rm e}}{\vare_{0}m_{\parallel}}}$ 
and $\omega^{\perp}_{\rm p} = \sqrt{\dfrac{e^2n_{\rm e}}{\vare_{0}m_{\perp}}}$ 
being the plasma frequencies in respective directions.  
The first term of Eq.~(\ref{j0}) is the ordinary longitudinal current 
projected in the $\hat{\bm \alpha}$ direction, and the second term is the perpendicular component; 
the latter includes the effect of the RSOI. 
The first term of Eq.~(\ref{s0}) represents the Edelstein effect \cite{Edelstein}, 
in which a spin polarization 
${\bm \sigma}_{\rm E}=\kappa_{\rm E}(\omega)\hat{\bm \alpha}\times{\bm E}$ 
is induced by the electric field because of the RSOI (Fig.~\ref{EE-IEE}(a)). 
The third term of Eq.~(\ref{j0}) represents the inverse Edelstein effect \cite{Raimondi14},  
where a charge current 
${\bm j}_{\rm IE} = \kappa_{\rm IE}(\omega)\hat{\bm \alpha}\times {\bm B}$ 
is induced by 
the time-varying magnetic field $i\omega{\bm B}$ as a result of the RSOI (Fig.~\ref{EE-IEE}(b)). 
Note that the frequency-dependent coeficient 
$\kappa_{\rm IE}(\omega)$ is related to $\kappa_{\rm E}(\omega)$ 
through the Onsager relation [Eq.~(\ref{kappa-IE})].  
It is apparent that the second term of Eq.~(\ref{j0}) is due to the combination of the direct and inverse Edelstein effects \cite{STKT16}, 
i.e., a non-equilibrium spin accumulation 
$\sim\hat{\bm \alpha}\times {\bm E}$, 
which is induced by the Edelstein effect, 
is subsequently converted to a charge current density 
${\bm j}_{\rm IE\cdot E}\sim \hat{\bm \alpha}\times{\bm \sigma}_{\rm EE}$, 
which is due to the inverse Edelstein effect (Fig.~\ref{EE-IEE}(a)). 
The second term in brackets in Eq.~(\ref{s0}) 
represents the Onsager reciprocal to this behavior,  
where a current $\sim\hat{\bm \alpha}\times {\bm B}$, 
which is induced by the inverse Edelstein effect, 
is subsequently converted to a spin polarization 
${\bm \sigma}_{\rm E\cdot IE}\sim \hat{\bm \alpha}\times{\bm j}_{\rm IEE}$ 
by the Edelstein effect (Fig.~\ref{EE-IEE}(b)). 

\subsection{Optical conductivity}

The optical conductivity is obtained by substituting  
Eqs.~(\ref{cfs0}) into Eq.~(\ref{oc-1}). 
For $\qv={\bm 0}$ and ${\bm M}={\bm 0}$, 
we obtain 
\begin{align}
\sigma^{(0)}_{ij}(\omega) &\equiv
\frac{-ie^2}{\omega+i0}
\left[
 \chi^{jj}_{ij}({\bm 0},\omega,{\bm 0})-\frac{n_{\rm e}}{m_{\perp}}
\delta^{\perp}_{ij} - \frac{n_{\rm e}}{m_{\parallel}}\hat{\alpha}_{i}\hat{\alpha}_{j}\right],
\nonumber\\
&= \sigma_{\rm D}^{\parallel}(\omega)\hat{\alpha}_{i}\hat{\alpha}_{j}+\sigma_{\rm D}^{\perp}(\omega)\delta^{\perp}_{ij}, 
\label{oc-HM}
\end{align}
which yields the first and second terms 
of Eq.~(\ref{j0}).  
It is apparent that the optical conductivity tensor is symmetric 
and has diagonal components, which take a uniaxially anisotropic form. 
This anisotropic property is equivalent to that of a hyperbolic metamaterial \cite{PQE15}. 
Thus, this material is expected to exhibit unusual electromagnetic wave propagation phenomena known as negative refraction and backward waves.  
Details of this behavior are presented in Sec.~VII.

In the first order of $\qv$, 
the optical conductivity $\sigma^{\rm E-IE}_{ij}(\qv, \omega)$ 
originates from the Edelstein effect 
given in the first term of Eq.~(\ref{s0}), along with the inverse Edelstein effect given in the third term of Eq.~(\ref{j0}).  
Substituting  Eqs.~(\ref{cfs0}) into Eq.~(\ref{oc-1}), 
we obtain 
\begin{align}
&\sigma^{\rm E-IE}_{ij}(\qv,\omega) \nonumber\\
&\equiv 
\frac{e\hbar\gamma}{\omega+i0}\left\{
\chi^{js}_{ij}({\bm 0},\omega,{\bm 0})\vare_{lkj}+\vare_{ikl}\chi^{sj}_{lj}({\bm 0},\omega,{\bm 0})
\right\}q_{k},\nonumber\\
&= i\hbar \gamma \kappa_{\rm E}(\omega)
(q_{ i }\hat{\alpha}_{j}-\hat{\alpha}_{i}q_{ j }). 
\label{oc-EIE}
\end{align} 
This tensor is antisymmetric and 
has $\qv$-linear off-diagonal components.   
The induced current ${\bm j}_{\rm E-IE}$  
is given by 
\begin{align}
{\bm j}_{\rm E-IE} &= 
\kappa_{\rm IE}(\omega) \hat{\bm \alpha}\times {\bm B}
+(-i\hbar\gamma \qv)\times \kappa_{\rm E}(\omega)
\hat{\bm \alpha}\times {\bm E},\nonumber\\
&= 
i\hbar\kappa_{\rm E}(\omega) {\bm B}_{\rm eff}(\qv)\times {\bm E}, 
\end{align}
where ${\bm B}_{\rm eff}(\qv)= \hat{\bm \alpha}\times \qv$ 
is a $\qv$-induced effective magnetic field. Thus, it is apparent that 
an electron flowing in the ${\bm E}$ direction experiences a Lorentz force, and that the current is bent around ${\bm B}_{\rm eff}(\qv)$. 
This indicates that the ${\bm E}$ of a linearly polarized wave  
rotates around ${\bm B}_{\rm eff}(\qv)$ (Fig.~\ref{EE-IEE}(c)). 
Hence, 
the electric field acquires a component parallel to $\qv$, 
which may be called ``Rashba-induced birefringence.''  
Details are presented in Sec.~VIII-B.

\section{Microscopic calculation: Ferromagnetic Rashba conductor}
In this section, we consider the effect of the exchange interaction 
and the RSOI on the current and spin responses.  

\subsection{Response functions at first orders in $\qv$ and ${\bm M}$}

We first evaluate the response functions at $\qv={\bm 0}$ and for the first order of ${\bm M}$. 
The results are obtained from the following expressions (see Appendices B and C):  
\begin{subequations}
\label{cor-1}
\begin{align}
\chi^{jj,(1)}_{ij}(\omega,{\bm M})
&\equiv
M_{k}\left.\frac{\partial}{\partial M_{k}}\chi^{jj}_{ij}({\bm 0},\omega,{\bm M})\right|_{{\bm M}={\bm 0}},\nonumber\\
&=
-i\frac{n_{\rm e}}{m_{\perp}}\frac{M}{\epsilon_{\rm F}}D(\omega)
\vare_{ijk}\hat{M}_{\parallel,k}, \\
\label{chi-js-1}
\chi^{js,(1)}_{ij}(\omega,{\bm M})
&\equiv
M_{k}\left.\frac{\partial}{\partial M_{k}}\chi^{js}_{ij}({\bm 0},\omega,{\bm M})\right|_{{\bm M}={\bm 0}},\nonumber\\
&=i\frac{\hbar}{\alpha}
\frac{n_{\rm e}}{m_{\perp}}
\frac{M}{\epsilon_{\rm F}}
\left(
N_{i}(\omega)\hat{\alpha}_{j}-\hat{\bm \alpha}\cdot{\bm N}(\omega)\delta_{ij}
\right), \\
\label{chi-sj-1}
\chi^{sj,(1)}_{ij}(\omega,{\bm M})
&\equiv
M_{k}\left.\frac{\partial}{\partial M_{k}}\chi^{sj}_{ij}({\bm 0},\omega,{\bm M})\right|_{{\bm M}={\bm 0}},\nonumber\\
&=
-i\frac{\hbar}{\alpha}
\frac{n_{\rm e}}{m_{\perp}}
\frac{M}{\epsilon_{\rm F}}
\left(
\hat{\alpha}_{ i }N_{j}(\omega)-\hat{\bm \alpha}\cdot{\bm N}(\omega)\delta_{ij}
\right), 
\\
\chi^{ss,(1)}_{ij}(\omega,{\bm M})
&\equiv
M_{k}\left.\frac{\partial}{\partial M_{k}}\chi^{ss}_{ij}({\bm 0},\omega,{\bm M})\right|_{{\bm M}={\bm 0}},\nonumber\\
&=
-i\frac{\hbar^2}{\alpha^2}\frac{n_{\rm e}}{m_{\perp}}\vare_{ijk}N_{k}(\omega), 
\label{ss}
\end{align}
\end{subequations}
where  
\begin{align}
\label{N}
{\bm N}(\omega) = D(\omega)\hat{\bm  M}_{\parallel} 
+ \tilde{\alpha}E_{3}(\omega)\hat{\bm  M}_{\perp}, 
\end{align}
with  
$\hat{\bm  M} = {\bm M}/|{\bm M}|$ being a unit vector 
in the magnetization direction 
and $\hat{\bm  M}_{\parallel} = (\hat{\bm \alpha}\cdot \hat{\bm  M})\hat{\bm \alpha}$ and $\hat{\bm  M}_{\perp}= \hat{\bm  M}-\hat{\bm  M}_{\parallel}$. 
The $D(\omega)$ and $E_{3}(\omega)$ coefficients are complex functions of 
$\omega$, which are plotted in Fig.~\ref{CDE}(b--e).  
 Their analytic expressions are given in Appendix C. 
Note that the frequency depdendence of $D(\omega)$ and that of $E_{3}(\omega)$ differ significantly. 
That is, $D(\omega)$ has cusps at transition edges $\omega_{\pm}$ 
(Fig.~\ref{CDE}(b)), whereas $E_{3}(\omega)$ diverges at $\omega_{\pm}$ (Fig.~\ref{CDE}(e)). 

For response functions at the first order of $\qv$ and ${\bm M}$, 
we require only $\chi^{jj}_{ij}$, which is given by 
\begin{align}
&\chi^{jj,(11)}_{ij}(\qv,\omega,{\bm M})\nonumber\\
&\equiv
q_{k}M_{l}\left.\frac{\partial^2}{\partial q_{k}\partial M_{l}}\chi^{jj}_{ij}({\bm q},\omega,{\bm M})\right|_{
\qv={\bm 0},
{\bm M}={\bm 0}},
\nonumber\\
&=\frac{n_{\rm e}}{m_{\perp}}\frac{M}{k_{\rm F}\epsilon_{\rm F}}
\left[
E_{1}(\omega)\hat{\cal {\bm T}} \cdot\qv \delta^{\perp}_{ij}
+\frac{1}{2}E_{2}(\omega)\left\{
\hat{\cal {T}} _{ i }q_{\perp,j}+q_{\perp,i}\hat{\cal {T}} _{ j }
\right\}
\right], 
\label{chi-jj-11}
\end{align}
where 
$\hat{\cal {\bm T}} = \hat{\bm \alpha}\times\hat{\bm M}$ 
and 
$\qv_{\perp} = \qv - \hat{\bm \alpha}(\qv\cdot\hat{{\bm \alpha}})$. 
Thus, we see that only the perpendicular components of ${\bm M}$  
contribute to the ${\bm M}$- and ${\bm q}$-induced current response. 
As mentioned above (see the last paragraph of Sec. II), the vector 
$\hat{\cal {\bm T}}=\hat{\bm \alpha}\times \hat{\bm M}$ 
can be regarded as a toroidal moment \cite{Spaldin08}.

\subsection{Induced current and spin}

Substituting Eq.~(\ref{cor-1}) into Eqs.~(\ref{jp-1}) and (\ref{s-1}),  
we obtain 
the current and spin densities  ${\bm j}^{(1)}(\qv,\omega)$ and 
${\bm \sigma}^{(1)}(\qv,\omega)$, respectively, 
to the first order of ${\bm M}$: 
\begin{align}
\label{j1}
{\bm j}^{(1)}&= 
\sigma_{\rm AH}(\omega)\hat{\bm M}_{\parallel}\times {\bm E}+
\kappa_{\rm IEM}(\omega)\hat{\bm \alpha}\times({\bm N}(\omega)\times{\bm B}),
\\
\label{s1}
{\bm \sigma}^{(1)} 
&=
\kappa_{\rm EM}{\bm N}(\omega)\times(\hat{\bm \alpha}\times{\bm E})
+\chi_{M}{\bm N}(\omega)\times{\bm B}, 
\end{align}
where 
\begin{subequations}
\begin{align}
&\sigma_{\rm AH}(\omega)= 
\dfrac{i\vare_{0}(\omega_{\rm p}^{\perp})^2}{\omega+i0}
\dfrac{M}{\epsilon_{\rm F}}D(\omega), \\
&\kappa_{\rm EM} (\omega)= \dfrac{e}{\omega+i0}\dfrac{\hbar}{\alpha}\dfrac{n_{\rm e}}{m_{\perp}}
\dfrac{M}{\epsilon_{\rm F}},\\
&\kappa_{\rm IEM}(\omega)= i\hbar\gamma\omega\kappa_{\rm EM} (\omega) ,\\
&\chi_{M} = -i\hbar\gamma\frac{\hbar^2}{\alpha^2}\frac{n_{\rm e}}{m_{\perp}}M. 
\end{align}
\end{subequations}
The first term of Eq.~(\ref{j1}) represents the anomalous Hall effect 
\cite{{Inoue},{Kovalev},{BRV11},{Titov16}}. 
Further, the second term of Eq.~(\ref{j1}) and the first term of Eq.~(\ref{s1}) 
represent the electromagnetic cross-correlation effects 
due to the RSOI and ${\bm M}$, 
which are reciprocal to each other. 
The contributions of these terms to the total induced current, which consists of 
the polarization and magnetization currents, is of first order in both 
$\qv$ and ${\bm M}$. 
Thus, this is the same order of magnitude as the current density obtained 
from $\chi^{jj,(11)}_{ij}$ in Eq.~(\ref{chi-jj-11}) (see Sec. VII-C).  

\subsection{Optical conductivity}
\begin{figure*}
\includegraphics[width=16.8cm]{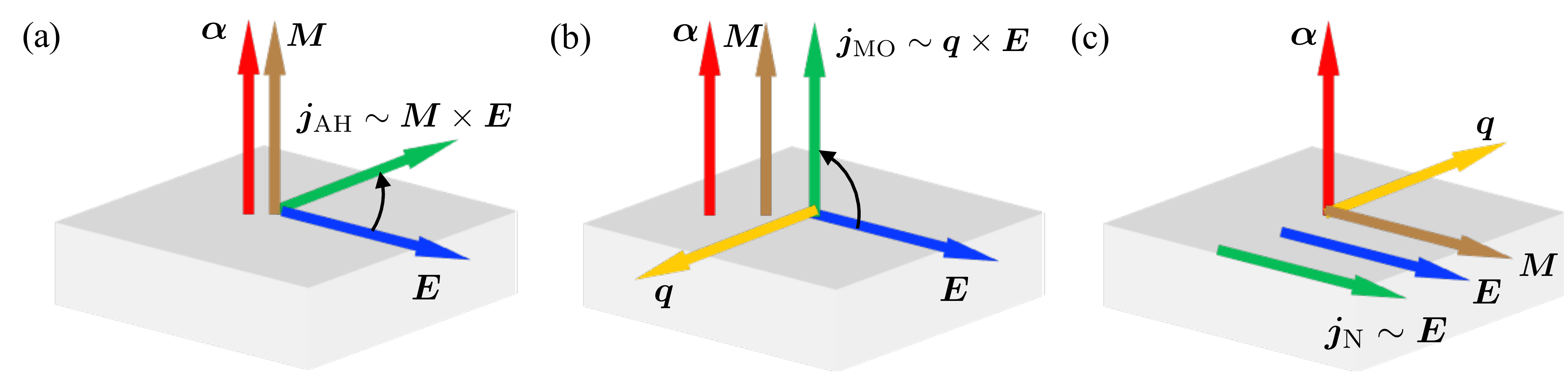}
\caption{
Schematic illustration showing current responses in ferromagnetic Rashba conductor. 
(a) Anomalous Hall current ${\bm j}_{\rm AH}$ and (b, c) $\qv$- and ${\bm M}$-induced currents. 
(a) An anomalous Hall current ${\bm j}_{\rm AH}\sim {\bm M}\times {\bm E}$ 
is induced by the Rashba spin orbit interaction, the parallel component of the magnetization 
${\bm M}_{\parallel}=(\hat{\bm \alpha}\cdot{\bm M})\hat{\bm \alpha}$, and an external electric field ${\bm E}$. 
As the electron orbits are bent in the plane perpendicular to ${\bm M}$ (or ${\bm \alpha}$), 
the ${\bm E}$ of a linearly polarized wave rotates around ${\bm M}$, 
yielding Faraday rotation. 
(b) A $\qv$-dependent charge current 
${\bm j}_{\rm MO} \sim {\cal {\bm Q}}^{\parallel}\cdot(\qv \times {\bm E})$ 
is induced by the parallel ``quadrupole'' moment dyadic 
${\cal {\bm Q}}^{\parallel} = \hat{\bm \alpha}{\bm M}_{\parallel}+ {\bm M}_{\parallel}\hat{\bm \alpha}$. 
As this moment causes a rotation of the plarization plane of the linearly polarized wave around 
$\qv$, ${\bm M}$-induced optical activity can occur. 
(c) A $\qv$-dependent charge current is induced by 
the ``toroidal'' moment $\hat{\cal {\bm T}} = \hat{\bm \alpha}\times \hat{\bm M}$ and 
the perpendicular ``quadrupole'' moment dyadic ${\cal {\bm Q}}=\hat{\bm \alpha}{\bm M}_{\perp} + 
{\bm M}_{\perp}\hat{\alpha}$. 
This current depends on the direction of $\qv$ and ${\bm M}$ and is nonreciprocal, 
generating a difference in absorption for counterpropagating electromagnetic waves 
relative to $\qv$ or ${\bm M}$. 
}
\label{AH-MO-NDD}
\end{figure*}
The optical conductivities in the presence of ${\bm M}$ can be 
obtained by substituting Eqs.~(\ref{cor-1}) and (\ref{chi-jj-11}) into Eq.~(\ref{oc-1}). 
For $\qv = {\bm 0}$ and at the first order of ${\bm M}$, 
we obtain 
\begin{align}
\sigma^{\rm AH}_{ij}(\omega,{\bm  M}_{\parallel}) 
&\equiv \frac{-ie^2}{\omega+i0}\chi^{jj,(1)}_{ij}(\omega,{\bm M}), \nonumber\\
&=-\sigma_{\rm AH}(\omega)\vare_{ijk}\hat{M}_{\parallel,k}
\label{oc-AH}. 
\end{align}
The induced current 
${\bm j}_{\rm AH}=\sigma_{\rm AH}(\omega)\hat{\bm M}_{\parallel}\times{\bm E}$ 
represents the anomalous Hall current (Fig.~\ref{AH-MO-NDD}(a)) [Eq. (\ref{Sec1-AHc})Recent optical spectroscopy measurements performed on a BiTeI semiconductor 
subjected to a static magnetic field have revealed a cusp structure in the corresponding optical conductivity 
$\sigma_{\rm AH}(\omega)$ \cite{Lee-Tokura11}; this finding qualitatively agrees with our result. 
Because of the anomalous Hall effect, the electron flow in the 
${\bm E}$ direction is bent (Fig.~\ref{AH-MO-NDD}(a)), 
and the ${\bm E}$ of a linearly polarized wave rotates around 
$\hat{\bm M}_{\parallel}$ in the Rashba conductor; this is called ``Faraday rotation'' and is studied in Sec.~IX-A. 

For the first-order terms in $\qv$ and ${\bm M}$, 
it is convenient to decompose the optical conductivity into two parts  
related to the parallel and perpendicular components of ${\bm M}$, respectively. 
The former is given by 
\begin{align}
\label{sM-para}
&\sigma^{M,\parallel}_{ij}(\qv,\omega,{\bm M}_{\parallel})\nonumber\\ &\equiv
\frac{e\hbar\gamma}{\omega+i0}
\left\{
\chi^{js,(1)}_{ij}(\omega,{\bm M}_{\parallel})\vare_{lkj}+\vare_{ikl}\chi^{sj}_{lj}(\omega,{\bm M}_{\parallel})
\right\}q_{k},
\nonumber\\
&=\sigma_{M}(\omega)\frac{D(\omega)}{\tilde{\alpha}}
\left(
{\cal Q}^{\parallel}_{il}\vare_{lkj}+\vare_{ilk}{\cal Q}^{\parallel}_{lj}
\right)q_{\perp,k}, 
\end{align}
where
\begin{align}
\sigma_{M}(\omega) = 
\dfrac{i\vare_{0}(\omega_{\rm p}^{\perp})^2}{\omega+i0}
\dfrac{1}{2k_{\rm F}^{\perp}}\dfrac{M}{\epsilon_{\rm F}}, 
\end{align}
and we have replaced the gyromagnetic ratio $\gamma$ by $\dfrac{e}{2m_{\perp}}$ 
and 
introduced the ``quadrupole'' moment \cite{Spaldin08} 
${\cal Q}^{\parallel}_{ij} = \hat{\alpha}_{i}\hat{M}_{\parallel,j}+\hat{\alpha}_{j}
\hat{M}_{\parallel,i}$. 
Thus, the induced current ${\bm j}_{\rm MO}$  
is given by 
\begin{align}
&{\bm j}_{\rm MO}(\qv,\omega) = \kappa_{\rm MO}(\omega)
\left\{{\cal \bm Q}^{\parallel}\cdot (\qv_{\perp}\times {\bm E})
-\qv_{\perp}\times({\cal Q}^{\parallel}\cdot{\bm E})
\right\}, 
\label{j-MO}
\end{align}
where $\kappa_{\rm MO}(\omega) = \sigma_{M}D(\omega)/\tilde{\alpha}$. 
The ${\bm j}_{\rm MO}$
induced by ${\cal Q}^{\parallel}_{ij}$ arises 
when ${\bm E}$ and ${\bm q}_{\perp}$ are noncollinear 
(Fig.~\ref{AH-MO-NDD}(b)) and, 
thus, the optical conductivity tensor has off-diagonal components linear in $\qv$.  
This result indicates that the ``quadrupole'' moment 
${\cal {\bm Q}}^{\parallel}$ causes a rotation of the ${\bm E}$ 
of the linearly polarized wave around $\qv_{\perp}$; thus, a similar phenomenon to natural optical activity 
is expected to occur, as mentioned in Sec. IV. 
However, the signal will be considerably weaker than that of the anomalous Hall effect, as 
$\kappa_{\rm MO}(\omega)$ has a small parameter 
$\omega^{\parallel}/(ck_{\rm F}^{\perp})$, 
which is $\sim 10^{-3}$ for BiTeI.  
In this study, we do not pursue this phenomenon further. 

The remaining ``perpendicular'' component originates from $\chi^{jj,(11)}_{ij}$ in Eq.~(\ref{chi-jj-11}) and 
the electromagnetic cross-correlation effects obtained from Eqs.~(\ref{chi-js-1}) and (\ref{chi-sj-1}) 
are given by 
\begin{widetext}
\begin{align}
\label{sM-perp}
\sigma^{M,\perp}_{ij}(\qv,\omega,{\bm M}_{\perp}) &=
\frac{-ie^2}{\omega+i0}\chi^{jj,(11)}_{ij}(\qv,\omega,{\bm M})+
\frac{e\hbar\gamma}{\omega+i0}
\left\{
\chi^{js,(1)}_{ij}(\omega,{\bm M}_{\perp})\vare_{lkj}+\vare_{ikl}\chi^{sj}_{lj}(\omega,{\bm M}_{\perp})
\right\}q_{k},
\nonumber\\ 
&=-\sigma_{M}\bigg[
\left\{\left(2E_{1}(\omega)+E_{3}(\omega)\right)\delta^{\perp}_{ij}
+E_{3}(\omega)\hat{\alpha}_{i}\hat{\alpha}_{j}\right\}
{\cal {\bm T}}\cdot \qv, \nonumber\\
&+\left(E_{2}(\omega)-\frac{1}{2}E_{3}(\omega)\right)
\left({\cal T}_{ i }q_{\perp,j}+ {\cal T}_{j}q_{\perp,i}\right)
-\frac{1}{2}E_{3}(\omega)
\left(
{\cal Q}^{\perp}_{il}\vare_{lkj}+\vare_{ilk}{\cal Q}^{\perp}_{lj}
\right)q_{\perp,k}
\bigg] . 
\end{align} 
\end{widetext}
 Again, we have replaced $\gamma$ by $\dfrac{e}{2m_{\perp}}$ 
in the second term. 
In addition, we have introduced the ``quadrupole'' moment  as 
${\cal Q}^{\perp}_{ij} = \hat{\alpha}_{i}\hat{M}_{\perp,j}+\hat{\alpha}_{j}
\hat{M}_{\perp,i}$. 
Contrary to $\sigma^{M,\parallel}_{ij}$ [Eq.~(\ref{sM-para})],  
$\sigma^{M,\perp}_{ij}$ can have a $\qv$-linear term 
in the diagonal components. 
From Eq.~(\ref{sM-perp}), 
the induced current is given by  
\begin{align}
{\bm j}_{\rm N} &= (\hat{\cal {\bm T}}\cdot\qv)
\left\{
\kappa^{\rm N}_{1}(\omega)\hat{\bm \alpha}\times(\hat{\bm \alpha}\times{\bm E})
+\kappa^{\rm N}_{2}(\omega)(\hat{\bm \alpha}\cdot{\bm E})\hat{\bm \alpha}
\right\}\nonumber\\
&+\kappa^{\rm N}_{3}(\omega)\left\{
\hat{\cal {\bm T}}(\qv_{\perp}\cdot{\bm E}) + \qv_{\perp}(\hat{\cal {\bm T}}\cdot {\bm E})
\right\}
\nonumber\\
&+\kappa^{\rm N}_{4}(\omega)
\left\{{\cal \bm Q}^{\perp}\cdot (\qv_{\perp}\times {\bm E})
-\qv_{\perp}\times({\cal Q}^{\perp}\cdot{\bm E})
\right\}, \label{j-NDD}
\end{align}
where 
$\kappa^{\rm N}_{1}(\omega) = \sigma_{M}(2E_{1}(\omega)+E_{3}(\omega))$, 
$\kappa^{\rm N}_{2}(\omega) = -\sigma_{M}E_{3}(\omega)$, 
$\kappa^{\rm N}_{3}(\omega) = -\sigma_{M}(E_{2}(\omega)-E_{3}(\omega)/2)$, 
and 
$\kappa^{N}_{4}(\omega) = \sigma_{M}E_{3}(\omega)/2$. 
It is apparent that, 
for the configuration with $\hat{\cal {\bm T}} \parallel \qv$ and $\hat{\bm M}_{\perp} \parallel {\bm E}$ 
 (Fig.~\ref{AH-MO-NDD}(c)),
the current is efficiently induced by the toroidal and quadrupole moments. Then, the diagonal components 
satisfy $\sigma^{M,\perp}_{ii}(-\qv,\omega,{\bm M}_{\perp})
=\sigma^{M,\perp}_{ii}(\qv,\omega,-{\bm M}_{\perp})
=-\sigma^{M,\perp}_{ii}(\qv,\omega,{\bm M}_{\perp})$. 
Thus, the induced longitudinal current depends on the direction of $\qv$ or ${\bm M}$; this nonreciprocal current flow induces 
a difference in absorption between the two counterpropagating electromagnetic waves 
($\qv$ and $-\qv$), or 
between the two opposite magnetization directions 
(${\bm M}$ and $-{\bm M}$).   
As transport coefficients contain $E_{\mu}(\omega)\, (\mu = 1,2,3)$, which 
diverge at $\omega_{\pm}$, 
significant enhancement is expected in the 
anisotropic propagation, which is called ``nonreciprocal directional dichroism.'' 
This phenomenon is considered in detail in Sec.~IX-B.

\section{Wave propagations in nonmagnetic Rashba conductor} 

 We now examine wave propagations in a bulk Rashba conductor   
based on the wave equation (\ref{weq-0}) 
and the $\vare_{ij}$ given by Eq.~(\ref{expand-epsilon}), 
combined with the microscopic results obtained in the previous sections. 
 In this section, we focus on the nonmagnetic case (${\bm M}={\bm 0}$); the effects of ${\bm M}$ are considered in the next section. 
In what follows, 
it is convenient to choose the coordinate system $({\bm e}_{x},{\bm e}_{y},{\bm e}_{z})
=(\hat{\qv}_{\perp}, \hat{\bm \alpha}\times \hat{\bm \qv}_{\perp}, \hat{\bm \alpha})$, 
where $\hat{\qv}_{\perp}=\qv_{\perp}/|\qv_{\perp}|$ with 
$\qv_{\perp}=\qv-(\hat{\bm \alpha}\cdot\qv)\hat{\bm \alpha}$ 
being a unit vector perpendicular to $\hat{\bm \alpha}$. 
In this coordinate system, 
$\qv$ is expressed as $\qv=q_{x}{\bm e}_{x}+q_{z}{\bm e}_{z}$ and  
the wave equation is given by Eq.~(\ref{weq-a}). 

\subsection{Effects of anisotropy in $\vare^{(0)}_{ij}$: \\  Negative refraction and backward waves}

Let us first consider the anisotropy of $\vare^{(0)}_{ij}(\omega)$. 
Substituting $\sigma^{(0)}_{ij}(\omega)$ [Eq.~(\ref{oc-HM})] 
into Eq.~(\ref{dielectric}), we obtain 
\begin{align}
\label{dielectric-0-R}
\vare^{(0)}_{ij}(\omega) &= 
\delta_{ij} + \frac{1}{\vare_{0}}\frac{i}{\omega}\sigma^{(0)}_{ij}(\omega),\nonumber\\
&=
\begin{pmatrix}
\vare_{x}(\omega)&0&0\\
0&\vare_{x}(\omega)&0\\
0&0&\vare_{z}(\omega)\\
\end{pmatrix},
\end{align}
where 
\begin{align}
\label{vare-perp}
&\vare_{x}(\omega) = 
1-\frac{(\omega_{\rm p}^{\perp})^2}{\omega(\omega+i\eta)}(1+C(\omega)), \\
\label{vare-para}
&\vare_{z}(\omega)=1-\frac{(\omega_{\rm p}^{\parallel})^2}{\omega(\omega+i\eta)}. 
\end{align}
Note that we have replaced the infinitesimal $0$ with a finite damping parameter 
$\eta$ in the dielectric functions. 
In the following, the $C(\omega)$, $D(\omega)$, 
$E_{1}(\omega)$, $E_{2}(\omega)$, and $E_{3}(\omega)$ coefficients 
are also evaluated with this finite $\eta$. (Details are provided in Appendix C.) It is apparent that the $\vare^{(0)}_{ij}$ tensor has the same uniaxial symmetry as that given in Eq.~(\ref{dielectric-0-1}), 
and $\hat{\bm \alpha}={\bm e}_{z}$ gives the optic axis. 
Thus, there exist two types of wave solution: a linearly polarized ordinary wave and 
an extraordinary wave. 
 Here, we concentrate on the extraordinary wave. 
The dispersion relation is given by 
\begin{align}
\frac{q_{x}^2}{\vare_{z}}+\frac{q_{z}^2}{\vare_{x}}=\frac{\omega^2}{c^2}  . 
\label{dispersion-ew}
\end{align}

\begin{figure}[t]
\begin{center}
\resizebox{8.4cm}{!}{\includegraphics[angle=0]{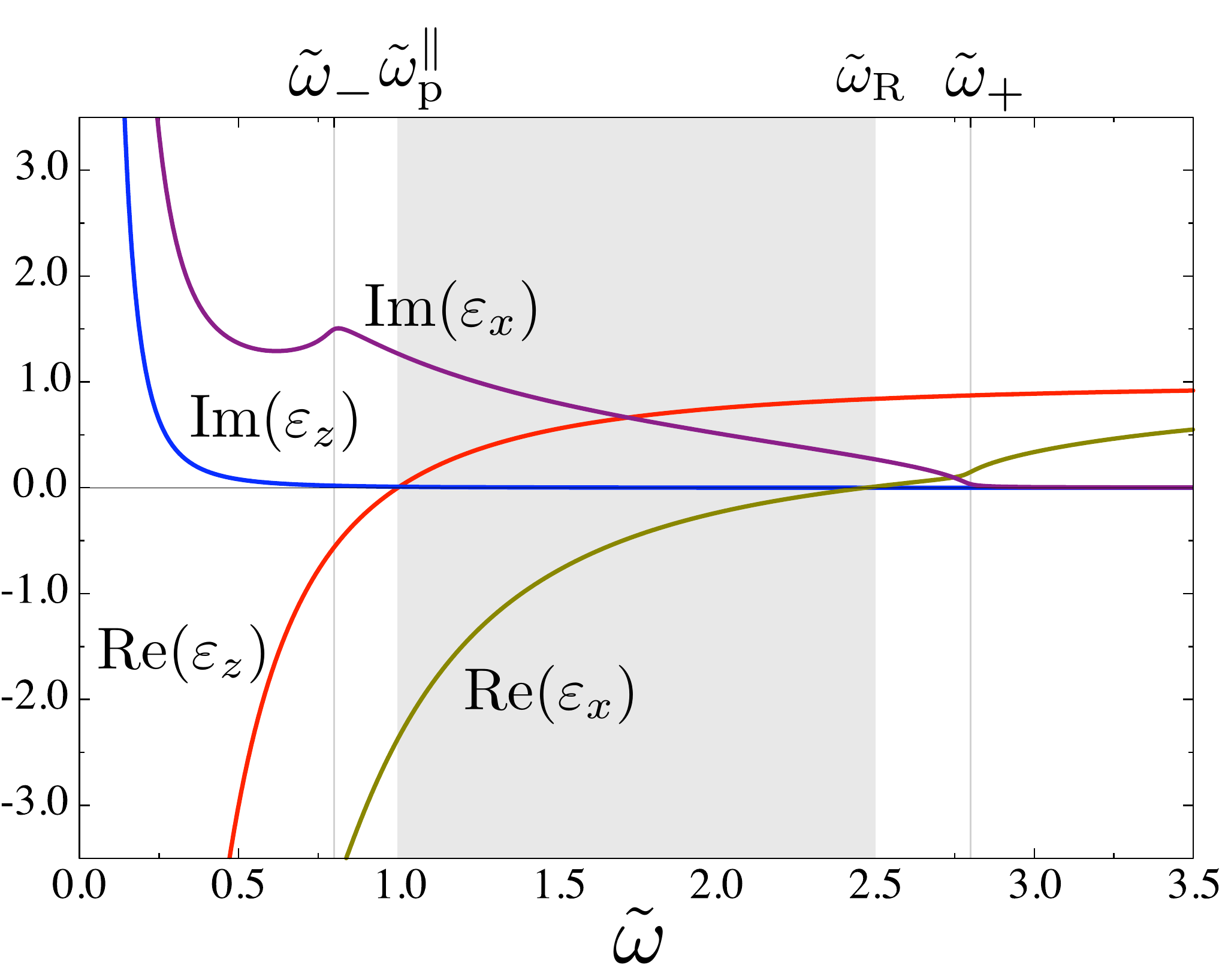}}
\end{center}
\caption{Real and imaginary parts of $\vare_{z}$ and $\vare_{x}$ 
as functions of  $\tilde{\omega}={\omega}/\omega^{\parallel}_{\rm p}$ for 
$\tilde{\alpha}=0.67$ and 
$\tilde{\eta} = \eta/\omega_{p}^{\parallel}=0.01$. 
The gray shaded region represents the frequency range  
$\omega^{\parallel}_{\rm p} < \omega < \omega_{\rm R}=\tilde{\omega}_{\rm R}\omega^{\parallel}_{\rm p}$,   
in which ${\rm Re}(\vare_{z})>0$ and ${\rm Re}(\vare_{x})<0$.}
\label{epsilon}
\end{figure}

\begin{figure}[t]
\includegraphics[width=8.4cm]{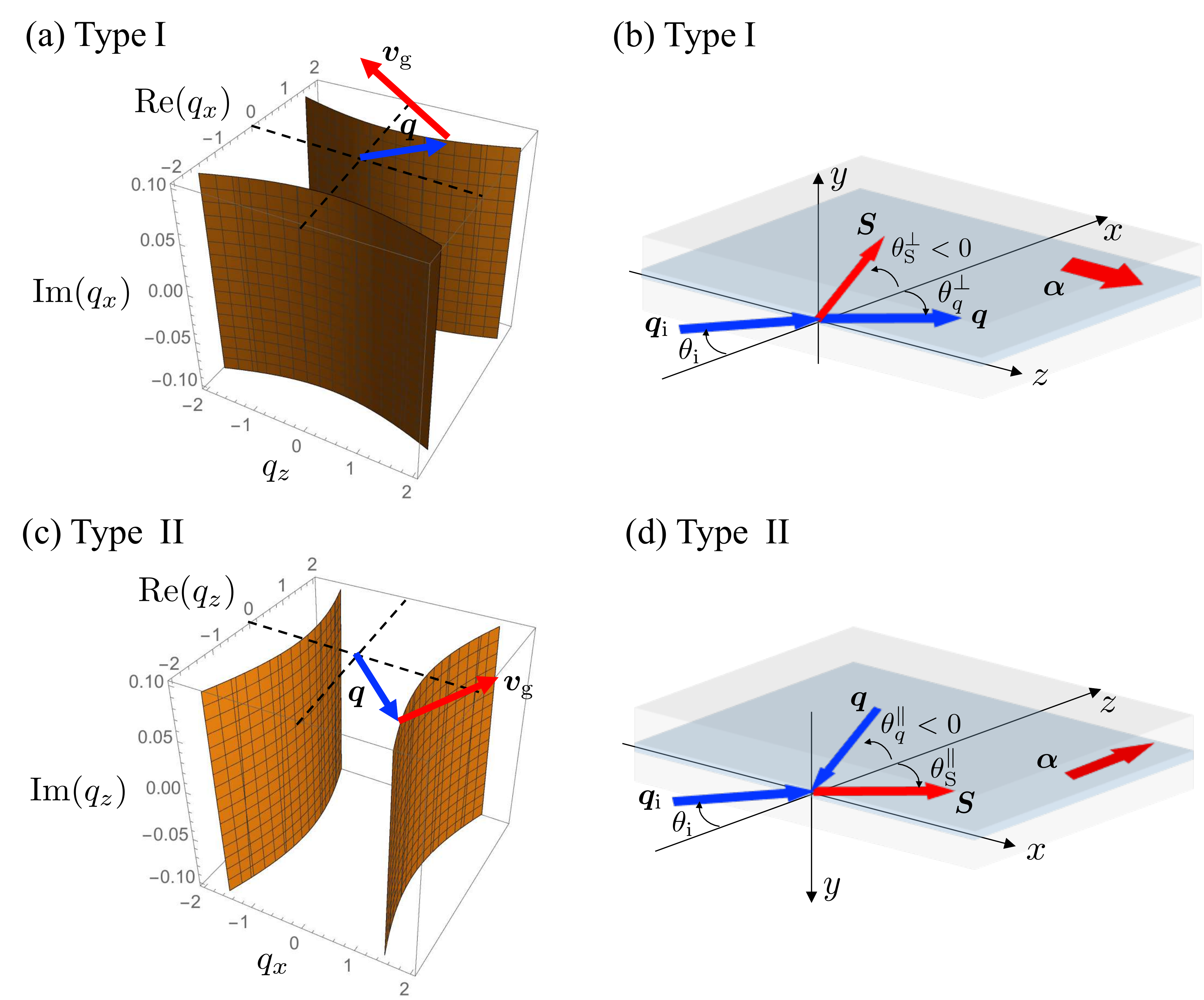}
\caption{Equifrequency dispersion curve in hyperbolic frequency region for 
$\omega_{\rm p}^{\parallel}< \omega<\omega_{\rm R}$  
and schematic diagram of refraction of extraordinary plane waves 
obliquely incident from vacuum on semi-infinite Rashba conductor.  
For the Type-I and --II cases, $q_{z}$ and $q_{x}$ are taken to be real, respectively. 
(b) For the Type-I hyperboloid, the energy flow is refracted to the negative side at the interface  
for all incident angles $\theta_{\rm i}$.    
(d) On the other hand, 
for the Type-II hyperboloid, it is possible that the normal component of the wave vector can be taken to be negative 
within the positive energy flow direction. 
}
\label{Sec7-TypeI-II}
\end{figure}

To illustrate the dispersion curve of Eq.~(\ref{dispersion-ew}) for the extraordinary wave,  
we consider the frequency dependence of $\vare_{x}$ and $\vare_{z}$ 
in Eqs. (\ref{vare-perp}) and (\ref{vare-para}), respectively.  
The real and imaginary parts of $\vare_{z}$ and 
$\vare_{x}$ as functions of $\omega$ are plotted in Fig.~\ref{epsilon}. 
It is apparent that there is a region $\omega^{\parallel}_{\rm p} < \omega < \omega_{\rm R}$, 
in which ${\rm Re}(\vare_{z})>0$ and ${\rm Re}(\vare_{x})<0$. 
Here, $\omega_{\rm R}$ is the modified plasma frequency due to the RSOI and 
is determined by  ${\rm Re}(\vare_{x}(\omega_{\rm R})) = 0$. 
In this frequency region, the equifrequency dispersion surface is hyperbolic  
and is further classified into two types (Fig.~\ref{Sec7-TypeI-II}(a, c)). 

The Type-I hyperboloid has a surface gap in the $x$-direction when $q_{z}$ is real, 
which can be realized by the experimental configuration illustrated in 
Fig.~\ref{Sec7-TypeI-II}(b), 
where the incident plane includes the $z$-axis.  
In this case, 
the transverse component of the group velocity ${\bm v}_{\rm g}=\nabla_{\qv}\omega(\qv)$,  
which is equivalent to ${\bm S}$,   
has the opposite sign to that of $\qv$ (Fig.~\ref{Sec7-TypeI-II}(a)).  
This implies that ${\bm S}$ is refracted to the negative side 
and $\qv$ is on the positive side 
with respect to the interface normal ($x$-axis in Fig.~\ref{Sec7-TypeI-II}(b)).  
Thus, for the Type-I hyperboloid, a negative refraction is expected 
at the interface\cite{{Lindell01},{Smith03},{Belov03}}. 
The Type-II hyperboloid has a surface gap in the $z$-direction 
for real $q_{x}$, 
which can be realized by the experimental configuration illustrated 
in Fig.~\ref{Sec7-TypeI-II}(d), 
where the incident plane includes the $x$-axis.  
In this case, 
it is possible that the direction of the group velocity is positive and 
the normal component of $\qv$, i.e., $q_{x}$, is negative 
(Fig.~\ref{Sec7-TypeI-II}(c)).  
This implies that the energy flow is refracted to the positive side with respect to the interface normal (the $z$-axis in Fig.~\ref{Sec7-TypeI-II}(d)) 
and that $\qv$ points in the negative direction in the Rashba conductor. 
Thus, for a Type-II hyperboloid,  
it is expected that a backward wave (i.e., a wave with negative phase velocity) is realized
\cite{{Lindell01},{Smith03},{Belov03}}. 
Further details of the above behaviors are presented in the following subsections.  

For BiTeI, $\omega_{\rm R} \simeq 1.4 \times 10^{15}~{\rm rad /s}
~(\hbar\omega_{\rm R}\simeq 0.9~{\rm eV})$, 
which is larger than the bare plasma frequency $\omega_{\rm p}^{\perp}\simeq 
1.2\times10^{15}~{\rm rad/s}$.   
This yields the hyperbolic frequency region 
$\omega^{\rm \parallel}_{\rm p}< \omega  < \omega_{\rm R}$ 
with $\omega_{\rm p}^{\parallel} \simeq 5.5 \times 10^{14}~{\rm rad /s} ~(\hbar\omega_{\rm p}^{\parallel}\simeq 0.36~{\rm eV})$, 
in which ${\rm Re}(\vare_{z})>0$ and 
${\rm Re}(\vare_{x})<0$ (Fig.~\ref{epsilon}); this covers the infrared region.  
Thus, in this hyperbolic frequency region, 
the Rashba conductor is insulating in the $z$-direction and metallic in the $x$-direction.  
 
Note that the increase in plasma frequency 
($\omega_{\rm p}^{\perp} \to \omega _{\rm R})$  
is due to the RSOI on top of the electron-mass anisotropy. 
When 
$\omega_{\rm p}^{\perp} = \sqrt{m_{\parallel}/m_{\perp}}\omega_{\rm p}^{\parallel}$ is larger than the
$\omega_{0}$ 
at which ${\rm Re}(C(\omega_{0}))=0$, 
${\rm Re}(C(\omega_{\rm R}))$ becomes positive (Fig.~\ref{CDE}(a)).  
As the threshold is given by $m_{\parallel}/m_{\perp} \simeq 2.4$, 
the RSOI contributes to the expansion of the hyperbolic frequency region in the case of BiTeI ($m_{\parallel}/m_{\perp} \simeq 5)$. 
In contrast, 
in the isotropic mass model 
($m_{\perp}=m_{\parallel}$) \cite{STKT16}, 
$\omega_{\rm R}$ is always smaller than $\omega^{\perp}_{\rm p}(=\omega^{\parallel}_{\rm p})$. 
Thus, in the hyperbolic region, the metallic and insulating directions are interchanged.

\subsubsection{Negative refraction in Type-I hyperboloid}

Here, we demonstrate that the Type-I hyperboloid exhibits negative refraction. 
Let us consider a semi-infinite Rashba conductor in vacuum (Fig.~\ref{Sec7-TypeI-II}(b)), with the refraction of a plane wave at oblique incidence. 
The incident wave vector is 
$\qv_{\rm i} = (\omega/c)(\sin \theta_{\rm i}{\bm e}_{x} + \cos \theta_{\rm i}{\bm e}_{z})$, 
where $\theta_{\rm i}$ is the angle of incidence. 
We first specify the direction of $\qv = q_{x}{\bm e}_{x} + q_{z}{\bm e}_{z}$ 
in the Rashba conductor from the energy flow perspective. 
From Fig.~\ref{Sec7-TypeI-II}(a), 
the conservation of the tangential component of $\qv$,  
which can be taken to be real and positive, 
yields $q_{z}=(\omega/c)\sin\theta_{\rm i}$. 
Substituting this expression into Eq.~(\ref{dispersion-ew}), we have
\begin{align}
q_{x} = \pm\frac{\omega}{c}
\left(\sqrt{\frac{\vare'_{\rm I}+|\vare_{\rm I}|}{2}}
+i{\rm sgn}(\vare''_{\rm I})\sqrt{
\frac{-\vare'_{\rm I}+|\vare_{\rm I}|}{2}}\right), 
\label{q-perp}
\end{align}
where 
\begin{align}
&\vare_{\rm I}=\vare_{\rm I}'+i\vare''_{\rm I} = 
\vare_{z}-\frac{\vare_{z}}{\vare_{x}}\sin^2\theta_{\rm i}, 
\end{align}
with $\vare'_{\rm I}$ and $\vare''_{\rm I}$ being the real and imaginary 
parts of $\vare_{\rm I}$, respectively. 
Further, ${\rm sgn}(x)$ is a sign function, with ${\rm sgn}(x>0) =1$ and 
${\rm sgn}(x<0) = -1$.  
Note that ${\rm sgn}(\vare''_{\rm I})=1$ in the hyperbolic region,  
as $\vare''_{\rm  I}={\rm Im}\vare_{z}-|\vare_{x}|^{-2}(
{\rm Re}(\vare_{x}){\rm Im}(\vare_{z})-{\rm Re}(\vare_{z}){\rm Im}(\vare_{x}))\sin^{2}\theta_{\rm i}>0$ (see Fig.~\ref{epsilon}). 
However, at this stage, 
we cannot determine the $\pm$ selection in Eq.~(\ref{q-perp}). 
To overcome this problem, we must consider the energy flow in the Rashba conductor \cite{{Belov03},{Tatara13}}.

The time-averaged Poynting vector is given by (see Appendix D) 
\begin{align}
{\bm S}(\rv)
=& \frac{\omega\vare_{0}}{2}\left\{{\rm Re}\left(
\frac{\vare_{z}}{q_{x}}\right)|E_{z}|^2{\bm e}_{x}
+
{\rm Re}\left(\frac{\vare_{x}}{q_{z}}\right)
|E_{x}|^2{\bm e}_{z}
\right\}\nonumber\\
&\times
e^{-2{\rm Im}\qv\cdot\rv}. 
\label{S-B}
\end{align}
As the energy must flow into the medium from the interface (Fig.~\ref{Sec7-TypeI-II}(b)), 
${\bm S}({\bm 0})\cdot{\bm e}_{x}=S_{x}({\bm 0})  >0$, or 
\begin{align}
{\rm Re}\left(\frac{\vare_{z}}{q_{x}}\right) =
\frac{{\rm Re}(\vare_{z}){\rm Re}(q_{x})
+{\rm Im}(\vare_{z}){\rm Im}(q_{x})}{|q_{x}|^2}> 0. 
\label{c-I}
\end{align}
Further, as ${\rm Re}(\vare_{z})>0$ and ${\rm Im}(\vare_{z})>0$ 
in the hyperbolic region (Fig.~\ref{epsilon}), 
we must choose the ``+'' option for Eq.~(\ref{q-perp}). 
Thus, both the real and imaginary parts of $q_{x}$ are positive in the hyperbolic region 
(${\rm Re}(q_{z})>0$ and ${\rm Im}(q_{z})>0$), which exhibits a forward wave with loss. 
On the other hand, the tangential component of the Poynting vector 
at $\rv={\bm 0}$ is negative, as   
${\bm S}({\bm 0})\cdot{\bm e}_{z}=S_{z}({\bm 0}) \propto {\rm Re} (\vare_{x}/q_{z}) 
=c{\rm Re}(\vare_{x})/(\omega\sin\theta_{\rm i})<0$ (Fig.~\ref{epsilon}).   
Thus, for the Type-I hyperboloid, we conclude that the inequalities ${\rm Re}(q_{x})>0$ and $S_{z}({\bm 0}) < 0$ 
are satisfied in the hyperbolic region for all $\theta_{\rm i}$; in the hyperbolic region, a forward wave and negative refraction are exhibited (Fig.~\ref{Sec7-TypeI-II}(b)) \cite{Belov03}. 

\begin{figure*}[t]
\begin{center}
\resizebox{16.8cm}{!}{\includegraphics[angle=0]{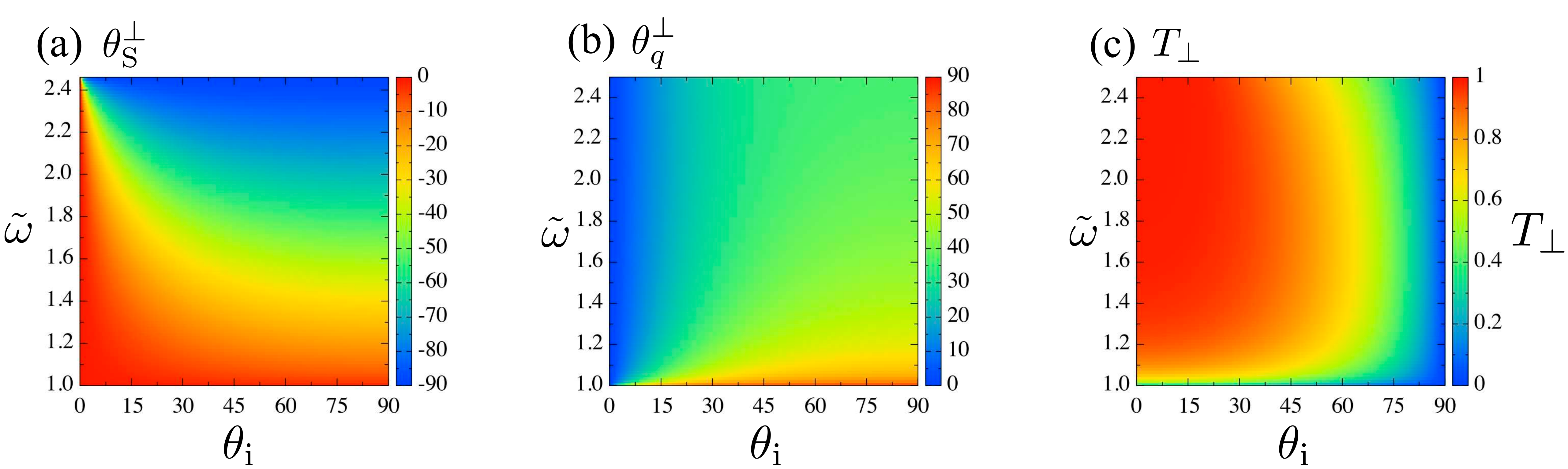}}
\end{center}
\caption{Density plots of (a) Poynting-vector angle of refraction $\theta_{S}^{\perp}$,  
(b) wave-vector angle of refraction $\theta_{q}^{\perp}$, 
and  
(c) transmittance $T_{\perp}$ of Rashba interface, as functions of $\theta_{\rm i}$ and $\tilde{\omega}$.}
\label{Density-Plots-I}
\end{figure*}

To confirm this finding, 
we evaluate the angles of refraction of ${\bm S}$ and $\qv$, $\theta_{S}^{\perp}$ and $\theta_{q}^{\perp}$, respectively, along with the transmittance $T_{\perp}$ at the interface. 
 These features are given by  
\begin{align}
\label{TS-perp}
&\tan\theta_{S}^{\perp} = \frac{S_{z}({\bm 0})}{S_{x}({\bm 0})}
=\frac{{\rm Re}(1/\vare_{x})}{{\rm Re}(q_{x}
/\vare_{z})}\frac{\omega}{c}\sin\theta_{\rm i}, \\
&\tan\theta_{q}^{\perp} = \frac{q_{z}}{{\rm Re}(q_{x})}=\frac{1}{{\rm Re}(q_{x})}\frac{\omega}{c}
\sin\theta_{\rm i}, \\
&T_{\perp}= \frac{S_{x}({\bm 0})}{S_{{\rm i},x}}=
\dfrac{4\dfrac{\omega}{c}{\rm Re}\left(\dfrac{\vare_{z}}{q_{x}}\right)\cos\theta_{\rm i}}{\left|
1+\dfrac{\omega}{c}\dfrac{\vare_{z}}{q_{x}}\cos\theta_{\rm i}
\right|^2}, 
\label{Tperp}
\end{align}
where 
$S_{{\rm i},x}$ is the normal component of 
the time-averaged Poynting vector of an incident plane wave in vacuum. 
The calculations of $\theta_{S}^{\perp}$ and $T_{\perp}$ are presented in Appendix D. 
In addition, $\theta_{S}^{\perp}$, $\theta_{q}^{\perp}$, 
and $T_{\perp}$ 
are presented as functions of $\theta_{\rm i}$ and  
$\tilde{\omega}=\omega/\omega_{\rm p}^{\parallel}$ in Fig.~\ref{Density-Plots-I}.  
It is apparent that 
$\theta_{S}^{\perp}$ (Fig.~\ref{Density-Plots-I}(a)) 
and $\theta_{q}^{\perp}$ (Fig.~\ref{Density-Plots-I}(b))  
are negative and positive, respectively, for all $\theta_{\rm i}$ and 
in the hyperbolic frequency region 
($\omega_{\rm p}^{\parallel} < \omega < \omega_{\rm R}$). 
Note that, for a given frequency $\tilde{\omega}$,
the negative $\theta_{S}^{\perp}$ 
takes constant values and 
the absolute value of the angle increases with increasing $\omega$.  
This focusing effect \cite{STKT16} is stronger for larger frequencies close to 
$\tilde{\omega}_{\rm R}$. 
However, 
as the transmission window appears for small $\theta_{\rm i}$,  
the focusing effect is observable for $\theta_{\rm i} < \sim 30^{\circ}$.

\subsubsection{Backward wave in Type-II hyperboloid} 

Here, we focus on a wave propagation in the Type-II configuration and 
demonstrate the backward wave phenomenon.  
As shown in Fig.~\ref{Sec7-TypeI-II}(d),  
the conservation of the tangential component of  $\qv$ 
yields $q_{x} = (\omega/c)\sin\theta_{\rm i}$. 
Substituting this expression into Eq.~(\ref{dispersion-ew}), 
we have 
\begin{align}
q_{z} = 
\pm\frac{\omega}{c}
\left(\sqrt{\frac{\vare'_{\rm II}+|\vare_{\rm II}|}{2}}
+i{\rm sgn}(\vare''_{\rm II})\sqrt{
\frac{-\vare'_{\rm II}+|\vare_{\rm II}|}{2}}\right),
\label{q-para}
\end{align}
where 
\begin{align}
\vare_{\rm II}=\vare_{\rm II}'+i\vare''_{\rm II} = 
\vare_{x}-\frac{\vare_{x}}{\vare_{z}}\sin^2\theta_{\rm i}, 
\end{align}
with $\vare'_{\rm II}$ and $\vare''_{\rm II}$ being the real and imaginary 
parts of $\vare_{\rm II}$, respectively. 
To determine the $\pm$ selection in Eq.~(\ref{q-para}), 
we consider the energy flow, as previously. 
As the energy must flow away from the interface (Fig.~\ref{Sec7-TypeI-II}(d)), 
the inequality $S_{z}({\bm 0})>0$, or 
\begin{align}
{\rm Re}\left(\frac{\vare_{x}}{q_{z}}\right) 
=\frac{
{\rm Re}(\vare_{x}){\rm Re}(q_{z})+{\rm Im}(\vare_{x}){\rm Im}(q_{z})
}{|q_{z}|^2}
> 0, 
\label{c-I}
\end{align}
must be satisfied. 
Further, as ${\rm Re}(\vare_{x})<0$ and ${\rm Im}(\vare_{x})>0$ 
in the hyperbolic region ($\omega_{\rm p}^{\parallel}<\omega<\omega_{\rm R}$), 
the inequality ${\rm Re}(\vare_{x}){\rm Re}(q_{z})+{\rm Im}(\vare_{x}){\rm Im}(q_{z} )>0$ 
is satisfied for  
(i) a  forward wave with loss, corresponding to 
${\rm Re}(q_{z})>0$ and ${\rm Im}(q_{z})>0$ for ${\rm Im}(\vare_{\rm II} )>0 $; 
(ii) an evanescent wave, corresponding to ${\rm Re}(q_{z}) = 0$  
and ${\rm Im}(q_{\parallel} )>0$; 
and 
(iii) a backward wave with loss, corresponding to ${\rm Re}(q_{z} )<0$ 
and ${\rm Im}(q_{z} )>0$ for ${\rm Im}(\vare_{\rm II}) <0$. 
For (i), we confirm this response numerically. As regards (ii) and (iii), these conclusions are obvious, as
${\rm Re}(\vare_{x})<0$ and ${\rm Im}(\vare_{x})>0$. 
On the other hand, 
the tangential component of the Poynting vector
in Eq.~(\ref{S-B}) is always positive, as 
${\rm Re} (\vare_{z}/q_{x}) = 
c{\rm Re}\vare_{z}/(\omega \sin\theta_{\rm i})>0$ (Fig.~\ref{Sec7-TypeI-II}(d)); this indicates that the energy flow is always refracted to the positive side with respect to the interface normal ($x$-axis). 
Therefore, we conclude that a backward wave and positive refraction are expected for a Type-II hyperboloid \cite{Belov03}. 

To confirm the validity of the above argument, 
we evaluate the angles of refraction of ${\bm S}$ and $\qv$, $\theta_{S}^{\parallel}$ and $\theta_{q}^{\parallel}$, respectively, along with the transmittance $T_{\parallel}$ at the interface.  
These features are respectively given by 
\begin{align}
\label{TS-para}
&\tan\theta_{S}^{\parallel} 
= \frac{S_{x}({\bm 0})}{S_{z}({\bm 0})}
=\frac{{\rm Re}(1/\vare_{z})}{{\rm Re}(q_{z}/\vare_{x})}\frac{\omega}{c}\sin\theta_{\rm i}, \\
\label{Tq-para}
&\tan\theta_{q}^{\parallel} =\frac{q_{x}}{{\rm Re}(q_{z})}= \frac{1}{{\rm Re}(q_{z})}\frac{\omega}{c}\sin\theta_{\rm i}, \\
&T_{\parallel}= 
\frac{S_{z}({\bm 0})}{S_{{\rm i},z}}=
\dfrac{4\dfrac{\omega}{c}{\rm Re}\left(\dfrac{\vare_{x}}{q_{z}}\right)\cos\theta_{\rm i}}{\left|
1+\dfrac{\omega}{c}\dfrac{\vare_{x}}{q_{z}}\cos\theta_{\rm i}
\right|^2}.  
\label{Tpara}
\end{align}
The calculations of 
$\theta_{S}^{\parallel}$ and $T_{\parallel}$ are presented in Appendix D. 
Futher, Fig. \ref{Density-Plots-II} shows 
$\theta_{S}^{\parallel}$, 
$\theta_{q}^{\perp}$, and 
$T_{\parallel}$ as functions of all $\theta_{\rm i}$  
and 
$\tilde{\omega}\equiv\omega/\omega_{\rm p}^{\parallel}$ 
in the hyperbolic frequency region ($1.0 <\tilde{\omega} < 2.37$).  
It is apparent that the energy flow is always in the positive direction 
(Fig.~\ref{Density-Plots-II}(a)), whereas  
$\qv$ changes from negative to positive directions  
($-90^{\circ} < \theta_{q}^{\perp} <  90^{\circ}$) 
with increasing $\theta_{\rm i}$ (Fig.~\ref{Density-Plots-II}(b)). 
We find that the energy of the electromagnetic wave 
propagates along the Rashba interface around the 
${\rm Re}(q_{z})=0$ region (Fig.~\ref{Density-Plots-II}(a, c)). 
The existence of such a surface wave can be understood from 
the dispersion relation perspective. 
For a given $\omega$, the Type-II hyperboloid allows a complex wave vector  
$\qv = i({\rm Im}q_{z}){\bm e}_{x} + q_{x}{\bm e}_{x}$, 
which corresponds to ${\rm Re}(\vare_{\rm II})<0$ and ${\rm Im}(\vare_{\rm I})=0$. 
Thus, this wave can propagate along the interface 
and evanesce in the Rashba conductor. 
Furthermore, 
Fig.~\ref{Density-Plots-II}(c) shows that 
the transmission at the interface is almost complete, i.e., $T_{\parallel} \simeq 1$, 
covering broad ranges of $\omega$ and $\theta_{\rm i}$. 
\begin{figure*}[t]
\begin{center}
\resizebox{16.8cm}{!}{\includegraphics[angle=0]{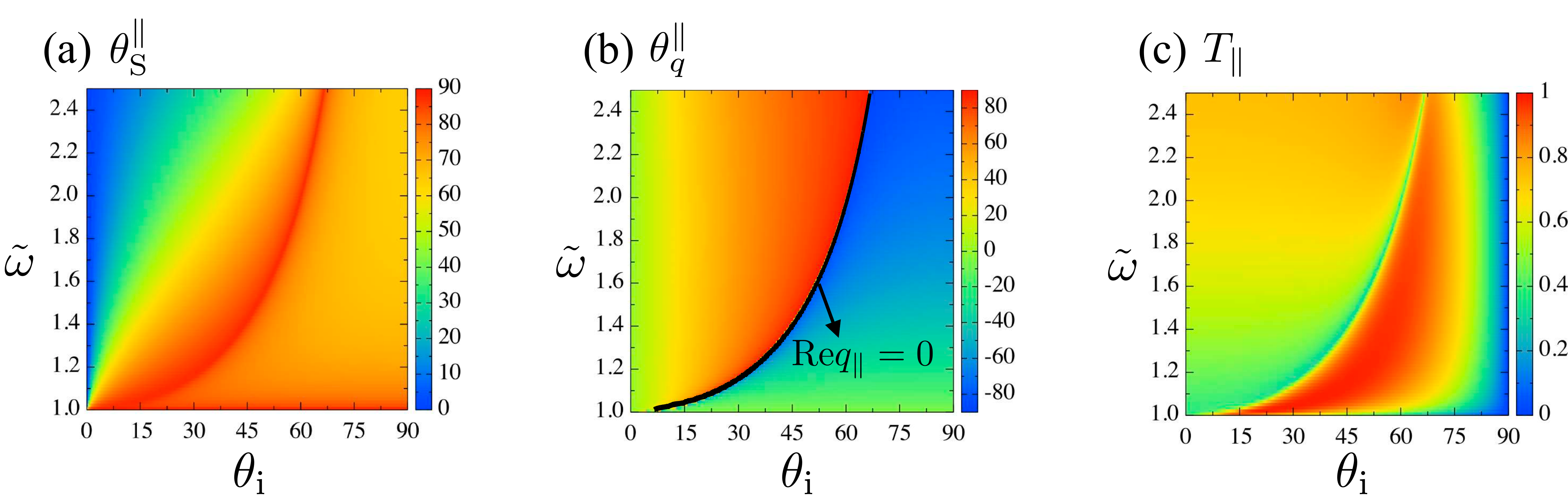}}
\end{center}
\caption{Density plots of (a) Poynting-vector angle of refraction $\theta_{S}^{\parallel}$,  
(b) wave-vector angle of refraction $\theta_{q}^{\parallel}$, 
and  
(c) transmittance $T_{\parallel}$ of Rashba interface,  
as functions of $\theta_{\rm i}$ and $\tilde{\omega}$. The thick black line in (b) separates the forward 
$({\rm Re}q_{\parallel}>0)$ and backward $({\rm Re}q_{\parallel} < 0)$ waves. }
\label{Density-Plots-II}
\end{figure*}
\subsection{Effects of $\alpha_{ijk}(\omega)q_{k}$: Rashba-induced birefringence}

As the Rashba conductor breaks space inversion symmetry, 
the $\qv$-linear term $\alpha_{ijk}(\omega)q_{k}$ can exist in the dielectric tensor.  
This term originates from the electromagnetic cross-correlation effects, i.e., 
the combination of the Edelstein and inverse Edelstein effects. 
Substituting the optical conductivity 
$\sigma^{\rm E-IE}_{ij}(\qv,\omega)$ [Eq.~(\ref{oc-EIE})] into Eq.~(\ref{dielectric}), 
we obtain  
\begin{align}
\label{dielectric1-1}
\alpha_{ijk}q_{k} 
&= \frac{1}{\vare_{0}}\frac{i}{\omega}\sigma^{\rm E-IE}_{ij}(\qv,\omega),
\nonumber\\
&= \begin{pmatrix}
0&0&-i\alpha_{\rm R}(\omega)cq_{x}/\omega\\
0&0&0\\
i\alpha_{\rm R}(\omega)cq_{x}/\omega
\end{pmatrix}, 
\end{align}
where 
\begin{align}
\label{alpha-R}
\alpha_{\rm R}(\omega)
= \frac{1}{2\tilde{\alpha}}\frac{\omega^{\perp}_{\rm p}}{ck^{\perp}_{\rm F}}\frac{\omega^{\perp}_{\rm p}}{\omega+i\eta}C(\omega). 
\end{align}
It is apparent that 
the dielectric tensor of Eq.~(\ref{dielectric1-1}) contains only off-diagonal components and has the same form as Eq.~(\ref{dielectric-lb}). 
Thus, linear birefringence is expected, which we call ``Rashba-induced birefringence.'' 

\begin{figure}[h]
\centering
\resizebox{8.4cm}{!}{\includegraphics{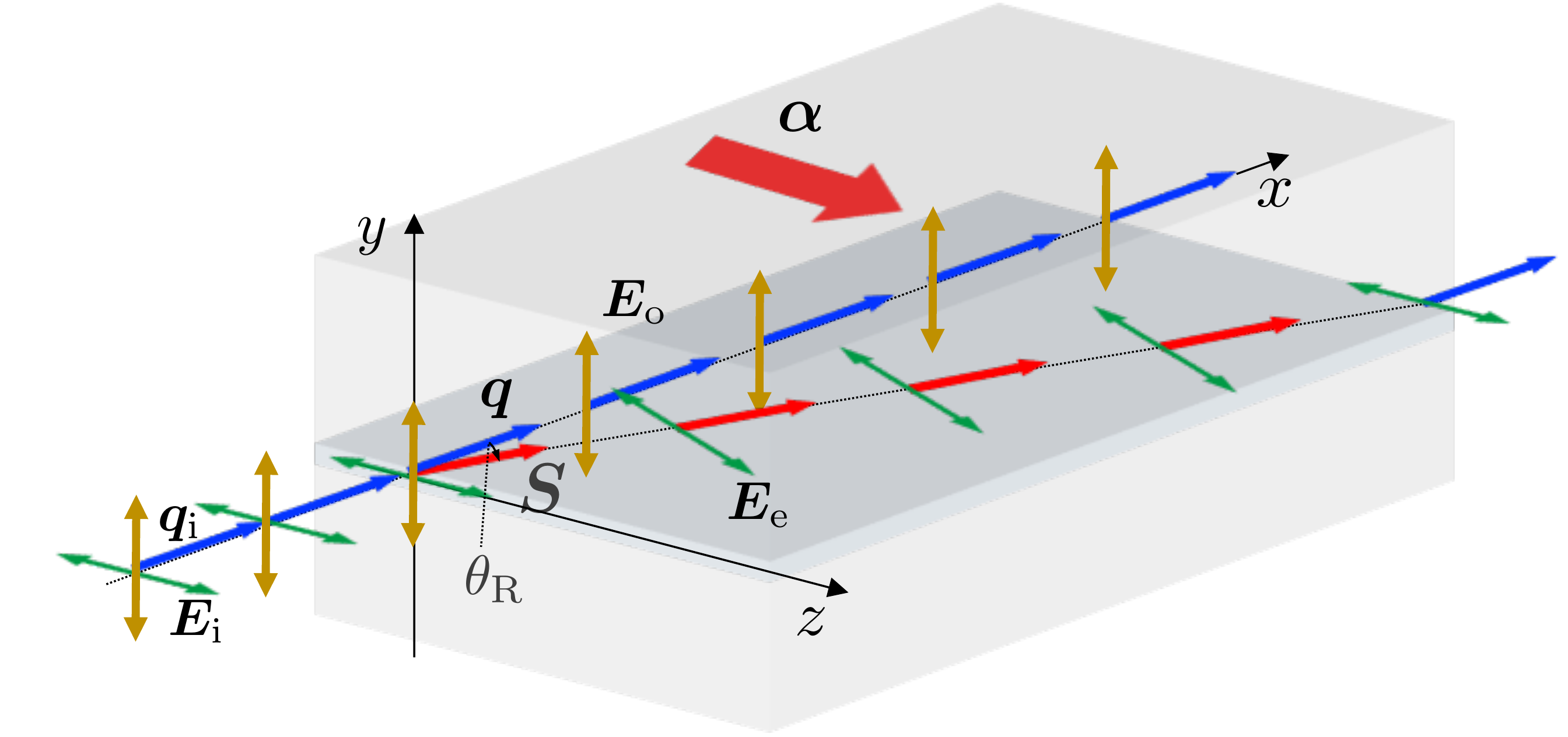}}
\caption{Schematic illustration of Rashba-induced linear birefringence, 
in which a linearly polarized wave is refracted at the interface between a vacuum and a Rashba conductor.}
\label{Sec7-RB}
\end{figure}

\begin{figure}[h]
\centering
\resizebox{8.4cm}{!}{\includegraphics{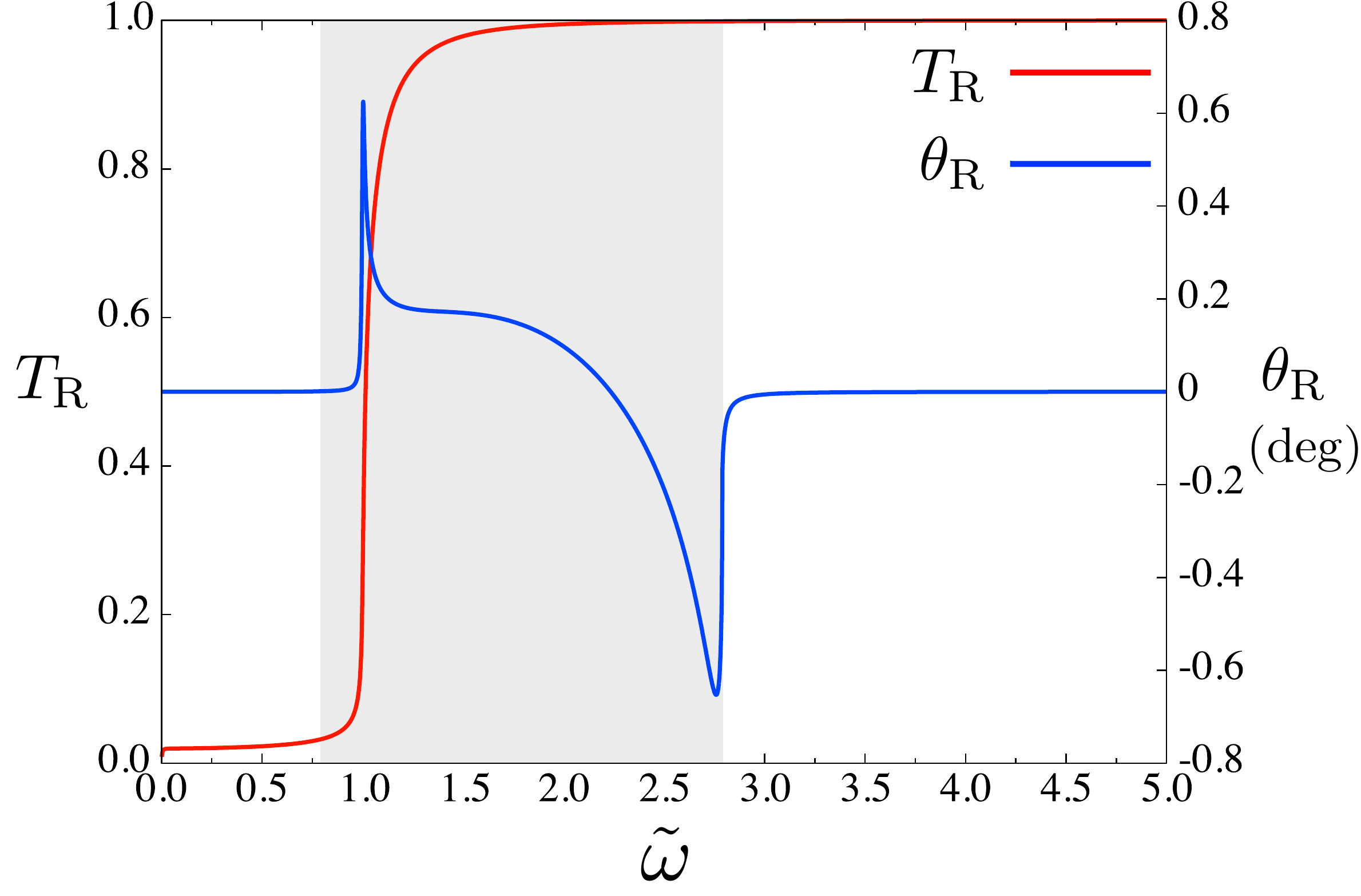}}
\caption{Transmittance $T_{\rm R}$ and tilt angle $\theta_{\rm R}$ 
as functions of $\tilde{\omega}$ 
for Rashba-induced linear birefringence. }
\label{Sec7-T-Theta}
\end{figure}

Let us consider the normal wave incidence shown in Fig.~\ref{Sec7-RB}. 
where the direction of the incident-wave electric field is along the 
$z$-axis: ${\bm E}_{\rm i}(\rv,t)=E_{\rm i}{\bm e}_{z}e^{i\omega(x/c-t)}$. 
In this configuration, by using the result of Sec IV-B-2, 
we obtain two types of wave solution. 
One is an ordinary wave with dispersion relation 
$cq_{\rm o}(\omega) = \omega\sqrt{\vare_{x}(\omega)}$ 
and eigen vector 
${\bm E}_{\rm o}=E_{0}{\bm e}_{y}$. 
The other is an extraordinary wave, 
the dispersion relation of which is given by $q = q_{\rm R}(\omega)$, where 
\begin{align}
q_{\rm R} 
= \frac{\omega}{c}\sqrt{\frac{\vare_{z}\vare_{x}}{\vare_{x} + \alpha_{\rm R}^2}}\simeq \frac{\omega}{c}\sqrt{\vare_{z}}, 
\end{align} 
assuming $|\vare_{x}| \gg |\alpha^2_{\rm R}|$. 
The electric field 
${\bm E}_{\rm e} = E_{0x}{\bm e}_{x} + E_{0z}{\bm e}_{z}$  
acquires a longitudinal component, such that 
\begin{align} 
E_{0x} = \frac{i\alpha_{\rm R}}{\vare_{x}}\sqrt{\vare_{z}}E_{0z},  
\label{eigen-Rb}
\end{align}
at first order in $\alpha_{\rm R}$. 
Let us evaluate the transmittance $T_{\rm R}$ and 
tilt angle $\theta_{\rm R}$ due to the Rashba-induced birefringence, 
which is determined by the angle between $\qv$ and ${\bm S}$. 
The time-averaged Poynting vector of the extraordinary wave is given by 
\begin{align}
{\bm S}(\rv) = \frac{\vare_{0}c}{2}|E_{0z}|^{2}
\left[{\rm Re}(\sqrt{\vare_{z}}){\bm e}_{x} 
- {\rm Re}\left(\frac{i\alpha_{\rm R}}{\vare_{x}}\right)
|\vare_{z}|{\bm e}_{z}\right]e^{-2{\rm Im}q_{\rm R}z}
,
\end{align}
which has a component perpendicular to ${\bm q}$.  
The transmittance is given by 
\begin{align}
T_{\rm R} = 
\frac{{\bm S}(0)\cdot{\bm e}_{x}}{{\bm S}_{\rm i}\cdot{\bm e}_{x}}=
\frac{4{\rm Re}(\sqrt{\vare_{z}})}{\left|1+\sqrt{\vare_{z}}\right|^2},
\end{align}
where ${\bm S}_{\rm i} = \dfrac{\vare_{0}c}{2}|E_{\rm i}|^2{\bm e}_{x}$ is the time-averaged Poynting vector of the incident wave. 
The tilt angle at the interface (Fig.~\ref{Sec7-RB}) is 
\begin{align}
\tan\theta_{\rm R} = \frac{{\bm S}(0)\cdot{\bm e}_{z}}{{\bm S}(0)\cdot{\bm e}_{x}}=
-\left[
{\rm Re}\left(\frac{i\alpha_{C}}{\vare_{x}}\right)
\frac{{\rm Re}(\sqrt{\vare_{z}})}{|\vare_{z}|}
\right]. 
\end{align}
We present $T_{\rm R}$ and $\theta_{\rm R}$ as functions of $\omega$ in Fig.~\ref{Sec7-T-Theta}. 
There are two peaks in $\theta_{\rm R}$ 
at $\omega_{\rm p}^{\parallel}$ and $\omega_{+}$. 
As the Rashba conductor is transparent above $\omega_{\rm p}^{\parallel}$, 
a peak is observable at $\omega_{+}$ (Fig.~\ref{Sec7-T-Theta}). 

\section{Wave propagations in ferromagnetic Rashba conductor}

In the presence of ${\bm M}$, 
the system breaks both the space inversion and time reversal symmetries.  
Here, we consider wave propagations for 
the Faraday configuration ($\qv \parallel {\bm M}$) with 
$\hat{\bm \alpha} \cdot {\bm M} \equiv M_{\parallel} \neq 0$, 
and for the Voigt configuration  ($\qv \perp {\bm M}$) 
with $\qv\cdot({\bm \alpha}\times{\bm M})\neq 0$. 

\subsection{Effects of $\beta_{ijk}(\omega)M_{k}$: Faraday and Kerr effects}
\begin{figure}[t]
\resizebox{8.4cm}{!}{\includegraphics{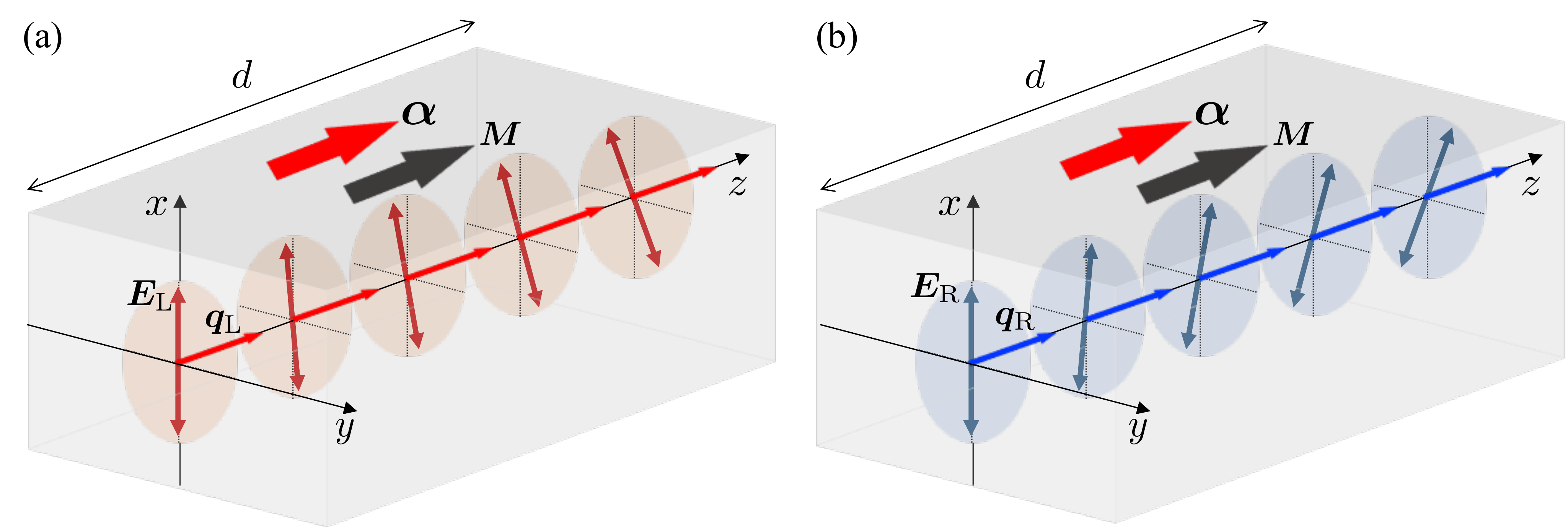}}
\caption{Schematic illustration of Faraday effect. 
(a) Left- and (b) right-handed circularly polarized waves normally incident 
on Rashba conductor slab of thickness $d$ 
in Faraday configuration, where ${\bm M}$ is 
parallel to $\qv_{\rm L,R}$ and ${\bm \alpha}$. 
}
\label{MCD-config}
\end{figure}

The optical conductivity 
$\sigma^{\rm AH}_{ij}(\omega)$ [Eq.~(\ref{oc-AH})]
due to the anomalous Hall effect contributes to the dielectric tensor 
according to
\begin{align}
\label{dielectric2-1}
\beta_{ijk}(\omega)M_{k}&=\frac{1}{\vare_{0}}\frac{i}{\omega}\sigma^{\rm AH}_{ij},\nonumber\\
&=\begin{pmatrix}
0&\vare_{\rm AH}(\omega)&0\\
-\vare_{\rm AH}(\omega)&0&0\\
0&0&0
\end{pmatrix}_{ij},
\end{align}
where 
\begin{align}
\label{vare-AH}
\vare_{\rm AH}(\omega)=
-\frac{(\omega_{\rm p}^{\perp})^2}{\omega(\omega+i\eta)}
\frac{M_{\parallel}}{\epsilon_{\rm F}}iD(\omega).  
\end{align}
Equation (\ref{dielectric2-1}) contains the off-diagonal components only and
has the same form as Eq.~(\ref{tensor-Faraday}). 
Thus, as an ${\bm M}$-induced optical phenomenon, 
the Faraday effect is expected. 
Setting $\qv=q_{z}{\bm e}_{z}$ and ${\bm M} = M{\bm e}_{z}$ in Eq.~(\ref{weq-0}), 
we obtain the dispersion relations $q_{z} = q_{\rm L,R}(\omega)$ and 
\begin{align}
q_{\rm L,R}(\omega) = \frac{\omega}{c}\sqrt{\vare_{\rm L,R}(\omega)}, 
\label{qLR}
\end{align}
where 
\begin{align}
\vare_{\rm L,R}(\omega) &= 
\vare_{x}(\omega)\pm i \vare_{\rm AH}(\omega),\nonumber\\
&= 1-\frac{(\omega_{\rm p}^{\perp})^2}{\omega(\omega+i\eta)}
\left(1+C(\omega) \pm \frac{M}{\epsilon_{\rm F}}D(\omega)\right),
\end{align}
with eigen vectors 
${\bm E}_{\rm L,R} = ({\bm e}_{x} \pm i {\bm e}_{y})E_{0}/\sqrt{2}$ 
corresponding to the left- $(+)$ and right-handed $(-)$ circularly polarized waves.

\begin{figure*}
\resizebox{16.8cm}{!}{\includegraphics[angle=0]{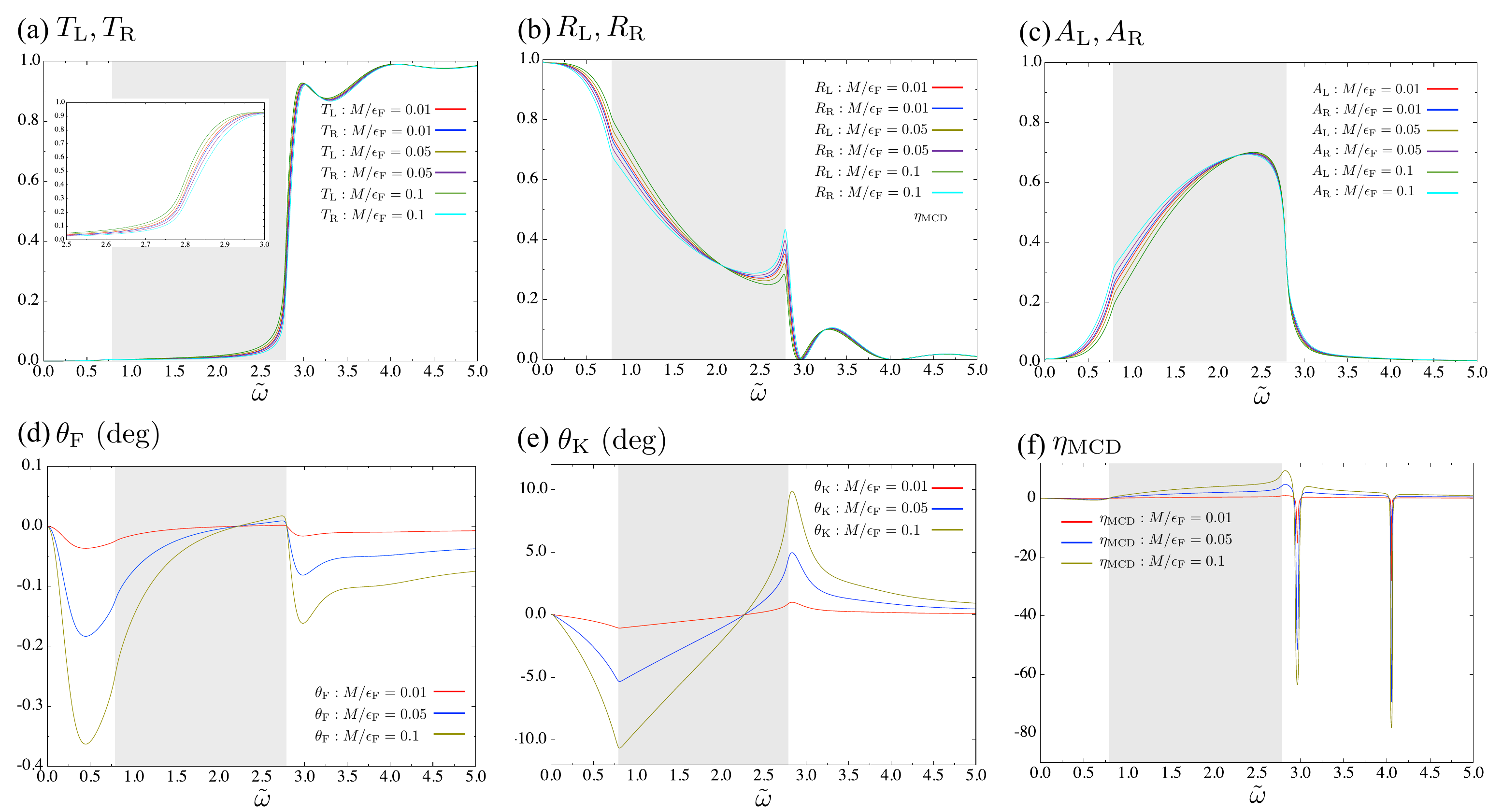}}
\caption{(a) Transmittance $T_{\rm L, R}$, 
(b) reflectance $R_{\rm L, R}$, and (c) absorbance $A_{\rm L,R}$ 
for left- and right-handed circularly polarized waves. 
(d) Faraday rotation angle $\theta_{\rm F}$, 
(e) Kerr rotation angle $\theta_{\rm K}$, 
and 
(f) MCD ratio $\eta_{\rm MCD}$.  
All graphs are plotted as functions of $\tilde{\omega}$ 
for $M/\epsilon_{\rm F} = 0.01, 0.05$, and $0.1$, and thickness $d = 1\,{\rm \mu m}$. }
\label{TRA-MCD}
\end{figure*}

The Faraday rotation angle and the MCD are given by Eqs.~(\ref{Sec3-Faraday-angle}) and~(\ref{sec-3-MCD}), repspectively. 
Here, we evaluate these values more precisely.  
Let us consider an electromagnetic wave propagation normally incident on 
a ferromagnetic Rashba slab with thickness $d$ (Fig.~\ref{MCD-config}).  
When the incident wave is linearly polarized, the electric field vector rotates; this effect is called Faraday or Kerr rotation for the transmitted 
or reflected waves, respectively. 
The corresponding rotation angles are given by 
\begin{align}
&\theta_{\rm F}(\omega) = \frac{1}{2}
{\rm arg}\left(\frac{{\cal T}_{\rm L}(\omega)}{{\cal T}_{\rm R}(\omega)}\right), 
\\
&\theta_{\rm K}(\omega) =\frac{1}{2}
{\rm arg}\left(\frac{{\cal R}_{\rm L}(\omega)}{{\cal R}_{\rm R}(\omega)}\right)
,\end{align}
respectively, where ${\rm arg}(z)$ is the argument of a complex number $z$ 
and 
${\cal T}_{\rm L(R)}(\omega)$ and ${\cal R}_{\rm L(R)}(\omega)$ 
are the transmission and reflection amplitudes, respectively, for each circularly polarized wave (see Appendix D). Here, 
\begin{align}
&{\cal T}_{\rm L,R}(\omega) = \frac{
(1-\chi_{\rm L,R}^2(\omega))e^{i(q_{\rm L,R}(\omega)-\omega/c)d}}{1-\chi^2_{\rm L,R}(\omega)e^{2iq_{\rm L,R}(\omega)d}}, \\
&{\cal R}_{\rm L,R}(\omega) = \frac{\chi_{\rm L,R}(\omega)(1-e^{2iq_{\rm L,R}(\omega)d})}{1-\chi^2_{\rm L,R}(\omega)e^{2iq_{\rm L,R}(\omega)d}}, 
\end{align}
with 
\begin{align}
\chi_{\rm L,R}(\omega) = \frac{1-\sqrt{\vare_{\rm L,R}(\omega)}}{1+\sqrt{\vare_{\rm L,R}(\omega)}}. 
\end{align}
The magnitude of the MCD is defined as the difference in absorption rate between the left- and right-handed circularly polarized waves, i.e.,
\begin{align}
\eta_{\rm MCD}(\omega) = \frac{A_{\rm L}(\omega)-A_{\rm R}(\omega)}{A_{\rm L}(\omega)+A_{\rm R}(\omega)}, 
\end{align}
where $A_{\rm L,R}(\omega)$ is the absorbance. 
The absorbance is obtained from $A_{\rm L,R}(\omega)= 1-T_{\rm L,R}(\omega)-R_{\rm L,R}(\omega)$, 
where $T_{\rm L,R}(\omega)=|{\cal T}_{\rm L,R}(\omega)|^2$ and $R_{\rm L,R}(\omega) = |{\cal R}_{\rm L,R}(\omega)|^2$ 
are the transmittance and reflectance, respectively. 
In Fig.~\ref{TRA-MCD}, $T_{\rm L,R}(\omega)$, $R_{\rm L,R}(\omega)$, $A_{\rm L,R}(\omega)$, 
$\theta_{\rm F}(\omega)$, $\theta_{\rm K}(\omega)$, and ${\eta}_{\rm MCD}(\omega)$ are presented as functions of $\tilde{\omega}=\omega /\omega_{\rm p}^{\parallel}$ 
for $M/\epsilon_{\rm F} = 0.01, 0.05$, and $0.1$ and for $d = 1\,{\rm \mu m}$. 
It is apparent that higher values are obtained in all cases with increasing $M/\epsilon_{\rm F}$. 
For $\omega < \omega_{-}$, the incident wave is almost perfectly reflected at 
the surface of the Rashba conductor. 
Thus, $\theta_{\rm K}(\omega)$ is observable near $\omega_{-}$. 
For $\omega > \omega_{+}$,
as the electromagnetic wave can pass through the Rashba conductor, 
$\theta_{\rm F}(\omega)$ is observable near $\omega_{+}$.  
Note that the two sharp peaks of $\eta_{\rm MCD}$, which appear at $\omega > \omega_{+}$,   
are due to vanishing of $R_{\rm L, R}(\omega)$ at these frequencies; thus, these peaks are unobservable.  
Therefore, the MCD signal can be detected in the reflected wave for 
$\omega < \omega_{-}$ 
and in the transmitted wave for $\omega > \omega_{+}$. 

\subsection{Effects of $\gamma_{ijkl}(\omega)q_{k}M_{l}$: \\  Nonreciprocal directional dichroism}
\begin{figure}[t]
\resizebox{8.4cm}{!}{\includegraphics{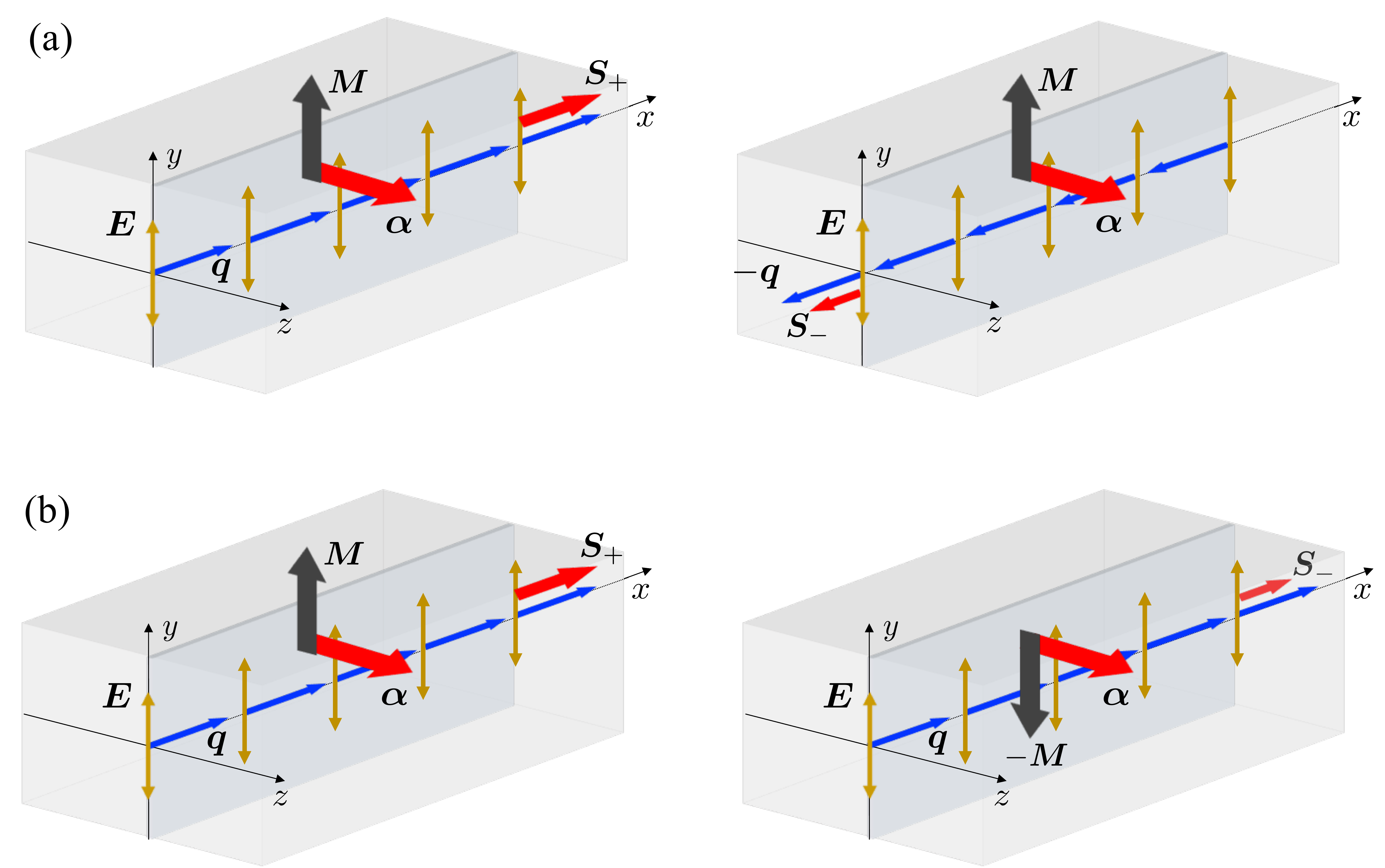}}
\caption{Schematic illustration of nonreciprocal directional dichroism 
for propagation of linearly polarized waves. 
(a) The optical absorption rates differ between the two 
counterpropagating waves ($\qv$ and $-\qv$). 
(b) The optical absorption rates differ between the two cases with 
mutually opposite magnetization (${\bm M}$ and $-{\bm M}$). }
\label{Sec8-NDD}
\end{figure}

Finally, let us consider the nonreciprocal directional dichroism. 
This effect is observable when the system is in the Voigt configuration 
($\qv\perp{\bm M}$) and when $\qv\cdot({\bm \alpha}\times {\bm M})$ is nonzero. 
Here, we set $\qv = q_{x} {\bm e}_{x}$ and ${\bm M} = M{\bm e}_{y}$, as illustrated 
in Fig.~\ref{Sec8-NDD}. 
Thus, the toroidal moment is given by 
$\hat{\cal {\bm T}}=\hat{\bm \alpha}\times\hat{\bm M}=-{\bm e}_{x}$ 
and the quadrupole moment is 
${\cal Q}_{ij}^{\perp} = \hat{\alpha}_{i}\hat{M}^{\perp}_{j}+
\hat{\alpha}_{j}\hat{M}^{\perp}_{i}=\delta_{iz}\delta_{jy}+\delta_{iy}\delta_{jz}$. 
With the optical conductivity $\sigma^{M,\perp}_{ij}(\qv,\omega,{\bm M}_{\perp})$ [Eq.~(\ref{sM-perp})], 
the dielectric tensor is given by 
\begin{align}
\gamma_{ijkl}(\omega)q_{k}M_{l} 
&=\begin{pmatrix}
-2\gamma_{12}(\omega)Mcq_{x}/\omega&0&0\\
0&-2\gamma_{13}(\omega)Mcq_{x}/\omega&0\\
0&0&0
\end{pmatrix}, 
\end{align}
where 
$\gamma_{12}(\omega) = \gamma_{1}(\omega)+\gamma_{2}(\omega)$ and 
$\gamma_{13}(\omega) = \gamma_{1}(\omega)+\gamma_{3}(\omega)$, 
with 
\begin{align}
\label{gamma-E}
\gamma_{\mu}(\omega) = \frac{1}{2}\frac{\omega^{\perp}_{\rm p}}{ck_{\rm F}^{\perp}}\frac{1}{\epsilon_{\rm F}}\frac{\omega^{\perp}_{\rm p}}{\omega+i\eta}E_{\mu}(\omega),~~(\mu= 1,2,3). 
\end{align}

\begin{figure*}
\resizebox{18.8cm}{!}{\includegraphics[angle=0]{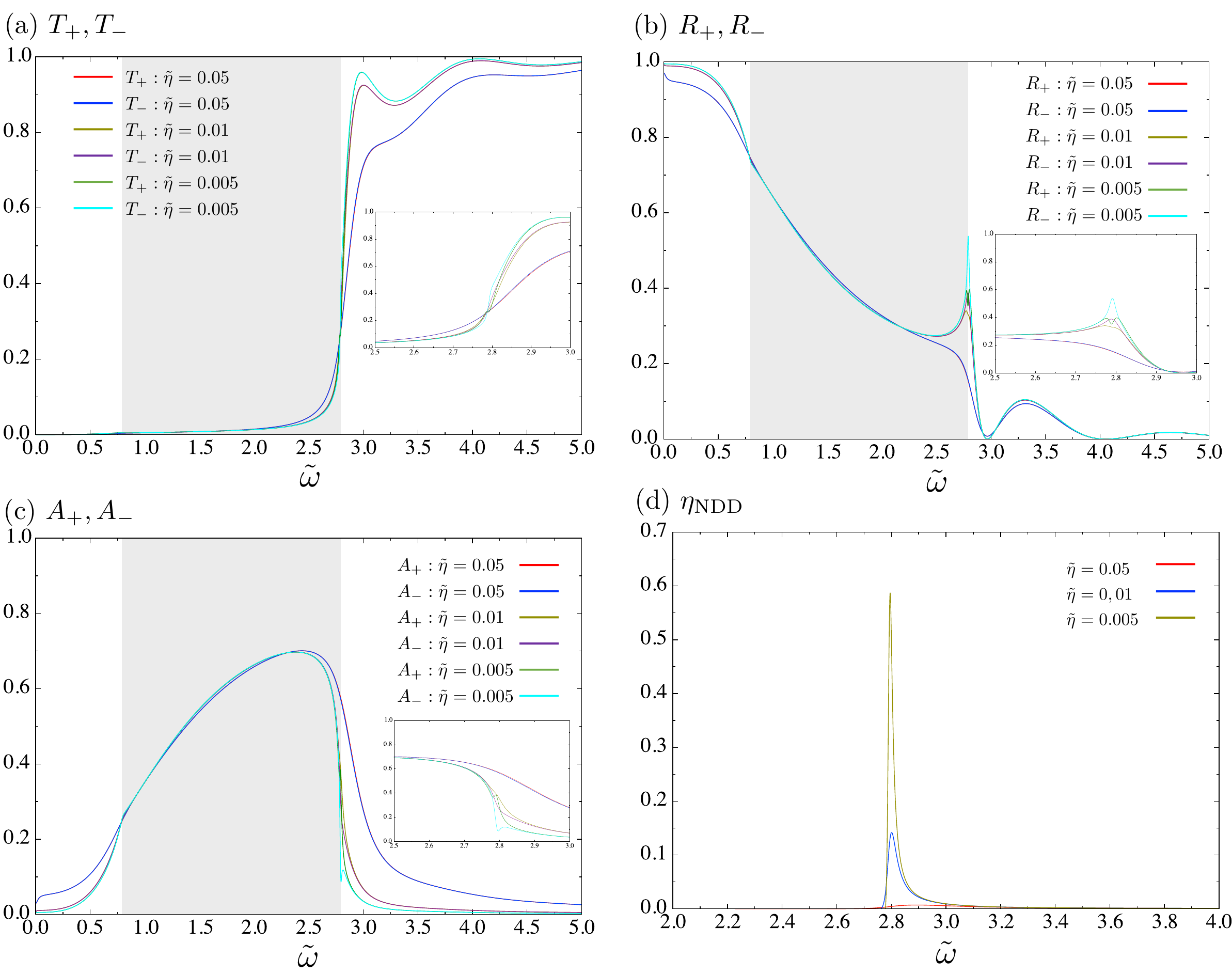}}
\caption{
(a) Transmittance $T_{\pm}$, 
(b) reflectance $R_{\pm}$, 
and (c) absorbance $A_{\pm}$ 
for linearly polarized waves 
propagating in positive and negative directions 
in Rashba conductor with thickness $d = 1.0~{\rm \mu m}$. 
(d) NDD ratio $\eta_{\rm NDD}$. 
All graphs are plotted as functions of $\tilde{\omega}$ 
for $\tilde{\eta}= 0.05, 0.01$, and $0.005$.}
\label{NDD}
\end{figure*}

\begin{widetext}
The wave equation (\ref{weq-0}) is given by 
\begin{align}
\begin{pmatrix}
-\omega^2\vare_{x}(\omega)+2\omega \gamma_{12}(\omega)Mcq_{x}
& 0
&i\omega\alpha_{C}(\omega) cq_{x}\\
0
& c^2q_{x}^2-\omega^2\vare_{x}+ 2\omega \gamma_{13}(\omega)Mcq_{x} 
& 0\\
-i\omega\alpha_{C}(\omega)  cq_{x}
& 0
& c^2q_{x}^2-\omega^2\vare_{z}(\omega)
\end{pmatrix}
\begin{pmatrix}
E_{x}\\
E_{y}\\
E_{z}
\end{pmatrix}
=0, 
\label{weq-3}
\end{align}
which yields the characteristic equation  
\begin{align}
(c^2q^2_{x}-\omega^2\vare_{x}(\omega)+2\omega \gamma_{13}(\omega)Mcq_{x} )
\left\{
(c^2q^2_{x}-\omega^2\vare_{z}(\omega))(\omega^2\vare_{x}(\omega)-2\omega\gamma_{12}(\omega)
Mcq)+\omega^2\alpha^2_{C}(\omega)c^2q^2_{x}
\right\}=0. 
\end{align}
\end{widetext}
For simplicity, 
we concentrate on the wave solution ${\bm E}=E_{y}{\bm e}_{y}$ with linear polarization.   
The characteristic equation 
\begin{align}
c^2q^2_{x}+2\omega \gamma_{13}(\omega)Mcq_{x}-\omega^2\vare_{x}(\omega) =0, 
\label{eigen-eq-NDD}
\end{align}  
contains the linear term with respect to $q_{x}$ and $M$.  
This indicates that, for the replacement $\qv \to -\qv$ or ${\bm M} \to {\bm M}$,  
the sign of the second term of Eq.~(\ref{eigen-eq-NDD}) changes. 
Directional dichroism is expected between the two counterpropagating waves 
(Fig.~\ref{Sec8-NDD}(a)) 
or between the two opposite magnetization directions (Fig.~\ref{Sec8-NDD}(b)).   
The dispersion relation $q_{x}=q_{\pm}(\omega)$ 
for a wave propagating in the positive 
(corresponding to $+\qv$ or $+{\bm M}$) and negative 
(corresponding to $-\qv$ or $-{\bm M}$) directions is given by 
\begin{align}
q_{\pm}(\omega)&=  \pm \frac{\omega}{c}\gamma_{13}(\omega)M+ 
\frac{\omega}{c}\sqrt{\vare_{x}(\omega)+\gamma_{13}^2(\omega)M^2}. 
\label{qpm}
\end{align}
The magnitude of the nonreciprocal directional dichroism (NDD) is defined by the difference in the absorption rate of the waves 
propagating in the positive ($q_{+}$) and negative ($q_{-}$) directions, such that  
\begin{align}
\eta_{\rm NDD}(\omega) = \frac{A_{+}(\omega)-A_{-}(\omega)}{A_{+}(\omega)+A_{-}(\omega)}, 
\end{align}
where $A_{\pm}(\omega)$ is the absorbance for each direction. 
In a similar manner to the MCD discussed in the previous subsection,   
the transmission and reflection amplitudes 
${\cal T}_{\pm}(\omega)$ and ${\cal R}_{\pm}(\omega)$ 
are given by replacing $q_{\rm L,R}(\omega)$ [Eq.~(\ref{qLR})] with $q_{\pm}(\omega)$ [Eq.~(\ref{qpm})], such that 
\begin{align}
&{\cal T}_{\pm}(\omega) = \frac{(1-\chi_{\pm}^2(\omega))e^{i(q_{\pm}(\omega)-\omega/c)d}}{1-\chi^{2}_{\pm}(\omega)e^{2iq_{\pm}(\omega)d}}, \\
&{\cal R}_{\pm}(\omega) = \frac{\chi_{\pm}(\omega)(1-e^{2iq_{\pm}(\omega)d})}{1-\chi^2_{\pm}(\omega)e^{2iq_{\pm}(\omega)d}},\\
&\chi_{\pm}(\omega)=\frac{1\mp\gamma_{13}(\omega)M-\sqrt{\vare_{x}(\omega)+\gamma_{13}^2(\omega)M^2}}
{1\pm\gamma_{13}(\omega)M+\sqrt{\vare_{x}(\omega)+\gamma_{13}^2(\omega)M^2}}. 
\end{align}
In Fig.~\ref{NDD}, the transmittance $T_{\pm}(\omega)$, reflectance $R_{\pm}(\omega)$,  
absorbance $A_{\pm}(\omega)$, and magnitude of $\eta_{\rm NDD}(\omega)$ are presented as 
functions of 
$\tilde{\omega}=\omega /\omega_{\rm p}^{\parallel}$ 
for $M/\epsilon_{\rm F} = 0.1$ 
and 
$\tilde{\eta} \equiv \eta/ \omega_{\rm p}^{\parallel} = 0.05, 0.01$, and $0.005$. 
It is apparent that the absorbance has a sharp peak at 
$\omega_{+}$, 
which strongly depends on $\tilde{\eta}$. 
For $\tilde{\eta}=0.005$, the NDD ratio is $\eta_{\rm NDD}\sim 0.5$ at $\omega=\omega_{+}$. 
This enhanced NDD at the transition edge is attributed to the singular behavior of the $\gamma_{13}(\omega)\propto E_{1}(\omega) + E_{3}(\omega)$ coefficients, 
which originates from the expansion of the correlation functions 
in Eq.~(\ref{gamma-ijkl}) 
with respect to $\qv$ and ${\bm M}$. 
This expansion can be related to the second derivative of $C(\omega)$ for 
$\omega$, 
with $C(\omega) \propto \theta(\omega_{+}-\omega)\sqrt{\omega_{+}-\omega}$ 
close to $\omega_{+}$ in the clean limit ($\eta \to 0$). 
Thus, $\gamma_{13}(\omega) $ behaves as 
$\gamma_{13}(\omega) \propto \dfrac{d^2}{d\omega^2}C(\omega) \sim (\omega_{+}-\omega)^{-3/2}$ and
 $\gamma_{13}(\omega)$ diverges at $\omega_{+}$ for $\eta=0$.  
For a finite $\eta$, 
the directional dichroism is strongly enhanced 
by a factor $\gamma_{13}(\omega) \sim \tilde{\eta}^{-3/2}$ 
at the transition edge. 
For the lower transition edge $\omega_{-}$, similar to the case of $\omega_{+}$, 
the coefficients $E_{1}(\omega)$ and $E_{3}(\omega)$ diverge at $\omega_{-}$ in the clean limit. 
However, the sharp peak is suppressed by a finite value of $\tilde{\eta}$ 
because of the cancellation 
(see Fig.~\ref{F-CDE} in Appendix C).  
Therefore, the NDD signal may be detected near $\omega_{+}$. 
\section{Summary}

We have investigated various types of electromagnetic wave propagation 
in a three-dimensional ferromagnetic Rashba conductor. 
In the first part of the paper, we derived the wave equation for an electric field 
and demonstrated the possible wave propagations in a ferromagnetic conducting medium 
based on the symmetry of the dielectric tensor, as expressed in Eq.~(\ref{expand-epsilon}). 
When the dielectric tensor $\vare_{ij}^{(0)}$ takes a uniaxial form, 
it is possible for the dispersion relation for the extraordinary wave to become hyperbolic. Then, the medium exhibits unusual optical properties known as negative refraction and backward waves. 
When the system breaks the space inversion symmetry, 
a wave-vector $\qv$-linear term ($\alpha_{ijk}q_{k}$) arises in the dielectric tensor. 
This contributes to the antisymmetric (off-diagonal) components, 
and yields $\qv$-induced rotation of the polarization vector ${\bm E}$  
that corresponds to optical activity or linear birefringence. 
In the presence of magnetization ${\bm M}$, 
the system breaks the time reversal symmetry and exhibits 
the ${\bm M}$-induced rotation of ${\bm E}$ 
that is the Faraday effect or the Cotton-Mouton effect. 
These phenomena are described by the 
${\bm M}$-linear term in the dielectric tensor, $\beta_{ijk}M_{k}$, 
which is antisymmetric and has off-diagonal components. 
Finally, when the system breaks both the space inversion and time reversal symmetries, a term appears in the symmetric part of the dielectric tensor  
(possiblly in diagonal components), $\gamma_{ijkl}q_{k}M_{l}$, 
which is linear in both $\qv$ and ${\bm M}$. 
This yields ${\bm M}$-induced spatial dispersion phenomena, namely, 
magneto-chiral dichroism and birefringence for the Faraday configuration, 
and nonreciprocal directional dichroism and birefringence for the Voigt configuration.

In the second part of the study, 
we considered a definite microscopic model, 
namely, a ferromagnetic Rashba conductor, 
and explicitly evaluated the current and spin response functions. 
These functions were combined in an optical conductivity, which was examined 
to the first order of $\qv$ and ${\bm M}$.  
In the absence of the magnetization (${\bm M}={\bm 0}$), 
we demonstrated the anisotropic property of the current response due to the 
electron-mass anisotropy and the direct and inverse Edelstein effects. 
We also demonstrated the current induced by 
the electromagnetic cross correlation effects as 
${\bm j}_{\rm E-IE}\sim ({\bm \alpha}\times \qv)\times{\bm E}$, which represents 
the combined effect of the Edelstein and inverse Edelstein effects. 
This result indicates that electrons flowing in the ${\bm E}$-direction experience 
a Lorentz force, with their orbits being bent by the ${\bm q}$-induced effective magnetic field 
${\bm B}_{\rm eff}(\qv) = {\bm \alpha}\times \qv$; hence, linear birefringence is generated.

In the presence of ${\bm M}$, 
the current and spin responses depend on the relative direction 
of ${\bm M}$ and ${\bm \alpha}$. 
When these components are parallel, 
there arise two types of currents: the 
anomalous Hall current, ${\bm j}_{\rm AH}=\sigma_{\rm AH}(\omega)
(\hat{\bm \alpha}\cdot\hat{\bm M})\hat{\bm \alpha}\times{\bm E}$, 
yields the Faraday effect or Cotton-Mouton effect, 
whereas ${\bm j}_{\rm MO} \sim (\hat{\bm \alpha}\cdot\hat{\bm M})\qv\times{\bm E}$ 
which is due to spatial dispersion, 
generates Rashba-induced magneto optical activity. 
When the above components are orthogonal (${\bm \alpha}\perp{\bm M}$), 
the current is induced by 
the ``toroidal'' and 
``quadrupole'' moments, $\hat{\cal {\bm T}}=\hat{\bm \alpha}\times\hat{\bm M}$ and ${\cal Q}^{\perp}_{ij}= \hat{\bm \alpha}_{i}\hat{M}^{\perp}_{j} + \hat{M}^{\perp}_{i}\hat{\alpha}_{j}$, respectively,
as given by Eq.~(\ref{j-NDD}). Further,  
the diagonal components of the optical conductivity 
are not invariant under $\qv \to -\qv$ or ${\bm M} \to -{\bm M}$. 
The resulting nonreciprocal current flow generates a difference in absorption 
for counter-propagating electromagnetic waves. 

Based on these explicit results, 
we then demonstrated electromagnetic wave propagations in a ferromagnetic Rashba conductor. 
We first demonstrated that a material with large Rashba spin-split bands 
is a good candidate for a hyperbolic medium that exhibits negative refraction and backward waves. 
 These are due to the (uniaxial) anisotropy of the dielectric tensor 
$\vare^{(0)}_{ij}(\omega)$, which is the first term in Eq.~(\ref{expand-epsilon}).  
Then, the Rashba-induced birefringence was demonstrated as a combined effect of the direct and inverse Edelstein effects.  
This effect is governed by the second term of Eq.~(\ref{expand-epsilon}), $\alpha_{ijk}(\omega)q_{k}$.   
In the presence of ${\bm M}$ and when the system is in the Faraday configuration 
(${\bm \alpha}\parallel{\bm M}$) and $\qv \parallel {\bm M}$, 
the medium induces Faraday rotation, Kerr rotation, 
and magnetic circular dichroism,
because of the third term of Eq.~(\ref{expand-epsilon}), $\beta_{ijk}(\omega)M_{k}$.   
For the Voigt configuration (${\bm \alpha}\perp{\bm M}$) and ${\bm \alpha} \perp \qv$,  
nonreciprocal directional dichroism can occur. 
 This is induced by ${\cal {\bm T}}$ 
and ${\cal Q}_{ij}^{\perp}$,  
and is governed by the fourth term 
of Eq.~(\ref{expand-epsilon}), $\gamma_{ijkl}^{\perp}(\omega)q_{k}M_{\perp,l}$.  
This effect is found to be enhanced strongly at the spin-split transition edge in the electron band.


Before ending this paper, 
let us briefly consider the optical properties of Weyl semimetals in our context. 
The Weyl semimetals constitute another type of momentum-dependent spin-orbit coupled system 
with broken spatial inversion and/or time reversal symmetries. 
The low-energy Hamiltonian is given by \cite{{Burkov12}, {Tewari13}} 
\begin{align}
{\cal H}_{\rm W} = \tau_{z}v_{\rm F}
\hat{\bm p}\cdot{\bm \sigma} + \tau_{z}b_{0} + {\bm b}\cdot{\bm \sigma},
\label{W-Hamiltonian}
\end{align}
where $v_{\rm F}$ is the Fermi velocity. 
 The two Weyl nodes are specified by the chirality variable, $\tau_{z}=\pm 1$. Further, $2b_{0}$ is the energy difference between the two Weyl nodes, 
which arises when the system breaks the space inversion symmetry, 
while
$2{\bm b}$ denotes the displacement vector in the momentum space 
separating the two Weyl nodes, which arises when the system breaks the time 
reversal symmetry. 
Based on linear response theory, 
the electromagnetic response of the Weyl semimetal is obtained from 
the Hamiltonian (\ref{W-Hamiltonian}) as 
\begin{align}
{\bm j} = \sigma^{\rm W}_{\rm AH}(\omega){\bm b}\times {\bm E} 
+ \sigma^{\rm W}_{\rm C}(\omega)b_{0}{\bm B}, 
\label{W-j}
\end{align}
where $\sigma^{\rm W}_{\rm AH}(\omega)$ and $\sigma^{\rm W}_{\rm C}(\omega)$ are frequency-dependent coefficients. 
The first and second terms represent the anomalous Hall effect and chiral magnetic effect, respectively\cite{Franz13}. 
When $b_{0}\neq 0$ and ${\bm b}={\bm 0}$, 
the system breaks the space inversion symmetry and 
the second term of Eq.~(\ref{W-j}) gives the  
${\bm q}$-induced components of the dielectric tensor 
\begin{align}
\alpha_{ijk}(\omega)q_{k} = \alpha_{\rm W}(\omega)b_{0}\vare_{ijk}q_{k}, 
\end{align}
where $\alpha_{\rm W}(\omega)=(-i/\vare_{0}\omega^2)\sigma^{\rm W}_{\rm C}(\omega)$.  
This type of off-diagonal component in the dielectric tensor yields the 
optical activity
\cite{{Ma15},{Zhong16},{Kawaguchi-Tatara16}} (see Sec.~IV-B-1).  
On the other hand, 
when $b_{0}=0$ and ${\bm b}\neq{\bm 0}$, 
the system breaks the time reversal symmetry 
and the corresponding dielectric tensor component is given by 
\begin{align}
\beta_{ijk}(\omega)M_{k}~~\to~~\beta_{\rm W}(\omega)\vare_{ijk}b_{k}, 
\end{align} 
where $\beta_{\rm W}(\omega)=(-i/\vare_{0}\omega)\sigma^{\rm W}_{\rm AH}(\omega)$. 
Thus, this term yields ${\bm M}$-induced optical phenomena, 
namely, the Faraday effect and the Cotton-Mouton effect \cite{{Kargarian15},{Kawaguchi-Tatara16}} (see Sec.~IV-C). 

Finally, when the system breaks both the space inversion and time reversal symmetries 
($b_{0}\neq 0$ and ${\bm b}\neq{\bm 0}$), 
${\bm M}$-induced spatial dispersion phenomena are expected to occur. 
The relevant part of the dielectric tensor is deduced from 
\begin{align}
\gamma_{ijkl}(\omega)q_{k}M_{l}~~\to~~
\gamma_{\rm W}(\omega)b_{0}\delta_{ij}({\bm b}\cdot \qv), 
\end{align}
where $\gamma_{\rm W}(\omega)$ is a frequency-dependent coefficient. 
Thus, this diagonal component of the dielectric tensor generates
magneto-chiral dicrhoism and birefringence (see Sec. IV-D). 
Further details will be reported in the future.

\acknowledgments
This work is supported by Grant-in-Aid for Scientific Research (No. 17912949) 
from Japan Society for the Promotion of Science.

\appendix
\begin{widetext}
\section{Proof of Onsager reciprocal relation}

In this appendix, we prove the Onsager reciprocal relation for the response functions 
used in this paper, i.e., in the presence of momentum-dependent spin orbit coupling 
and magnetization ${\bm M}$. 
Here, we assume the following Hamiltonian: 
\begin{align}
\label{app-H}
{\cal H}_{\kv} = \epsilon_{\kv}+{{\bm B}}(\kv)\cdot{\bm \sigma} -{\bm M}\cdot{\bm \sigma}, 
\end{align}
where $\epsilon_{\kv}$ is an electronic dispersion, which is assumed to be even for $\kv$, and 
${\bm B}(\kv)$ is an effective magnetic field originating from 
a momentum-dependent spin orbit interaction as a result of broken space inversion symmetry 
and, thus, satisfying ${\bm B}(-\kv)=-{\bm B}(\kv)$. The third term represents the exchange interaction between the electron spin and ${\bm M}$, 
which breaks the time reversal symmetry. 
The eigen energy of this Hamiltonian is given by 
\begin{align}
\epsilon^{s}_{\kv,{\bm M}} = \epsilon_{\kv} + s |{\cal {\bm S}}_{\kv,{\bm M}}|, 
\end{align}
where $s=\pm 1$ represents the spin-split upper ($s=+1$) and lower ($s=-1$) bands and 
\begin{align}
{\cal {\bm S}}_{\kv, {\bm M}} = {{\bm B}}(\kv) - {\bm M}. 
\end{align} 
From this Hamiltonian (\ref{app-H}), 
we can write the Green function 
$G_{\kv,{\bm  M}}(z) = (z+\epsilon_{\rm F}-{\cal H}_{\kv})^{-1}$ 
as 
\begin{align}
G_{\kv,{\bm  M}}(z) &= g^{0}_{\kv, {\bm M}}(z) \sigma^{0} + {\bm g}_{\kv,{\bm M}}(z)\cdot{\bm \sigma}
\equiv g^{\mu}_{\kv,{\bm  M}}(z)\sigma_{\mu}, 
\end{align}
where 
\begin{align}
&g^{0}_{\kv,{\bm  M}}(z) = \frac{1}{2}\sum_{s=\pm 1}
\frac{1}{z-\epsilon^{s}_{\kv,{\bm M}}+\epsilon_{\rm F}}, \\
&{\bm g}_{\kv,{\bm  M}}(z) = \frac{1}{2}\sum_{s=\pm 1}
\frac{s\hat{\cal {\bm S}}_{\kv,{\bm M}}}{1-\epsilon^{s}_{\kv,{\bm M}}+\epsilon_{\rm F}},
\end{align}
and 
$\sigma_{\mu}= (\sigma_{0}, \sigma_{i})$, 
with $\sigma_{0}$ being the $2\times2$ unit matrix ($\mu=0,x,y,z$) 
and $\hat{\cal {\bm S}}_{\kv,{\bm M}}={\cal {\bm S}}_{\kv,{\bm M}}/|{\cal {\bm S}}_{\kv,{\bm M}}|$. 
As ${\cal {\bm S}}_{-\kv,-{\bm M}}=-{\cal {\bm S}}_{\kv,{\bm M}}$ 
and $\epsilon^{s}_{-\kv,-{\bm  M}} = \epsilon^{s}_{\kv,{\bm  M}}$, we have 
\begin{align}
G_{-\kv,-{\bm  M}}(z) = g^{0}_{\kv,{\bm  M}}(z) - {\bm g}_{\kv,{\bm  M}}(z)
\cdot{\bm \sigma}\equiv g^{\mu}_{\kv,{\bm  M}}(z)\bar{\sigma}_{\mu}, 
\label{r-G} 
\end{align}
where $\bar{\sigma}_{\mu}= (\sigma_{0}, -\sigma_{i})$. 

Let us first prove the Onsager reciprocal relation for the spin-spin correlation function: 
\begin{align}
\chi^{ss}_{ji}(-\qv,\omega,-{\bm  M}) =\chi^{ss}_{ij}(\qv,\omega,{\bm  M}). 
\end{align} 
Here, $\chi^{ss}_{ij}(\qv,\omega,{\bm  M})$ is written using Green's function as 
\begin{align}
\label{app-ss1}
\chi^{ss}_{ij}(\kv,\qv,\omega,{\bm M}) =-\sum_{\kv}
\int_{-\infty}^{\infty}
\frac{d\vare}{2\pi i} {\rm tr}\left[
\sigma_{i}G_{\kv_{+},{\bm  M}}(\vare_{+})\sigma_{j}G_{\kv_{-},{\bm  M}}(\vare_{-})\right]^{<}, 
\end{align} 
where $\kv_{\pm} = \kv\pm \qv/2$ and $\vare_{\pm} = \vare \pm \omega/2$. 
For $\chi^{ss}_{ji}(-\qv,\omega,-{\bm  M})$, we can write 
\begin{align}
\chi^{ss}_{ji}(-\qv,\omega,-{\bm  M}) =
 -\sum_{\kv}\int_{-\infty}^{\infty}
\frac{d\vare}{2\pi i} {\rm tr}\left[
\sigma_{j}G_{\kv_{-},-{\bm  M}}(\vare_{+})\sigma_{ i }G_{\kv_{+},-{\bm  M}}(\vare_{-})\right]^{<}. 
\end{align}
Changing the integration variable to $-\kv$ and using Eq. (\ref{r-G}), we have 
\begin{align}
&\chi^{ss}_{ji}(-\qv,\omega,-{\bm  M})  = 
-\sum_{\kv}\int_{-\infty}^{\infty}\frac{d\vare}{2\pi i}\left[
g^{\mu}_{\kv_{+},{\bm  M}}(\vare_{+})g^{\nu}_{\kv_{-},{\bm  M}}(\vare_{-})\right]^{<}
{\rm tr}\left[
\sigma_{j}\bar{\sigma}_{\mu}\sigma_{i}\bar{\sigma}_{\nu}
\right].
\end{align}
Using the trace formula, we have 
\begin{align}
{\rm tr}[\sigma_{j}\bar{\sigma}_{\mu}\sigma_{i}\bar{\sigma}_{\nu}]
&=
{\rm tr}[\sigma_{j}\sigma_{i}]
-{\rm tr}[\sigma_{j}\sigma_{k}\sigma_{ i }]
-{\rm tr}[\sigma_{ j }\sigma_{i}\sigma_{l}]
+{\rm tr}[\sigma_{ j }\sigma_{k}\sigma_{i}\sigma_{l}],
\nonumber\\
&=
{\rm tr}[\sigma_{ i }\sigma_{ j }]
+{\rm tr}[\sigma_{ i }\sigma_{k}\sigma_{j}]
+{\rm tr}[\sigma_{ i }\sigma_{ j }\sigma_{l}]
+{\rm tr}[\sigma_{ i }\sigma_{k}\sigma_{j}\sigma_{l}],
\nonumber\\
&={\rm tr}[\sigma_{ i }\sigma_{\mu}\sigma_{j}\sigma_{\nu}].
\end{align}
Thus, we obtain 
\begin{align}
\label{app-O-ss}
\chi^{ss}_{ji}(-\qv,\omega,-{\bm  M})
&=-\sum_{\kv}\int_{-\infty}^{\infty}\frac{d\vare}{2\pi i}
\left[
g^{\mu}_{\kv_{+},{\bm  M}}(\vare_{+})g^{\nu}_{\kv_{-},{\bm  M}}(\vare_{-})\right]^{<}
{\rm tr}\left[
\sigma_{i}\sigma_{\mu}\sigma_{j}\sigma_{\nu}
\right],\nonumber\\
&=-\sum_{\kv}\int_{-\infty}^{\infty}\frac{d\vare}{2\pi i}
{\rm tr}\left[
\sigma_{ i }G_{\kv_{+},{\bm  M}}(\vare_{+})\sigma_{ j }G_{\kv_{-},{\bm  M}}
\right]^{<},\nonumber\\
&= \chi^{ss}_{ij}(\qv,\omega,{\bm  M}). 
\end{align} 

The current-spin and spin-current correlation functions are given by 
\begin{align}
\label{app-js1-1}
\chi^{js}_{ij}(\qv,\omega,{\bm M}) &=  -\sum_{\kv}\int_{-\infty}^{\infty}
\frac{d\vare}{2\pi i} {\rm tr}\left[
\tilde{v}_{\kv,i}G_{\kv_{+},{\bm  M}}(\vare_{+})\sigma_{j}G_{\kv_{-},{\bm  M}}(\vare_{-})\right]^{<},\\
\label{app-sj1-1}
\chi^{sj}_{ij}(\qv,\omega,{\bm M}) &=  -\sum_{\kv}\int_{-\infty}^{\infty}
\frac{d\vare}{2\pi i} {\rm tr}\left[
\sigma_{i}G_{\kv_{+},{\bm  M}}(\vare_{+})\tilde{v}_{\kv,j}G_{\kv_{-},{\bm  M}}(\vare_{-})\right]^{<}, 
\end{align}
where 
\begin{align}
\label{app-velocity}
\tilde{v}_{\kv,i} &= \dfrac{1}{\hbar}\dfrac{\partial {\cal H}_{\kv}}{\partial k_{i}}
\equiv v_{\kv,i} + v_{\kv,i}^{j}\sigma_{j}, 
\end{align}
with $v_{\kv, i} =\dfrac{1}{\hbar}\dfrac{\partial \epsilon_{\kv}}{\partial k_{i}}$ 
being a conventional velocity 
and $v_{\kv,,i}^{j} = \dfrac{1}{\hbar}\dfrac{\partial {B}_{j}(\kv)}{\partial k_{i}}$ 
being due to the momentum-dependent spin orbit interaction. 
From Eq. (\ref{app-velocity}), Eqs. (\ref{app-js1}) and (\ref{app-sj1}) are written as 
\begin{align}
\label{chi-js-A}
\chi^{js}_{ij}(\qv,\omega,{\bm M}) &= \phi^{js}_{ij}(\qv,\omega,{\bm M})+ 
\sum_{\kv}v^{k}_{\kv,i}\varphi^{ss}_{kj}(\kv, \qv,\omega,{\bm M}),\\
\label{chi-sj-A}
\chi^{sj}_{ij}(\qv,\omega,{\bm M}) &= \phi^{sj}_{ij}(\qv,\omega,{\bm M})+ 
\sum_{\kv}v^{k}_{\kv,j}\varphi^{ss}_{ik}(\kv, \qv,\omega,{\bm M}),
\end{align}
where 
\begin{align}
\label{app-js1-2}
&\phi^{js}_{ij}(\qv,\omega,{\bm M}) =  -\sum_{\kv}\int_{-\infty}^{\infty}
\frac{d\vare}{2\pi i} {\rm tr}\left[
{v}_{\kv,i}G_{\kv_{+},{\bm  M}}(\vare_{+})\sigma_{j}G_{\kv_{-},{\bm  M}}(\vare_{-})\right]^{<},\\
\label{app-sj1-2}
&\phi^{sj}_{ij}(\qv,\omega,{\bm M}) =  -\sum_{\kv}\int_{-\infty}^{\infty}
\frac{d\vare}{2\pi i} {\rm tr}\left[
\sigma_{i}G_{\kv_{+},{\bm  M}}(\vare_{+}){v}_{\kv,j}G_{\kv_{-},{\bm  M}}(\vare_{-})\right]^{<}, \\
&\varphi^{ss}_{ij}(\kv, \qv,\omega,{\bm M}) =  -\int_{-\infty}^{\infty}
\frac{d\vare}{2\pi i} {\rm tr}\left[
\sigma_{i}G_{\kv_{+},{\bm  M}}(\vare_{+})\sigma_{j}G_{\kv_{-},{\bm  M}}(\vare_{-})\right]^{<}. 
\end{align}
Noting ${\bm v}_{-\kv}=-{\bm v}_{\kv}$, 
we obtain 
\begin{align}
\phi^{js}_{ji}(-\qv,\omega,-{\bm  M}) 
&=-\sum_{\kv}\int_{-\infty}^{\infty}\frac{d\vare}{2\pi i} \left[
-v_{j}g^{\mu}_{\kv_{+},{\bm  M}}(\vare_{+})
g^{\nu}_{\kv_{-},{\bm  M}}(\vare_{-})\right]^{<}
{\rm tr}[\bar{\sigma}_{\mu}\sigma_{i}\bar{\sigma_{\nu}}],
\nonumber\\
&=-\sum_{\kv}\int_{-\infty}^{\infty}\frac{d\vare}{2\pi i} \left[
g^{\mu}_{\kv_{+},{\bm  M}}(\vare_{+})
v_{j}g^{\nu}_{\kv_{-},{\bm  M}}(\vare_{-})\right]^{<}
{\rm tr}[\sigma_{i}\sigma_{\mu}\sigma_{\nu}],
\nonumber\\
&=-\sum_{\kv}\int_{-\infty}^{\infty}\frac{d\vare}{2\pi i} {\rm tr}\left[
\sigma_{i}g^{\mu}_{\kv_{+},{\bm  M}}(\vare_{+})\sigma_{\mu}v_{j}
g^{\nu}_{\kv_{-},{\bm  M}}(\vare_{-})\sigma_{\nu}
\right]^{<},\nonumber\\
&=\phi^{sj}_{ij}(\qv,\omega,{\bm  M}). 
\end{align}
Noting $v^{j}_{-\kv,i} = v^{j}_{\kv,i}$ and 
$\chi^{ss}_{ij}(\qv,\omega,{\bm M}) = \sum_{\kv}\varphi^{ss}_{ij}(\kv,\qv,\omega,{\bm M})$, 
we can easily check that 
\begin{align}
\sum_{\kv}v^{k}_{\kv,j}\varphi^{ss}_{ki}(\kv,-\qv,\omega,-{\bm M})
&=\sum_{\kv}v^{k}_{-\kv,j}\varphi^{ss}_{ki}(-\kv,-\qv,\omega,-{\bm M}),\nonumber\\
&=\sum_{\kv}v^{k}_{\kv,j}\varphi^{ss}_{ik}(\kv,\qv,\omega,{\bm M}). 
\end{align}
Thus, we obtain 
\begin{align}
\chi^{js}_{ji}(-\qv,\omega,-{\bm M}) &=
 \phi^{js}_{ji}(-\qv,\omega,-{\bm M})+\sum_{\kv}v^{k}_{\kv,j}\varphi^{ss}_{ki}(\kv,-\qv,\omega,-{\bm M}),\nonumber\\
&=\phi^{sj}_{ji}(\qv,\omega,{\bm M}) +\sum_{\kv}v^{k}_{\kv,j}\varphi^{ss}_{ik}(\kv,\qv,\omega,{\bm M}),\nonumber\\
&= \chi^{sj}_{ij}(\qv,\omega,{\bm M}). 
\end{align}

Finally, the current-current correlation function $\chi^{{j}{j}}_{ij}$  is given by 
\begin{align}
\chi^{jj}_{ij}(\qv, \omega ,{\bm M}) &= 
-\sum_{\kv}\int_{-\infty}^{\infty}
\frac{d\vare}{2\pi i} {\rm tr}\left[
\tilde{v}_{\kv,i}G_{\kv_{+},{\bm  M}}(\vare_{+})\tilde{v}_{\kv,j}G_{\kv_{-},{\bm  M}}(\vare_{-})\right]^{<}
\nonumber \\
&=\phi^{jj}_{ij}(\qv, \omega ,{\bm M})+
\sum_{\kv}v^{k}_{\kv,j}\varphi^{js}_{ik}(\kv,\qv, \omega ,{\bm M})
+\sum_{\kv}v^{k}_{\kv,i}\varphi^{sj}_{kj}(\kv,\qv, \omega ,{\bm M}) 
+\sum_{\kv}v^{k}_{\kv,i}v^{l}_{\kv,j}\varphi^{ss}_{kl}(\kv,\qv,\omega,{\bm M}), 
\label{app-jj1-1}
\end{align}
where 
\begin{align}
\label{app-jj1-2}
&\phi^{jj}_{ij}(\qv,\omega,{\bm M}) 
=  -\sum_{\kv}\int_{-\infty}^{\infty}\frac{d\vare}{2\pi i} {\rm tr}\left[
v_{\kv,i}G_{\kv_{+},{\bm  M}}(\vare_{+})v_{\kv,j}G_{\kv_{-},{\bm  M}}(\vare_{-})\right]^{<},\\
&\varphi^{js}_{ij}(\kv,\qv,\omega,{\bm M}) 
=  -\int_{-\infty}^{\infty}\frac{d\vare}{2\pi i} {\rm tr}\left[
v_{\kv,i}G_{\kv_{+},{\bm  M}}(\vare_{+})\sigma_{j}G_{\kv_{-},{\bm  M}}(\vare_{-})\right]^{<},\\
&\varphi^{sj}_{ij}(\kv,\qv,\omega,{\bm M}) 
=  -\int_{-\infty}^{\infty}\frac{d\vare}{2\pi i} {\rm tr}\left[
\sigma_{i}G_{\kv_{+},{\bm  M}}(\vare_{+})v_{\kv,j}G_{\kv_{-},{\bm  M}}(\vare_{-})\right]^{<},\\
\end{align}
For $\phi^{jj}_{ji}(-\qv,\omega,-{\bm M})$, 
we obtain 
\begin{align}
\phi^{jj}_{ji}(-\qv,\omega,-{\bm M}) &= 
 -\sum_{\kv}\int_{-\infty}^{\infty}\frac{d\vare}{2\pi i}\left[
v_{j}g^{\mu}_{\kv_{+},{\bm  M}}(\vare_{+})v_{i}g^{\nu}_{\kv_{-},{\bm  M}}(\vare_{-})\right]^{<}
{\rm tr}[\bar{\sigma}_{\mu}\bar{\sigma}_{\nu}],\nonumber\\
&= -\sum_{\kv}\int_{-\infty}^{\infty}\frac{d\vare}{2\pi i}\left[
v_{i}g^{\mu}_{\kv_{+},{\bm  M}}(\vare_{+})v_{j}g^{\nu}_{\kv_{-},{\bm  M}}(\vare_{-})\right]^{<}
{\rm tr}[{\sigma}_{\mu}{\sigma}_{\nu}],\nonumber\\
&= -\sum_{\kv}\int_{-\infty}^{\infty}\frac{d\vare}{2\pi i} {\rm tr}\left[
v_{i}g^{\mu}_{\kv_{+},{\bm M}}(\vare_{+})\sigma_{\mu}v_{j}g^{\nu}_{\kv_{-},{\bm M}}(\vare_{-})\sigma_{\nu}
\right]^{<},\nonumber\\
&=\phi^{jj}_{ij}(\qv,\omega,{\bm M}), 
\end{align}
and 
we can easily check that 
\begin{align}
\sum_{\kv}v^{k}_{\kv,i}\varphi^{js}_{jk}(\kv,-\qv,\omega,-{\bm M}) &= 
\sum_{\kv}v^{k}_{-\kv,i}\varphi^{js}_{jk}(-\kv,-\qv,\omega,-{\bm M}),\nonumber\\
&=\sum_{\kv}v^{k}_{\kv,i}\varphi^{sj}_{kj}(\kv,\qv,\omega,{\bm M}). 
\end{align}
Thus, we obtain 
\begin{align}
\chi^{jj}_{ji}(-\qv,\omega,-{\bm M})&=\phi^{jj}_{ji}(-\qv, \omega ,-{\bm M})+
\sum_{\kv}v^{k}_{\kv,i}\varphi^{js}_{jk}(\kv, -\qv, \omega ,-{\bm M})\nonumber\\
&
+\sum_{\kv}v^{k}_{\kv,j}\phi^{sj}_{ki}(\kv,-\qv, \omega ,-{\bm M}) 
+\sum_{\kv}v^{k}_{\kv,j}v^{l}_{\kv,i}\varphi^{ss}_{kl}(\kv,-\qv,\omega,-{\bm M}),\nonumber\\
&=\phi^{jj}_{ij}(\qv, \omega ,{\bm M})+
\sum_{\kv}v^{k}_{\kv,i}\varphi^{sj}_{kj}(\kv,\qv, \omega ,{\bm M})
+\sum_{\kv}v^{k}_{\kv,j}\varphi^{js}_{ik}(\kv,\qv, \omega ,{\bm M}) 
+\sum_{\kv}v^{k}_{\kv,i}v^{l}_{\kv,j}\varphi^{ss}_{lk}(\qv,\omega,{\bm M}),\nonumber\\
&=\chi^{jj}_{ij}(\qv,\omega,{\bm M}). 
\end{align}
\section{Calculation details}

In this Appendix, 
we concretely calculate the correlation functions, 
$\chi^{jj}_{ij}$, $\chi^{js}_{ij}$, $\chi^{sj}_{ij}$, and $\chi^{ss}_{ij}$ 
defined in Eqs. (\ref{app-jj1-1}), (\ref{app-js1-1}), (\ref{app-sj1-1}), and (\ref{app-ss1}).   
Following Langreth's method \cite{{Langreth76},{Haug98}}, 
one can calculate 
the lesser component of $G(\vare_{+})G(\vare_{-})$ as 
\begin{align}
\left[G(\vare_{+})G(\vare_{-})\right]^{<} &= 
G^{<}(\vare_{+})G^{\rm A}(\vare_{-}) + G^{\rm R}(\vare_{+})G^{<}(\vare_{-}),\nonumber\\
&= 
f(\vare_{+})(G^{\rm A}(\vare_{+})-G^{\rm R}(\vare_{+}))G^{\rm A}(\vare_{-}) + f(\vare_{-})G^{\rm R}(\vare_{+})(G^{\rm A}(\vare_{-})-G^{\rm R}(\vare_{-})). 
\end{align}
Combining this with  
\begin{align}
G^{\rm A}_{\kv,{\bm M}}(\epsilon)-G^{\rm R}_{\kv,{\bm M}}(\epsilon) = 
i\pi\sum_{s=\pm 1}\delta(\epsilon-\epsilon^{s}_{\kv,{\bm M}}+\epsilon_{\rm F})
(1+s \hat{\cal {\bm S}}_{\kv,{\bm M}}\cdot{\bm \sigma}), 
\end{align}
and 
performing the $\vare$-integral, 
we can write the correlation functions 
$\phi^{jj}_{ij}$, $\phi^{js}_{ij}$, $\phi^{sj}_{ij}$, and $\chi^{ss}_{ij}$ 
in Eqs. (\ref{app-jj1-2}), (\ref{app-js1-2}), (\ref{app-sj1-2}), and (\ref{app-ss1}),  
as  
\begin{subequations}
\begin{align}
\label{chi-jj}
&\phi^{jj}_{ij}(\qv,\omega,{\bm  M}) =-\frac{1}{2}\sum_{\kv}\sum_{s,s'}
v_{\kv, i }v_{\kv, j }{A}^{s,s'}(\kv_{+},\kv_{-},{\bm  M})
{\cal L}^{s,s'}(\kv_{+},\kv_{-},{\bm  M}),\\
&\phi^{js}_{ij}(\qv,\omega,{\bm  M}) = -\frac{1}{2}\sum_{\kv}\sum_{s,s'}
v_{\kv, i }{B}^{s,s'}_{ j }(\kv_{+},\kv_{-},{\bm  M})
{\cal L}^{s,s'}(\kv_{+},\kv_{-},{\bm  M}),\\
&\phi^{sj}_{ij}(\qv,\omega,{\bm  M}) = -\frac{1}{2}\sum_{\kv}\sum_{s,s'}
{C}^{s,s'}_{ i }(\kv_{+},\kv_{-},{\bm  M})v_{ \kv,j }
{\cal L}^{s,s'}(\kv_{+},\kv_{-},{\bm  M}),\\
&\chi^{ss}_{ij}(\qv,\omega,{\bm  M}) = 
-\frac{1}{2}\sum_{\kv}\sum_{s,s'}
{D}^{s,s'}
_{ij}(\kv_{+},\kv_{-},{\bm  M})
{\cal L}^{s,s'}(\kv_{+},\kv_{-},{\bm  M}),
\end{align}
\end{subequations}
where 
\begin{subequations}
\begin{align}
\label{F1-1}
&{A}^{s,s'}(\kv_{+},\kv_{-},{\bm  M}) = 1+ss'
\hat{\cal {\bm S}}_{\kv_{+},{\bm  M}}\cdot\hat{\cal {\bm S}}_{\kv_{-},{\bm  M}}, 
\\\label{F2-1}
&{\bm B}^{s,s'}(\kv_{+},\kv_{-},{\bm  M}) =
s\hat{\cal {\bm S}}_{\kv_{+},{\bm  M}}+s'\hat{\cal {\bm S}}_{\kv_{-},{\bm  M}}
-iss'\hat{\cal {\bm S}}_{\kv_{+},{\bm  M}}\times\hat{\cal {\bm S}}_{\kv_{-},{\bm  M}}, 
\\
\label{F3-1}
&{\bm C}^{s,s'}(\kv_{+},\kv_{-},{\bm  M}) =
s\hat{\cal {\bm S}}_{\kv_{+},{\bm  M}}+s'\hat{\cal {\bm S}}_{\kv_{-},{\bm  M}}
+iss'\hat{\cal {\bm S}}_{\kv_{+},{\bm  M}}\times\hat{\cal {\bm S}}_{\kv_{-},{\bm  M}}, 
\\\label{F4-1}
&{D}_{ij}^{s,s'}(\kv_{+},\kv_{-},{\bm  M}) =
\delta_{ij}
-is\vare_{ i  j k}(\hat{\cal {\bm S}}_{\kv_{+},{\bm  M}})_{k}
+is'\vare_{ i  j k}(\hat{\cal {\bm S}}_{\kv_{-},{\bm  M}})_{k}\nonumber\\
&
+ss'\left\{
(\hat{\cal {\bm S}}_{\kv_{+},{\bm  M}})_{ i }(\hat{\cal {\bm S}}_{\kv_{-},{\bm  M}})_{ j }
+
(\hat{\cal {\bm S}}_{\kv_{+},{\bm  M}})_{ j }(\hat{S}_{\kv_{-},{\bm  M}})_{ i }
-\delta_{ij}(
\hat{\cal {\bm S}}_{\kv_{+},{\bm  M}}\cdot\hat{\cal{\bm S}}_{\kv_{-},{\bm  M}}
)
\right\},
\end{align}
\end{subequations}
and 
\begin{align}
\label{Lindhard}
{\cal L}^{s,s'}(\kv_{+},\kv_{-},{\bm  M}) = 
\frac{f^{s}_{\kv_{+},{\bm  M}}-f^{s'}_{\kv_{-},{\bm  M}}}{\epsilon^{s}_{\kv_{+},{\bm  M}}
-\epsilon^{s'}_{\kv_{-},{\bm  M}}-\hbar\omega-i0},
\end{align}
is the Lindhard function. 

In the following, we evaluate these in the case of the ferromagnetic Rashba conductor. 
Putting ${\cal {\bm S}}_{\kv,{\bm M}} = {\bm \alpha}\times \kv -{\bm M}$ and 
expanding 
$A^{s,s'}$, ${\bm B}^{s,s'}$, 
${\bm C}^{s,s'}$, $D^{s,s'}_{ij}$, and 
${\cal L}^{s,s'}(\kv_{+},\kv_{-},{\bm  M})$ 
with respect to $\qv$ and ${\bm M}$, 
we calculate $\phi^{jj}_{ij}$, $\phi^{js}_{ij}$, 
$\phi^{sj}_{ij}$, and $\chi^{ss}_{ij}$ 
up to first order in $\qv$ and ${\bm  M}$. 
Furthermore, in the $\kv$-integral, 
we here choose the cylindrical coordinate system in $\kv$-space, 
where the cylindrical axis is taken in the direction of the Rashba field 
$\hat{\bm \alpha}$. 
Thus the sum of $\kv$ is written as 
\begin{align}
\sum_{\kv}(\cdots)_{\kv} = \frac{1}{8\pi^3}\int_{0}^{2\pi}d\varphi\int_{0}^{\infty}dk_{\perp}k_{\perp}\int_{-\infty}^{\infty}dk_{\parallel}(\cdots)_{\kv}, 
\end{align}
where $\varphi$ is the azimuth around the $\hat{\bm \alpha}$-axis and 
$k_{\perp} = |\kv_{\perp}|$ is the radial distance measured from the $\hat{\bm \alpha}$-axis. 
In this cylindrical coordinate system, 
we can write $\gamma_{\kv} = |\kv\times{\bm \alpha}| = \alpha k_{\perp}$ and 
the $\varphi$-integral around the $\hat{\bm \alpha}$-axis 
is calculated as 
\begin{subequations}
\begin{align}
&\langle {k}^{\perp}_{ i }{k}^{\perp}_{ j }\rangle \equiv
\frac{1}{2\pi}\int_{0}^{2\pi}d\varphi~
{k}^{\perp}_{ i }{k}^{\perp}_{ j } 
=\frac{k_{\perp}^2}{2}\delta^{\perp}_{ij}, \\
&\langle ({\bm \gamma}_{\kv})_{ i }({\bm \gamma}_{\kv})_{ j } 
\rangle =\frac{k_{\perp}^2}{2}\delta^{\perp}_{ij}, \\
&\langle (\kv_{\perp})_{i}({\bm \gamma}_{\kv})_{ j }
\rangle =-\frac{k_{\perp}^2}{2}\vare_{ijk}\hat{\alpha}_{k}, \\
&\langle {k}_{ i }^{\perp}{k}_{ j }^{\perp}{k}_{k}^{\perp}{k}_{l}^{\perp}
\rangle =
\frac{k_{\perp}^4}{8}(\delta_{ij}^{\perp}\delta_{kl}^{\perp} 
+ 
\delta_{ i k}^{\perp}\delta_{ j l}^{\perp} 
+ \delta_{ i l}^{\perp}
\delta_{ j k}^{\perp}), 
\end{align}
\end{subequations}
where $\delta_{ij}^{\perp}= \delta_{ij}-\hat{\alpha}_{ i }\hat{\alpha}_{ j }$ 
and $\vare_{ijk}$ is the totally anti-symmetric tensor with $\vare_{xyz}=1$. 
After some complicated calculations, we obtain 
\begin{subequations}
\begin{align}
\label{c-jjf}
\phi^{jj}_{ij}(\qv,\omega, M)&=
I_{1}\left\{
(\hat{\bm \alpha}\times{\bm M})_{ i }(\qv_{\perp})_{j}+(\qv_{\perp})_{i}(\hat{\bm \alpha}\times{\bm M})_{ j }
\right\}\nonumber\\
&
+\frac{m_{\perp}}{m_{\parallel}}
I_{1}\left\{
(\hat{\bm \alpha}\times{\bm M})_{ i }(\qv_{\parallel})_{j}+(\qv_{\parallel})_{i}(\hat{\bm \alpha}\times{\bm M})_{ j }
\right\},
\end{align}
\begin{align}
\label{c-jsf}
\phi^{js}_{ij}(\qv,\omega, M)&= -i\frac{\hbar}{m_{\perp}} \sum_{\kv}\frac{\gamma_{\kv}s_{\kv}}{H_{\kv}}
(\qv\times\hat{\bm \alpha})_{ i }\hat{\alpha}_{ j }\nonumber\\
&+\frac{\hbar}{\alpha}
\left(
4I_{2}+\frac{I_{1}}{4}-3J_{1}+J_{2}
\right)
(\hat{\bm \alpha}\times {\bm M})_{ i }(\qv\times\hat{\bm \alpha})_{ j }\nonumber\\
&+\frac{\hbar}{\alpha}
\left(-\frac{3}{4}I_{1}+J_{1}+J_{2}
\right)(\qv_{\perp})_{i}M^{\perp}_{ j } \nonumber\\
&+\frac{\hbar}{\alpha}
\left(-\frac{I_{1}}{4}-J_{1}-J_{2}\right)
(\hat{\bm \alpha}\times{\bm M})\cdot \qv \vare_{ i  j k}\hat{\alpha}_{k}\nonumber\\
&+\frac{\hbar}{\alpha}4
\left(\frac{m_{\perp}}{m_{\parallel}}I_{2}-I_{3}
\right)(\qv_{\parallel})_{i}M^{\perp}_{ j }\nonumber\\
&+\frac{\hbar}{\alpha}\left(
-I_{1}-4I_{2}+4J_{2}
\right)(\qv_{\perp})_{i}M^{\parallel}_{ j }\nonumber\\
&-8\frac{\hbar}{\alpha}I_{3}(\qv_{\parallel})_{i}M^{\parallel}_{ j }
\end{align}
\begin{align}
\label{c-sjf}
\phi^{sj}_{ij}(\qv,\omega, M)&= 
i\frac{\hbar}{m_{\perp}} \sum_{\kv}\frac{\gamma_{\kv}s_{\kv}}{H_{\kv}}
\hat{\alpha}_{ i }(\qv\times\hat{\bm \alpha})_{ j }\nonumber\\
&+\frac{\hbar}{\alpha}\left(
4I_{2}+\frac{I_{1}}{4}-3J_{1}+J_{2}\right)
(\qv\times\hat{\bm \alpha})_{ i }(\hat{\bm \alpha}\times {\bm M})_{ j }\nonumber\\
&+\frac{\hbar}{\alpha}
\left(-\frac{3}{4}I_{1}+J_{1}+J_{2}\right)
M^{\perp}_{ i }(\qv_{\perp})_{j} \nonumber\\
&+\frac{\hbar}{\alpha}
\left(\frac{I_{1}}{4}+J_{1}+J_{2}\right)
(\hat{\bm \alpha}\times{\bm M})\cdot \qv \vare_{ i  j k}\hat{\alpha}_{k}\nonumber\\
&+\frac{\hbar}{\alpha}
4
\left(\frac{m_{\perp}}{m_{\parallel}}I_{2}-I_{3}
\right)M^{\perp}_{ i }(\qv_{\parallel})_{j}\nonumber\\
&+\frac{\hbar}{\alpha}\left(
-I_{1}-4I_{2}+4J_{2}
\right)M^{\parallel}_{ i }(\qv_{\perp})_{j}\nonumber\\
&-8\frac{\hbar}{\alpha}I_{3}M^{\parallel}_{ i }(\qv_{\parallel})_{j}
\end{align}
\begin{align}
\label{c-ssf}
\chi^{ss}_{ij}(\qv,\omega, M)
&=2 \sum_{\kv}\frac{\gamma_{\kv}s_{\kv}}{H_{\kv}}(
2\delta_{ij}-\delta^{\perp}_{ij})
\nonumber\\ 
& +i \hbar\omega \sum_{\kv}
\vare_{ijk}\left[
\frac{s_{\kv}}{\gamma_{\kv}H_{\kv}}(2M_{k}-M_{k}^{\perp})
+
\left\{
\frac{n_{\kv}'}{H_{\kv}}+\frac{8\gamma_{\kv}s_{\kv}}{H_{\kv}^2}\right\}M^{\perp}_{k}
\right]\nonumber\\
&+\frac{\hbar^2}{\alpha^2}\left(
-2I_{2}+2J_{1}+J_{4}\right)
\left\{
M^{\perp}_{ i }(\qv\times\hat{\bm \alpha})_{ j }
+
(\qv\times\hat{\bm \alpha})_{ i }M^{\perp}_{ j }
\right\}\nonumber\\
&+\frac{\hbar^2}{\alpha^2}
\left(2I_{2}-2J_{1}-3J_{4}\right)
(\hat{\bm \alpha}\times {\bm M})\cdot\qv \delta^{\perp}_{ij}
\nonumber\\
&+\frac{\hbar^2}{\alpha^2}\left(
-4I_{2}-2I_{4}+2J_{1}+2J_{3}
\right)\left\{
M^{\parallel}_{ i }(\qv\times \hat{\bm \alpha})_{ j }
+
(\qv\times \hat{\bm \alpha})_{ i }M^{\parallel}_{ j }
\right\}
\nonumber\\
&+\frac{\hbar^2}{\alpha^2}(-4J_{4})
(\hat{\bm \alpha}\times{\bm M})\cdot \qv \hat{\alpha}_{ i }\hat{\alpha}_{ j }, 
\end{align}
\end{subequations}
where 
\begin{subequations}
\begin{align}
\label{I1}
I_{1}&=\frac{1}{\hbar\omega+i0}
\frac{\hbar^2}{2m_{\perp}^2\alpha}
 \sum_{\kv}\gamma_{\kv}s_{\kv}',\\
 \label{I2}
I_{2}&=\frac{1}{\hbar\omega+i0}\frac{\alpha}{8m_{\perp}} \sum_{\kv}
n_{\kv}',\\
\label{I3}
I_{3}&=\frac{1}{\hbar\omega+i0}\frac{\hbar^2}{8m_{\parallel}^2\alpha} \sum_{\kv}
\gamma_{\kv}s'_{\kv}
\left(\frac{k_{\parallel}}{k_{\perp}}\right)^2,\\
\label{I4}
I_{4}&=\frac{1}{\hbar\omega+i0}\frac{\alpha^3}{4\hbar^2} \sum_{\kv}
\frac{s_{\kv}'}{\gamma_{\kv}},\\
\label{J1}
J_{1}&= \frac{\alpha\hbar\omega}{8m_{\perp}} \sum_{\kv}
\left(
\frac{n_{\kv}'}{H_{\kv}}+8\frac{\gamma_{\kv}s_{\kv}}{H_{\kv}^2}
\right),\\
\label{J2}
J_{2}&=  \frac{\alpha\hbar\omega}{8m_{\perp}} \sum_{\kv}
\frac{s_{\kv}}{\gamma_{\kv}H_{\kv}},\\
\label{J3}
J_{3}&= \frac{\alpha^3\omega}{4\hbar} \sum_{\kv}
\frac{s_{\kv}'}{\gamma_{\kv}H_{\kv}},\\
\label{J4}
J_{4}&= \frac{\alpha\hbar\omega}{m_{\perp}} \sum_{\kv}
\left(
\frac{2\gamma^2_{\kv}n'_{\kv}}{H_{\kv}^2}
+\frac{\gamma_{\kv}s_{\kv}}{H_{\kv}^2}
+16\frac{\gamma_{\kv}^3 s_{\kv}}{H_{\kv}^3}\right),
\end{align}
\end{subequations}
with 
\begin{subequations}
\begin{align}
&n_{\kv} = \sum_{s=\pm 1}f_{\kv}^{s}, \\
&s_{\kv} = \sum_{s=\pm 1}s f_{\kv}^{s}, \\
&n'_{\kv} = \sum_{s=\pm 1}(f_{\kv}^{s})',  \\
&s'_{\kv} = \sum_{s=\pm 1}s (f_{\kv}^{s})',\\
&f_{\kv}^{s} = f(\epsilon_{\rm F}-\vare_{\kv}- s \gamma_{\kv}),~~f(\epsilon)\equiv \frac{1}{1+e^{\epsilon/(k_{\rm B}T)}}, \\
&\displaystyle{(f_{\kv}^{s})' 
= \left.\frac{d}{d\vare}f(\vare)\right|_{\epsilon_{\rm F}-\vare_{\kv}- s \gamma_{\kv}}},\\
&H_{\kv} = (\hbar\omega+i0)^2-4\gamma^2_{\kv}.
\end{align}
\end{subequations}
The integrals $I_{\mu}~(\mu=1,2,3,4)$ stem from intraband transitions 
$(s \to s)$.  
On the other hand, 
the integrals $J_{\mu}~(\mu=1,2,3,4)$ 
stem from interband transition between 
the Rashba-split bands ($s \to -s$). 
Substituting these into Eqs.~(\ref{chi-js-A}), (\ref{chi-sj-A}) and (\ref{app-jj1-1}),  
we obtain the correlation function $\chi^{jj}_{ij}$, 
which is evaluated to the first order of $\qv$ and ${\bm M}$, 
along with $\chi^{js}_{ij}$ and $\chi^{sj}_{ij}$, which are 
evaluated for $\qv={\bm 0}$ and 
to the first order of ${\bm M}$, as 
\begin{subequations}
\label{chi-jj-js-sj-A}
\begin{align}
\chi^{{j}{j}}_{ij}(\qv,\omega, M) 
&\simeq-\frac{n_{\rm e}}{m_{\perp}}C(\omega)\delta^{\perp}_{ij}
 -i\frac{n_{\rm e}}{m_{\perp}}\frac{1}{\epsilon_{\rm F}}
 D(\omega)(\hat{\bm \alpha}\cdot{\bm M})
\vare_{ijk}\hat{\alpha}_{k}
\nonumber\\
&+\frac{n_{\rm e}}{m_{\perp}}\frac{1}{k_{\rm F}^{\perp}\epsilon_{\rm F}}
\left[E_{0}(\omega)\left\{
(\hat{\bm \alpha}\times {\bm M})_{ i }(\qv_{\parallel})_{j} + 
(\qv_{\parallel})_{i}(\hat{\bm \alpha}\times{\bm M})_{ j }
\right\}+E_{1}(\omega)
(\hat{\bm \alpha}\times{\bm M} ) \cdot \qv \delta^{\perp}_{ij}
\right.\nonumber\\
&\left.
+\frac{E_{2}(\omega)}{2} \left\{
(\hat{\bm \alpha}\times {\bm M})_{ i }(\qv_{\perp})_{j} + 
(\qv_{\perp})_{i}(\hat{\bm \alpha}\times{\bm M})_{ j }
\right\}\right], \\
\chi^{{j}s}_{ij}(\qv,\omega, M) 
&\simeq
\frac{\hbar}{\alpha}\frac{n_{\rm e}}{m_{\perp}}C(\omega)
\vare_{ijk}\hat{\alpha}_{k}+i\frac{\hbar}{\alpha}\frac{n_{\rm e}}{m_{\perp}}\frac{M}{\epsilon_{\rm F}}
\left(N_{ i }(\omega)\hat{\alpha}_{ j }
-\hat{\bm \alpha}\cdot{\bm N}(\omega)\delta_{ij}\right),\\
\chi^{s{j}}_{ij}(\qv,\omega, M)
&\simeq 
-\frac{\hbar}{\alpha}\frac{n_{\rm e}}{m_{\perp}}C(\omega)
\vare_{ijk}\hat{\alpha}_{k}
-i\frac{\hbar}{\alpha}\frac{n_{\rm e}}{m_{\perp}}
\frac{M}{\epsilon_{\rm F}}
\left(\hat{\alpha}_{ i }N_{ j }(\omega)
-\hat{\bm \alpha}\cdot{\bm N}(\omega)\delta_{ij}\right),
\end{align}
\end{subequations}
where 
\begin{subequations}
\begin{align}
C(\omega) &= -\frac{4\tilde{\alpha}^2}{n_{\rm e}}\epsilon_{\rm F}
\sum_{\kv}\frac{\gamma_{\kv}s_{\kv}}{H_{\kv}}, 
\label{B-C}
\\
D(\omega) &= -\frac{4\tilde{\alpha}^2}{n_{\rm e}}\epsilon_{\rm F}^2\hbar\omega
\sum_{\kv}\frac{s_{\kv}}{\gamma_{\kv}H_{\kv}}
\label{B-D1}, \\
E_{0}(\omega) &=\frac{m_{\perp}}{n_{\rm e}}k_{\rm F}^{\perp}\epsilon_{\rm F}
\frac{m_{\perp}}{m_{\parallel}}
\left( I_{1}+4I_{2}-4\frac{m_{\parallel}}{m_{\perp}}I_{3}\right) \nonumber\\
&= 
\frac{1}{2n_{\rm e}\tilde{\alpha}}
\frac{\epsilon_{\rm F}}{\hbar\omega+i0}
 \sum_{\kv}\left[
\gamma_{\kv}s'_{\kv}\left\{
1-\frac{m_{\perp}}{m_{\parallel}}\left(\frac{k_{\parallel}}{k_{\perp}}\right)^2
\right\}+\frac{m\alpha^2}{\hbar^2}n'_{\kv}
\right], 
\label{B-E0}
\end{align}
\begin{align}
E_{1}(\omega)&=\frac{m_{\perp}}{n_{\rm e}}k_{\rm F}^{\perp}\epsilon_{\rm F}\left(
\frac{I_{1}}{2}+2I_{2} -2J_{2}-3J_{4}\right) 
\nonumber\\
&= 
\frac{1}{4n_{\rm e}\tilde{\alpha}}
\frac{\epsilon_{\rm F}}{\hbar\omega+i0}
 \sum_{\kv}\left(
\gamma_{\kv}s'_{\kv}+\frac{m_{\perp}\alpha^2}{\hbar^2}n'_{\kv}
\right)\nonumber\\
&+
\frac{\tilde{\alpha}}{2n_{\rm e}}\epsilon_{\rm F}^2\hbar\omega
 \sum_{\kv}
\left\{
\frac{s_{\kv}}{\gamma_{\kv}H_{\kv}}
-12\left(
 \frac{2\gamma^2_{\kv}n'_{\kv}}{H_{\kv}^2}
+\frac{\gamma_{\kv}s_{\kv}}{H_{\kv}^2}
+16\frac{\gamma_{\kv}^3 s_{\kv}}{H_{\kv}^3}
\right)
\right\}, 
\label{B-E1}\\
E_{2}(\omega) &=
2\frac{m_{\perp}}{n_{\rm e}}k_{\rm F}^{\perp}\epsilon_{\rm F}
\left(\frac{I_{1}}{2}+2I_{2} -2J_{2}+J_{4}\right)\nonumber\\
&= 
\frac{1}{2n_{\rm e}\tilde{\alpha}}
\frac{\epsilon_{\rm F}}{\hbar\omega+i0}
 \sum_{\kv}\left(
\gamma_{\kv}s'_{\kv}+\frac{m_{\perp}\alpha^2}{\hbar^2}n'_{\kv}
\right)\nonumber\\
&+
\frac{\tilde{\alpha}}{n_{\rm e}}\epsilon_{\rm F}^2\hbar\omega
 \sum_{\kv}
\left\{
\frac{s_{\kv}}{\gamma_{\kv}H_{\kv}}
+4\left(
 \frac{2\gamma^2_{\kv}n'_{\kv}}{H_{\kv}^2}
+\frac{\gamma_{\kv}s_{\kv}}{H_{\kv}^2}
+16\frac{\gamma_{\kv}^3 s_{\kv}}{H_{\kv}^3}
\right)
\right\}, 
\label{B-E2}\\
E_{3}(\omega) &= 
-\frac{2\tilde{\alpha}}{n_{\rm e}}\epsilon^{2}_{\rm F}\hbar\omega
\sum_{\kv}\left(
\frac{n'_{\kv}}{H_{\kv}}
+\frac{s_{\kv}}{\gamma_{\kv}H_{\kv}}
+\frac{8\gamma_{\kv}s_{\kv}}{H_{\kv}^{2}}
\right)
\label{B-E3}
\end{align}
\end{subequations}
with
\begin{align}
{\bm N}(\omega) &= 
D(\omega)\hat{{\bm M}}^{\parallel}
+\tilde{\alpha}E_{3}(\omega)\hat{{\bm M}}^{\perp},
\label{B-N}\\
\hat{{\bm M}}^{\parallel}&= (\hat{\bm \alpha}\cdot\hat{\bm M})\hat{\bm \alpha},\\
\hat{\bm M}^{\perp} &= \hat{\bm M}-\hat{\bm M}^{\parallel}.
\end{align}

\section{$\kv$-integrals}
In this section, 
we perform the $\kv$-integrals of Eqs.~(\ref{B-C})--(\ref{B-E3}). 
Because of the rotational invariance around the 
$\kv_{\parallel}$-axis,  we can write the sum of $\kv$ as  
\begin{align}
\sum_{\kv}(\cdots)_{\kv}= \frac{1}{4\pi^2}\int_{0}^{\infty}dk_{\perp}k_{\perp}
\int_{-\infty}^{\infty}dk_{\parallel}(\cdots)_{\kv}. 
\end{align}
At zero temperature, we can set $-(f_{\kv}^{s})'$ as 
\begin{align}
-(f_{\kv}^{s})' &= \delta(\vare_{\rm F}-\epsilon^{s}_{\kv}),
\nonumber\\
 &= 
\frac{\sqrt{m_{\perp}m_{\parallel}}}{\hbar^2}
\frac{\delta(k_{\parallel}-\kappa_{s})+\delta(k_{\parallel}+\kappa_{s})}{
\sqrt{({k}_{s}-{k}_{\perp})({k}_{-s}+{k}_{\perp})}}
\theta(k_{s}-k_{\perp}), 
\label{delta-f}
\end{align}
where $\delta(x)$ and $\theta(x)$ are the delta and step functions, respectively, and   
\begin{align}
&\kappa_{s}= 
\sqrt{\frac{m_{\parallel}}{m_{\perp}}}\sqrt{(k_{s}-k_{\perp})(k_{-s}+k_{\perp})},\\
&k_{s} = k_{{\rm F},\perp}\tilde{k}_{s},\\
&\tilde{k}_{s} = -s\tilde{\alpha} + \sqrt{1+\tilde{\alpha}^2}. 
\end{align}
Note that $\tilde{k}_{s}\tilde{k}_{-s} = 1$ and $\tilde{k}_{s}
-\tilde{k}_{-s} = -2s\tilde{\alpha}$. 
We first calculate the electron density 
$n_{\rm e}=\sum_{s=\pm 1}\sum_{\kv}f^{s}_{\kv}$. 
Performing integration by parts and using Eq.~(\ref{delta-f}), 
we have 
\begin{align}
n_{\rm e} &= \frac{(k_{\rm F}^{\perp})^2k_{\rm F}^{\parallel}}{2\pi^2}
\sum_{s=\pm 1}
\int_{0}^{\tilde{k}_{s}}d\tilde{k}_{\perp}\tilde{k}_{\perp}
\sqrt{(\tilde{k}_{s}-\tilde{k}_{\perp})(\tilde{k}_{-s}+\tilde{k}_{\perp})}, 
\end{align}
where $\tilde{k}_{\perp} = k_{\perp}/k_{\rm F}^{\perp}$. 
Performing the $\tilde{k}_{\perp}$-integral and taking the sum of $s$, 
we have 
\begin{align}
n_{\rm e} = n_{\rm e}^{(0)}(1+\Delta_{\tilde{\alpha}}), 
\end{align}
where 
$n_{\rm e}^{(0)} = \dfrac{(k_{\rm F}^{\perp})^2k_{\rm F}^{\parallel}}{3\pi^2}$ 
is the electron density in the absence of the RSOI and 
\begin{align}
\Delta_{\alpha} = \frac{3}{2}\tilde{\alpha}^2\left(1+\frac{1+\tilde{\alpha}^2}{\tilde{\alpha}}\tan^{-1}\tilde{\alpha}\right)
\end{align}
is the correction to the electron density. 

For $I_{1}$ in Eq.~(\ref{I1}), 
the $k_{\parallel}$ integration of $I_{1}$ is calculated as 
\begin{align}
I_{1} &=-\frac{1}{4\pi^2}
\frac{k_{\rm F}^{\perp}k_{\rm F}^{\parallel}}{\hbar\omega+i0}
\sum_{s}s
\int_{0}^{\tilde{k}_{s}}d\tilde{k}_{\perp}\frac{(\tilde{k}_{\perp})^2}{
\sqrt{(\tilde{k}_{s}-\tilde{k}_{\perp})(\tilde{k}_{-s}+\tilde{k}_{\perp})}
}, 
\end{align}
where $\tilde{k}_{\perp} = k_{\perp}/k_{\rm F}^{\perp}$. 
Performing the $\tilde{k}_{\perp}$-integral and taking the sum of $s$, 
we have 
\begin{align}
I_{1} = \frac{1}{4\pi^2m_{\perp}}
\frac{k_{\rm F}^{\perp}k_{\rm F}^{\parallel}}{\hbar\omega+i0}\left(
3\tilde{\alpha} + (1+3\tilde{\alpha}^2)\tan^{-1}\tilde{\alpha}
\right).
\end{align}
The $\kv$-integrations of 
$I_{2}$ and $I_{3}$ are calculated using the same procedure. 
The results are given by  
\begin{align}
I_{2}&=-
\frac{1}{8\pi^2m_{\perp}}
\frac{k_{\rm F}^{\perp}k_{\rm F}^{\parallel}}{\hbar\omega+i0}
(\tilde{\alpha}+\tilde{\alpha}^2\tan^{-1}\tilde{\alpha}),\\
I_{3}&=
\frac{1}{8\pi^2m_{\parallel}}
\frac{k_{\rm F}^{\perp}k_{\rm F}^{\parallel}}{\hbar\omega+i0}
\left(
\tilde{\alpha} + (1+\tilde{\alpha}^2)\tan^{-1}\tilde{\alpha}
\right).
\end{align}
As 
\begin{align}
I_{1} = -4I_{2} + 4\frac{m_{\parallel}}{m_{\perp}}I_{3}, 
\end{align}
the value of $E_{0}$ in Eq.~(\ref{B-E0}) vanishes. 

For the $\kv$-integration of $C(\omega), D(\omega)$, and $E_{\mu}(\omega)(\mu=1,2,3)$, 
we first calculate the following integrations as 
\begin{align}
&S \equiv -8\pi^2\alpha^3\sqrt{\frac{m_{\perp}}{m_{\parallel}}}\sum_{\kv}\frac{s_{\kv}}{\gamma_{\kv} H_{\kv}},
\label{C-S}\\
&N \equiv -8\pi^2\alpha^3\sqrt{\frac{m_{\perp}}{m_{\parallel}}}\sum_{\kv}\frac{n'_{\kv}}{H_{\kv}}. 
\label{C-N}
\end{align}
Using these expressions, we can write Eqs.(\ref{B-C})--(\ref{B-E3}) as 
\begin{subequations}
\begin{align}
C(\omega) &= C(0) + \frac{3}{4}\frac{\tilde{\alpha}^{-1}}{1+\Delta_{\tilde{\alpha}}}
\left(\frac{\hbar\omega}{4\epsilon_{\rm F}}\right)^2 S,
\label{C-C}\\
D(\omega) &=
\frac{3}{4}\frac{\tilde{\alpha}^{-1}}{1+\Delta_{\tilde{\alpha}}}
\frac{\hbar\omega}{4\epsilon_{\rm F}}S,
\label{C-D}\\
E_{1}(\omega) &= \frac{\tilde{\alpha}^{-1}}{16n_{\rm e}}\frac{4\epsilon_{\rm F}}{\hbar\omega+i0}\sum_{\kv}\left(\gamma_{\kv}s'_{\kv} + \frac{m\alpha^2}{\hbar^2}n'_{\kv}\right)\nonumber\\
&+
\frac{3}{8}\frac{\tilde{\alpha}^{-2}}{1+\Delta_{\tilde{\alpha}}}
\frac{\hbar\omega}{4\epsilon_{\rm F}}\left(
2S-\frac{3}{2}N+\frac{9}{4}\omega\frac{\partial S}{\partial \omega}
-\frac{3}{4}\omega\frac{\partial N}{\partial \omega}
+\frac{3}{8}\omega^2\frac{\partial^2 S}{\partial \omega^2}
\right),
\label{C-E1}
\\
E_{2}(\omega) &= \frac{\tilde{\alpha}^{-1}}{8n_{\rm e}}\frac{4\epsilon_{\rm F}}{\hbar\omega+i0}\sum_{\kv}\left(\gamma_{\kv}s'_{\kv} + \frac{m\alpha^2}{\hbar^2}n'_{\kv}\right)\nonumber\\
&+
\frac{3}{4}\frac{\tilde{\alpha}^{-2}}{1+\Delta_{\tilde{\alpha}}}
\frac{\hbar\omega}{4\epsilon_{\rm F}}
\left(
-S+\frac{1}{2}N
-\frac{3}{4}\omega\frac{\partial S}{\partial \omega}
+\frac{1}{4}\omega\frac{\partial N}{\partial \omega}
-\frac{1}{8}\omega^2\frac{\partial^2 S}{\partial \omega^2}
\right),
\label{C-E2}
\\
E_{3}(\omega) &=
\frac{3}{8}\frac{\tilde{\alpha}^{-2}}{1+\Delta_{\tilde{\alpha}}}
\frac{\hbar\omega}{4\epsilon_{\rm F}}
\left(
-S+N-\omega\frac{\partial S}{\partial \omega}
\right),
\label{C-E3}
\end{align}
\end{subequations}
where 
\begin{align}
C(0) = \frac{\tilde{\alpha}^2}{n_{\rm e}}\epsilon_{\rm F}\sum_{\kv}\frac{s_{\kv}}{\gamma_{\kv}}
=-\frac{1}{2}\frac{\Delta_{\tilde{\alpha}}}{1+\Delta_{\tilde{\alpha}}}. 
\end{align}
 
Let us calculate the $S$ of Eq. (\ref{C-S}) and 
the $N$ of Eq.  (\ref{C-N}). These terms are expressed as 
\begin{align}
S&=-2\alpha^2\frac{\hbar^2}{m_{\parallel}}
\sqrt{\frac{m_{\perp}}{m_{\parallel}}}
\sum_{s}s
\int_{0}^{\infty}dk_{\perp}
\int_{-\infty}^{\infty}dk_{\parallel} (k_{\parallel})^2
\frac{\delta(\epsilon_{\kv}+s\alpha k_{\perp}-\epsilon_{\rm F})}{(\hbar\omega+i0)^2-4\alpha^2(k_{\perp})^2}, 
\label{int-S1}\\
N&=2\alpha^3\sqrt{\frac{m_{\perp}}{m_{\parallel}}}
\sum_{s}\int_{0}^{\infty}dk_{\perp}k_{\perp}
\int_{-\infty}^{\infty}dk_{\parallel} 
\frac{\delta(\epsilon_{\kv} + s \alpha k_{\perp} -\epsilon_{\rm F})}{(\hbar\omega+i0)^2-4\alpha^2(k_{\perp})^2}
\label{int-N1}. 
\end{align}
Substituting Eq.~(\ref{delta-f}) into 
Eqs.~(\ref{int-S1}) and (\ref{int-N1}) and obtaining the $k_{\parallel}$-integrals,  
we have 
\begin{align}
S&=-
\sum_{s}s
\int_{0}^{\tilde{k}_{s}}d\tilde{k}_{\perp}
\frac{\sqrt{(\tilde{k}_{s}-\tilde{k}_{\perp})(\tilde{k}_{-s}+\tilde{k}_{\perp})}}{
(\nu+i0)^2-(\tilde{k}_{\perp})^2},
\label{int-S2}\\
N&=\tilde{\alpha}
\sum_{s}\int_{0}^{\tilde{k}_{s}}d\tilde{k}_{\perp}\frac{\tilde{k}_{\perp}}{
(\nu+i0)^2-(\tilde{k}_{\perp})^2} \frac{1}{\sqrt{(\tilde{k}_{s}-\tilde{k}_{\perp})(\tilde{k}_{-s}+\tilde{k}_{\perp})}}
\label{int-N2}, 
\end{align}
where $\nu= \dfrac{\hbar\omega}{2\alpha k_{\rm F}}$. 
Using $\dfrac{1}{\nu + i0}= {\cal P}\dfrac{1}{\nu} -i\pi\delta(\nu)$, 
where ${\cal P}$ denotes the Cauchy principal value 
and changing the variable $x= \sqrt{\dfrac{\tilde{k}_{s}-\tilde{k}_{\perp}}
{\tilde{k}_{-s}+\tilde{k}_{\perp}}}$, 
the real parts of $S$ and $N$ are given by 
\begin{align}
&{\rm Re}S = -2
\sum_{s}s\int_{0}^{\sqrt{\tilde{k}_{s}/\tilde{k}_{-s}}}dx
\left(\
\frac{1}{1+x^2}-\frac{\nu-\tilde{k}_{s}}{2\nu}\frac{1}{x^2+\dfrac{\nu-\tilde{k}_{s}}{\nu+\tilde{k}_{-s}}}
-\frac{\nu+\tilde{k}_{s}}{2\nu}\frac{1}{x^2+\dfrac{\nu+\tilde{k}_{s}}{\nu-\tilde{k}_{-s}}}
\right),\\
&{\rm Re}N = \tilde{\alpha}
\sum_{s}
\int_{0}^{\sqrt{\tilde{k}_{s}/\tilde{k}_{-s}}}dx
\left(
\frac{\nu-\tilde{k}_{s}}{\nu+\tilde{k}_{-s}}
\frac{1}{x^2+\dfrac{\nu-\tilde{k}_{s}}{\nu+\tilde{k}_{-s}}}
-\frac{\nu+\tilde{k}_{s}}{\nu-\tilde{k}_{-s}}
\frac{1}{x^2+\dfrac{\nu+\tilde{k}_{s}}{\nu-\tilde{k}_{-s}}}
\right). 
\end{align}
Noting that 
\begin{align}
&\frac{\nu\mp\tilde{k}_{s}}{\nu}=\frac{\overline{\omega}\mp\overline{\omega}_{-s}}{\overline{\omega}}, \\
&\frac{\nu\mp\tilde{k}_{s}}{\nu\pm\tilde{k}_{-s}}=
\frac{(\overline{\omega}\mp\overline{\omega}_{-s})^2}
{\overline{\omega}^2\pm2s\tilde{\alpha}^2\overline{\omega}-\tilde{\alpha}^2},
\end{align}
with $\overline{\omega} = \tilde{\alpha}\nu = \hbar\omega/(4\epsilon_{\rm F})$ 
being a dimensionless angular frequency normalized by 
the Fermi energy 
and 
$\overline{\omega}_{s} = \tilde{\alpha}(\sqrt{1+\tilde{\alpha}^2}+s\tilde{\alpha})$ 
being frequencies at the transition edges. 
Performing the $x$-integral, we finally obtain  
\begin{align}
&{\rm Re}S = 
\frac{1}{\overline{\omega}}\sum_{s}
(s-\overline{\omega})L(Q_{s})+2\tan^{-1}\tilde{\alpha}, 
\label{RS}\\
&{\rm Re}N = \tilde{\alpha}^2\sum_{s}\frac{s}{s-\overline{\omega}}
R(Q_{s}), 
\label{RN}
\end{align}
where 
\begin{align}
&Q_{s} 
=\frac{\overline{\omega}^2+2s\tilde{\alpha}^2\overline{\omega}-\tilde{\alpha}^2}{(s-\overline{\omega})^2}, 
\end{align}
\begin{align}
&L(x) 
=
\begin{cases}
\sqrt{x}\tan^{-1}
\left(\dfrac{\tilde{\alpha}}{\sqrt{x}}\right), & x>0 \\
&\\
\dfrac{1}{2}\sqrt{-x}\ln\left|\dfrac{\sqrt{-x}+\tilde{\alpha}}{\sqrt{-x}-\tilde{\alpha}}\right|,& x<0 
\end{cases}
\end{align}
\begin{align}
&R(x)=
\begin{cases}\dfrac{1}{\sqrt{x}}\tan^{-1}
\left(\dfrac{\tilde{\alpha}}{\sqrt{x}}\right), & x>0\\
&\\
-\dfrac{1}{2}\dfrac{1}{\sqrt{-x}}\ln\left|\dfrac{\sqrt{-x}+\tilde{\alpha}}{\sqrt{-x}-\tilde{\alpha}}\right|.& x<0
\end{cases}
\end{align}

For the imaginary parts of $S$ and $N$, we can easily obtain the $k_{\perp}$-integral due to the delta function $\delta(\nu-\tilde{k}_{\perp})$. The results are given by 
\begin{align}
&{\rm Im}S
=\frac{\pi}{2}\frac{1}{\overline{\omega}}
\sum_{s}s S_{s}(\bar{\omega}), 
\label{IS}\\
&    {\rm Im}N = -\frac{\pi}{2}\tilde{\alpha}^2
\sum_{s}T_{s}(\bar{\omega}), 
    \label{IN}
\end{align}
where 
\begin{align}
&S_{s}(\overline{\omega})=\theta(\overline{\omega}_{-s}-\overline{\omega})\sqrt{\tilde{\alpha}^2-2s\tilde{\alpha}^2\overline{\omega}-\overline{\omega}^2},\\
&T_{s}(\overline{\omega})=\theta(\overline{\omega}_{-s}-\overline{\omega})\frac{1}{\sqrt{\tilde{\alpha}^2-2s\tilde{\alpha}^2\overline{\omega}-\overline{\omega}^2}}.
\end{align}
From these results, we obtain 
\begin{subequations}
\begin{align}
\overline{\omega}\frac{\partial}{\partial \overline{\omega}}{\rm Re}S 
&=\tilde{\alpha}^2\sum_{s}\frac{s}{\overline{\omega}}R(Q_{s}),\\
\overline{\omega}^2\frac{\partial^2}{\partial \overline{\omega}^2}{\rm Re}S &=
-2\tilde{\alpha}^2\sum_{s}\frac{s}{\overline{\omega}}R(Q_{s})-\tilde{\alpha}^2(1+\tilde{\alpha}^2)\sum_{s}\frac{\overline{\omega}}{
(s-\overline{\omega})(\overline{\omega}^2+2s\tilde{\alpha}^2\overline{\omega}-\tilde{\alpha}^2)}R(Q_{s})\nonumber\\
&-\tilde{\alpha}^3\sum_{s}\frac{s-\overline{\omega}}{\overline{\omega}(\overline{\omega}^2
+2s\tilde{\alpha}^2\overline{\omega}-\tilde{\alpha}^2)},\\
\overline{\omega}\frac{\partial}{\partial \overline{\omega}}{\rm Re}N &= 
\tilde{\alpha}^2\sum_{s}\frac{s\overline{\omega}}{(s-\overline{\omega})^2}R(Q_{s})
-\tilde{\alpha}^2(1+\tilde{\alpha}^2)\sum_{s}\frac{\overline{\omega}^2}{(s-\overline{\omega})^2
(\overline{\omega}^2+2s\tilde{\alpha}^2\overline{\omega}-\tilde{\alpha}^2)}R(Q_{s})
\nonumber\\
&-\tilde{\alpha}^3\sum_{s}\frac{1}{\overline{\omega}^2+2s\tilde{\alpha}^2\overline{\omega}-\tilde{\alpha}^2},\\
\overline{\omega}\frac{\partial}{\partial \overline{\omega}}{\rm Im}S 
&= -{\rm Im}S + {\rm Im}N - \frac{\pi}{2}\sum_{s}s\overline{\omega}T_{s},\\
\overline{\omega^2}\frac{\partial^2}{\partial \overline{\omega}^2}{\rm Im}S &= 2{\rm Im}S -2{\rm Im}N
 +\frac{\pi}{2}\sum_{s}s\overline{\omega}T_{s}
-\frac{\pi}{2}\sum_{s}s\overline{\omega}(\overline{\omega}+s\tilde{\alpha}^2)^2
T^3_{s}, \\
\overline{\omega}\frac{\partial}{\partial \overline{\omega}}{\rm Im}N 
&=- \frac{\pi}{2}\tilde{\alpha}^2\sum_{s}\overline{\omega}(\overline{\omega}+s\tilde{\alpha}^2)T_{s}^3.
\end{align}
\end{subequations}
Substituting these results into Eqs.(\ref{B-C}) and (\ref{B-E3}), 
we obtain  
\begin{subequations}
\begin{align}
&{\rm Re}C(\bar{\omega}) 
= \frac{3}{4}\frac{\tilde{\alpha}^{-1}}{1+\Delta_{\tilde{\alpha}}}
\left[\overline{\omega}\sum_{s=\pm1}(s-\overline{\omega})
L(Q_{s})+2\overline{\omega}^2
\tan^{-1}\tilde{\alpha}\right] -\frac{1}{2} \frac{\Delta_{\tilde{\alpha}}}{1+\Delta_{\tilde{\alpha}}},
\label{V-RC}
\\
&{\rm Im}C(\bar{\omega}) 
=\frac{3\pi}{8} \frac{\tilde{\alpha}^{-1}}{1+\Delta_{\tilde{\alpha}}}
\overline{\omega}\sum_{s}s S_{s}(\bar{\omega}),
\label{V-IC}\\
&{\rm Re}D(\bar{\omega}) =\frac{3}{4}\frac{\tilde{\alpha}^{-1}}{1+\Delta_{\tilde{\alpha}}}
\left[\sum_{s=\pm1}
(s-\overline{\omega})L(Q_{s})
+2\overline{\omega}\tan^{-1}\tilde{\alpha}\right],
\label{V-RD}
\\
&{\rm Im}D(\bar{\omega}) 
= \frac{3\pi}{8}\frac{\tilde{\alpha}^{-1}}{1+\Delta_{\tilde{\alpha}}}
\sum_{s}s S_{s}(\bar{\omega}), 
\label{V-ID}
\end{align}
\begin{align}
&{\rm Re}E_{1}(\bar{\omega})=\frac{1}{4}
\frac{\Delta_{\tilde{\alpha}}\tilde{\alpha}^{-1}}{1+\Delta_{\tilde{\alpha}}}
\frac{1}{\overline{\omega}}
+\frac{3}{8}\frac{\tilde{\alpha}^{-2}}{1+\Delta_{\tilde{\alpha}}}
\left[
4\overline{\omega}\tan^{-1}\tilde{\alpha}
+2\sum_{s}(s-\overline{\omega})L(Q_{s})
+\frac{9}{8}\tilde{\alpha}^2\sum_{s}\frac{s-2\overline{\omega}}{s-\overline{\omega}}
s R(Q_{s})\right.
\nonumber\\
&\qquad~~\left.-\frac{3}{8}\tilde{\alpha}^2\sum_{s}
\frac{(s-3\overline{\omega})\{\tilde{\alpha}+\tilde{\alpha}^2R(Q_{s})\}}
{\overline{\omega}^2+2s\tilde{\alpha}^2\overline{\omega}-\tilde{\alpha}^2}
\right]
,\label{V-RE1}\\
&{\rm Im}E_{1}(\bar{\omega})=
-\frac{\pi}{4}
\frac{\Delta_{\tilde{\alpha}}\tilde{\alpha}^{-1}}{1+\Delta_{\tilde{\alpha}}}
\delta(\overline{\omega})
+
\frac{3\pi}{32}\frac{\tilde{\alpha}^{-2}}{1+\Delta_{\tilde{\alpha}}}
\left[\sum_{s}s S_{s}(\bar{\omega})
-\frac{15}{4}\overline{\omega}^2\sum_{s}s T_{s}(\bar{\omega})
+\frac{3}{4}\overline{\omega}^2(\tilde{\alpha}^4-\overline{\omega}^2)
\sum_{s}s T^{3}_{s}(\bar{\omega})
\right]
,\label{V-IE1}\\
&{\rm Re}E_{2}(\bar{\omega})=
\frac{1}{2}
\frac{\Delta_{\tilde{\alpha}}\tilde{\alpha}^{-1}}{1+\Delta_{\tilde{\alpha}}}
\frac{1}{\overline{\omega}}
-\frac{3}{4}\frac{\tilde{\alpha}^{-2}}{1+\Delta_{\tilde{\alpha}}}
\left[
2\overline{\omega}\tan^{-1}\tilde{\alpha}
+\sum_{s}(s-\overline{\omega})L(Q_{s})
+\frac{3}{4}\tilde{\alpha}^2\sum_{s}\frac{s-2\overline{\omega}}{s-\overline{\omega}}
s R(Q_{s})\right.\nonumber\\
&\qquad~~\left.
-\frac{1}{8}\tilde{\alpha}^2\sum_{s}
\frac{(s-3\overline{\omega})\{\tilde{\alpha}+\tilde{\alpha}^2R(Q_{s})\}}
{\overline{\omega}^2+2s\tilde{\alpha}^2\overline{\omega}-\tilde{\alpha}^2}
\right],\label{V-RE2}\\
&{\rm Im}E_{2}(\bar{\omega})=-\frac{\pi}{2}
\frac{\Delta_{\tilde{\alpha}}\tilde{\alpha}^{-1}}{1+\Delta_{\tilde{\alpha}}}
\delta(\overline{\omega})
-\frac{3\pi}{16}\frac{\tilde{\alpha}^{-2}}{1+\Delta_{\tilde{\alpha}}}
\left[\sum_{s}s S_{s}(\bar{\omega})
-\frac{5}{4}\overline{\omega}^2\sum_{s}s T_{s}(\bar{\omega})
+\frac{1}{4}\overline{\omega}^2(\tilde{\alpha}^4-\overline{\omega}^2)
\sum_{s}s T^3_{s}(\bar{\omega})
\right]
,\label{V-IE2}\\
&{\rm Re}E_{3}(\bar{\omega})=
\frac{3}{8}\frac{\tilde{\alpha}^{-2}}{1+\Delta_{\tilde{\alpha}}}
\left[
-2\overline{\omega}\tan^{-1}\tilde{\alpha}
-\sum_{s}(s-\overline{\omega})L(Q_{s})
-\tilde{\alpha}^2\sum_{s}\frac{s-2\overline{\omega}}{s-\overline{\omega}}
s R(Q_{s})
\right]
,\label{V-RE3}\\
&{\rm Im}E_{3}(\bar{\omega})=
\frac{3\pi}{16}\frac{\tilde{\alpha}^{-2}}{1+\Delta_{\tilde{\alpha}}}
\overline{\omega}^2\sum_{s}s T_{s}(\bar{\omega})
.\label{V-IE3}
\end{align}
\end{subequations}
Note that $E_{\mu}(\omega)$ diverges towards the transition edges 
$\bar{\omega}_{\pm}$ and at $\omega = 0$ (Fig.~\ref{CDE}). 
The former originates from the expansion of the 
correlation functions of Eq.~(\ref{chi-jj-js-sj-A}) with respect to $\qv$ and ${\bm M}$, which can be related to the derivative of 
$S(\omega)$ and $N(\omega)$ for $\omega$. 
The latter divergence comes from the delta-function singularity. 
Both divergences can be avoided by replacing the 
positive infinitesimal $0$ with the finite $\tilde{\eta} = \eta/(\hbar \omega^{\parallel}_{\rm p})$ 
in Eqs.~(\ref{int-S1}) and (\ref{int-N1}). 
We have also performed these calculations, and the results are shown in Fig.~\ref{F-CDE}. 
\begin{figure*}
\resizebox{16.8cm}{!}{\includegraphics[angle=0]{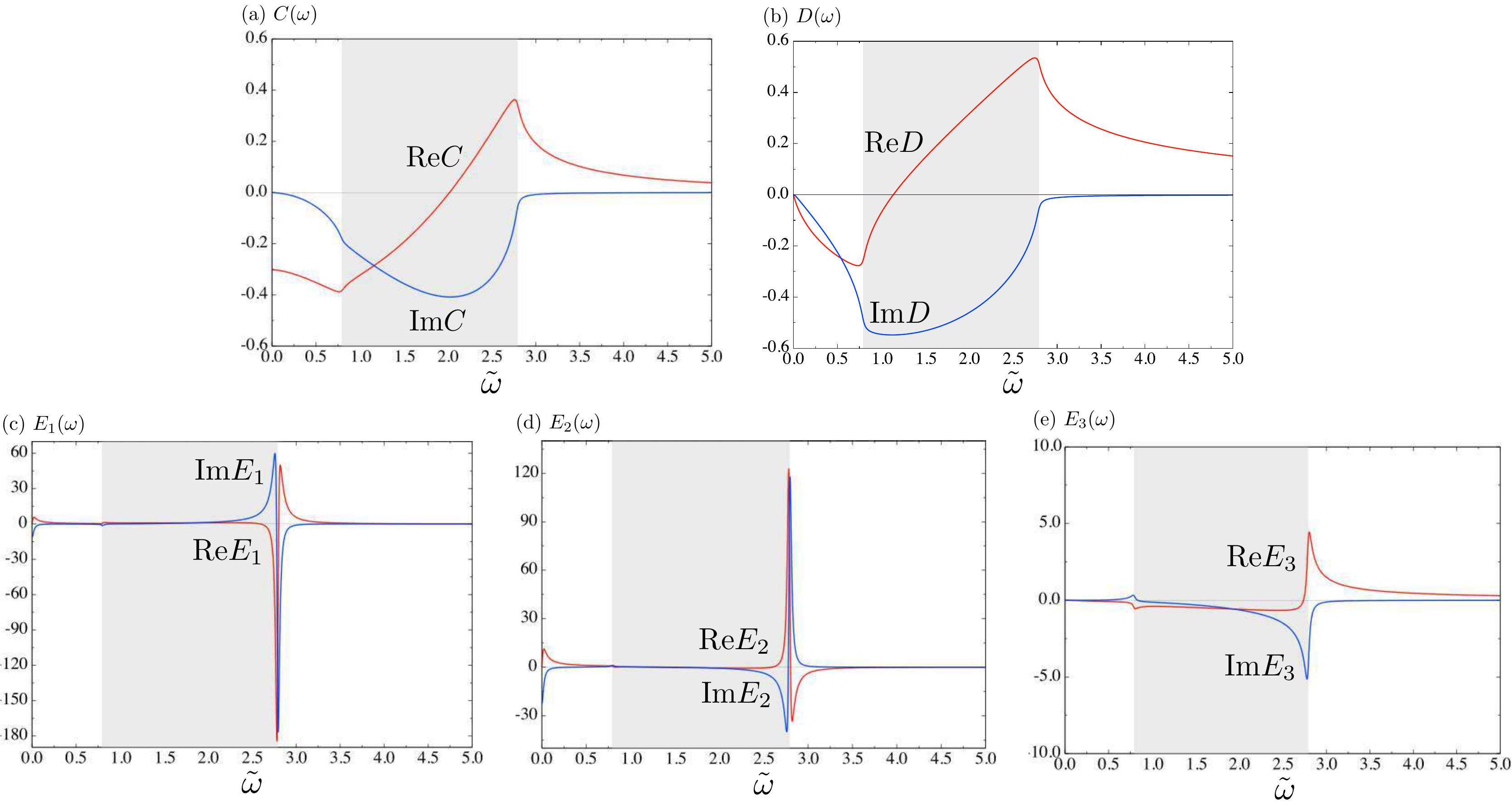}}
\caption{Real and imaginary parts of $C(\omega)$, $D(\omega)$, 
$E_{1}(\omega)$, $E_{2}(\omega)$, and $E_{3}(\omega)$ 
as functions of $\tilde{\omega}$ for $\tilde{\alpha} = 0.67$ 
and finite $\tilde{\eta} = 0.01$. 
For $C(\omega)$ and $D(\omega)$, 
cusps exist at the interband transition edges 
$\tilde{\omega}_{\pm} = \omega_{\pm} /\omega^{\parallel}_{\rm p}$ 
($\tilde{\omega}_{-} \simeq 0.8$ and $\tilde{\omega}_{+}\simeq 2.8$). 
For $E_{1}(\omega)$, $E_{2}(\omega)$, and $E_{3}(\omega)$, 
singularities exist at the interband transition edges 
$\tilde{\omega}_{\pm}$. 
The shaded region represents the hyperbolic frequency region (see Sec. VI-B). }
\label{F-CDE}
\end{figure*}
For $E_{\mu}(\omega) (\mu = 1,2,3)$, 
we see that the sharp peaks disappear at $\omega_{-}$ 
because of the term cancellation.

\section{Calculation of Poynting vector, angle of refraction, and transmittance}
\subsection{Calculation of Poynting vector ${\bm S}$ in Eq.~(\ref{S-B})}

In general, the time-averaged Poynting vector is given by \cite{Landau} 
\begin{align}
{\bm S}(\rv) &=\frac{\vare_{0}c^2}{2}
{{\rm Re}}({\bm E}e^{i\qv\cdot\rv}\times {\bm B}^{*}e^{-i\qv^{*}\cdot\rv}). 
\label{S-a1}
\end{align}
For an extraordinary wave propagating in the $x$-$z$ plane,  
the wave vector and the amplitude of the electric field are 
$\qv = q_{x}{\bm e}_{x} + q_{z}{\bm e}_{z}$ and 
${\bm E} =E_{x}{\bm e}_{x}+E_{z}{\bm e}_{z}$, respectively. 
Using these equations and Faraday's law, ${\bm B} = \qv \times {\bm E}/\omega$, 
equation (\ref{S-a1}) is solved as 
\begin{align}
{\bm S}&= \frac{\vare_{0}c^2}{2\omega}{\rm Re}\left\{
(|E_{z}|^2q^{*}_{x}-E_{x}^{*}E_{z}q^{*}_{z}){\bm e}_{x}
+ (
|E_{x}|^2q^{*}_{z}-E_{x}E^{*}_{\qv,3}q^{*}_{x}
){\bm e}_{z}
\right\}e^{-2i{\rm Im}\qv\cdot\rv}. 
\end{align} 
Substituting the dispersion relation for the extraordinary wave, 
$q^2_{x}/\vare_{z}+q^2_{z}/\vare_{x} = \omega^2/c^2$, 
given in Eq.~(\ref{dispersion-ew}) into 
the wave equation in Eq.~(\ref{weq-2}), 
we have 
\begin{align}
\vare_{x}q_{x}E_{x} + \vare_{z}q_{z}E_{z}=0,  
\label{extra1}
\end{align}
Using this and Eq.~(\ref{dispersion-ew}), 
we obtain
\begin{align}
{\bm S}(\rv)
&= \frac{\omega\vare_{0}}{2}\left\{
{\rm Re}\left(
\frac{\vare_{z}}{q_{x}}\right)
|E_{z}|^2
{\bm e}_{x}
+
{\rm Re}\left(\frac{\vare_{x}}{q_{z}}\right)|E_{x}|^2
{\bm e}_{z}\right\}
e^{-2i{\rm Im}{\bm q}\cdot\rv}. 
\label{S}
\end{align}

\subsection{Calculations of $\theta_{S}^{\perp}$ 
and $\theta_{S}^{\parallel}$ in 
Eqs.~(\ref{TS-perp}) and (\ref{TS-para})}

From Fig.~\ref{Sec7-TypeI-II}(a) and Eq.~(\ref{S}), 
the angle of refraction of the Poynting vector is given by 
\begin{align}
\tan \theta_{S}^{\perp} = 
\frac{{\bm S}({\bm 0})\cdot{\bm e}_{z}}{{\bm S}({\bm 0})\cdot{\bm e}_{x}}
=\frac{{\rm Re}\left(\dfrac{\vare_{x}}{q_{z}}\right)|E_{x}|^2}{{\rm Re}\left(
\dfrac{\vare_{z}}{q_{x}}\right)
|E_{z}|^2}. 
\label{App-TS-perp}
\end{align}
From Eq.~(\ref{extra1}), we have 
\begin{align}
\frac{|E_{x}|^2}{|E_{z}|^2}= \frac{|\vare_{z}q_{z}|^2}{|\vare_{x}q_{x}|^2}. 
\end{align}
Substituting this into Eq.~(\ref{App-TS-perp}), 
we obtain 
\begin{align}
\tan\theta_{S}^{\parallel} = 
\frac{{\rm Re}\left(\dfrac{q_{z}}{\vare_{x}}\right)}{{\rm Re}\left(\dfrac{q_{z}}{\vare_{x}}\right)}
=\frac{{\rm Re}(1/\vare_{x})}{{\rm Re}(q_{x}/\vare_{z})}\frac{\omega}{c}\sin\theta_{\rm i}, 
\end{align} 
where we have used $q_{z} = (\omega/c)\sin\theta_{\rm i}$
(Fig.~\ref{Sec7-TypeI-II}(b)). 

For $\theta_{S}^{\parallel}$, 
we can put $q_{x}= (\omega/c)\sin \theta_{\rm i}$ from Fig.~\ref{Sec7-TypeI-II}(b); thus, we obtain 
\begin{align}
\tan\theta_{S}^{\parallel} 
= \frac{{\bm S}({\bm 0})\cdot{\bm e}_{x}}{{\bm S}({\bm 0})\cdot{\bm e}_{z}}
=\frac{{\rm Re}\left(\dfrac{q_{x}}{\vare_{z}}\right)}{{\rm Re}\left(\dfrac{q_{x}}{\vare_{z}}\right)}
=\frac{{\rm Re}(1/\vare_{z})}{{\rm Re}(q_{z}/\vare_{x})}\frac{\omega}{c}\sin\theta_{\rm i}, 
\label{App-TS-para}
\end{align}
where we have used $q_{x} = (\omega/c)\sin\theta_{\rm i}$ (Fig.~\ref{Sec7-TypeI-II}(b)).

\subsection{Calculations of $T_{\perp}$ and $T_{\parallel}$ 
in Eqs.~(\ref{Tperp}) and (\ref{Tpara})}

\begin{figure}[t]
\resizebox{8.4cm}{!}{\includegraphics{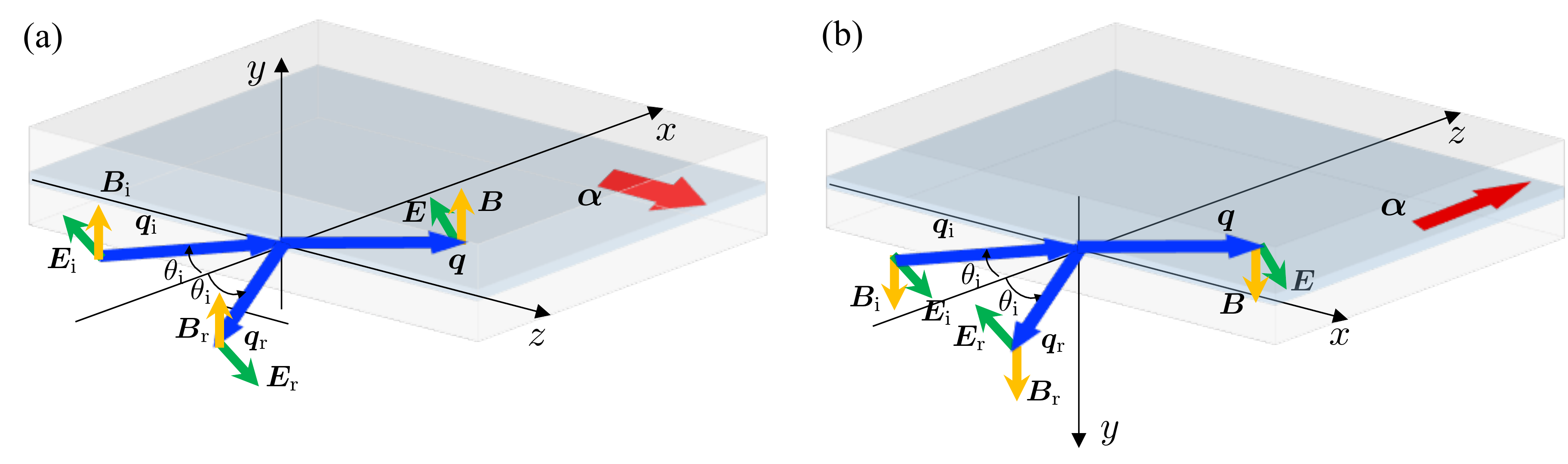}}
\caption{Schematic illustration of reflection and refraction 
of electromagnetic waves obliquely incident from vacuum on semi-infinite Rashba conductor. 
The polarization of the electric waves is parallel to the plane of incidence 
($x$-$z$ plane).  
The Rashba field ${\bm \alpha}$ direction is (a)
parallel and (b) perpendicular to the the vacuum-Rashba interface. 
The orientations of the magnetic field vectors are chosen in the positive $y$-direction. 
}
\label{config-ap}
\end{figure}

In the case of Fig.~\ref{config-ap}(a), 
the Poynting vector of the incident plane wave in vacuum, 
${\bm S}_{\rm i}$, is given by 
\begin{align}
{\bm S}_{\rm i} = \frac{\vare_{0}c^2}{2}|E_{\rm i}|^2( \cos\theta_{\rm i}{\bm e}_{x}
+\sin\theta_{\rm i}{\bm e}_{z}), 
\end{align}
where $E_{\rm i}$ is the amplitude of the incident-wave electric field. 
From this and Eqs.~(\ref{S}), we have 
\begin{align}
T_{\perp}\equiv
\frac{ {\bm S}({\bm 0})\cdot{\bm e}_{x}}{{\bm S}_{\rm i}\cdot{\bm e}_{x}} 
=
\frac{\omega}{c}\frac{
{\rm Re}\left({\vare_{z}}/{q_{x}}\right)}{\cos\theta_{\rm i}}
\frac{|E_{z}|^2}{|E_{\rm i}|^2}. 
\label{Tpara1}
\end{align}
To calculate $|E_{z}|^2/|E_{\rm i}|^2$, 
we consider the continuity conditions 
for the tangential components of the electromagnetic field  
at the interface \cite{Jackson} 
(Fig.~\ref{config-ap}(a)): 
\begin{align}
&({\bm E}_{\rm i}+{\bm E}_{\rm r})\cdot{\bm e}_{z}= {\bm E}\cdot{\bm e}_{z},\\
&({\bm B}_{\rm i} + {\bm B}_{\rm r})\cdot{\bm e}_{y} = {\bm B}\cdot{\bm e}_{y}, 
\label{conti-2}
\end{align}
where 
${\bm E}_{\rm i(r)}$ and ${\bm B}_{\rm i(r)}=\qv_{\rm i(r)}\times{\bm E}_{\rm i(r)}/\omega$
are the electric and magnetic fields of the incident (reflected) wave, respectively. 
Here, $\qv_{\rm i(r)}~(|\qv_{\rm i(r)}|=\omega/c)$ is the wave vector 
of the incident (reflected) wave. 
From Fig.~\ref{config-ap}(a), we have 
\begin{align}
&-(E_{\rm i} -E_{\rm r})\cos \theta_{\rm i}  = E_{z},\\
&E_{\rm i} +E_{\rm r} = \frac{c}{\omega}(q_{z}E_{x}-q_{x}E_{z}), 
\end{align}
By eliminating $E_{\rm r}$ and using Eqs.~(\ref{extra1}) and (\ref{dispersion-ew}), 
we have 
\begin{align}
\frac{E_{z}}{E_{\rm i}} = - \dfrac{2\cos\theta_{\rm i}}
{1+\dfrac{\omega}{c}\dfrac{\vare_{z}}{q_{x}}\cos\theta_{\rm i}}. 
\end{align}
Thus, we obtain 
\begin{align}
T_{\perp}=\dfrac{4\dfrac{\omega}{c}{\rm Re}\left(\dfrac{\vare_{z}}{q_{x}}\right)\cos\theta_{\rm i}}{\left|
1+\dfrac{\omega}{c}\dfrac{\vare_{z}}{q_{x}}\cos\theta_{\rm i}
\right|^2}. 
\end{align}

In the case of Fig.~\ref{config-ap}(b), 
the Poynting vector of the incident plane wave in vacuum, 
${\bm S}_{\rm i}$, is given by 
\begin{align}
{\bm S}_{\rm i} = \frac{\vare_{0}c^2}{2}|E_{\rm i}|^2( \sin\theta_{\rm i}{\bm e}_{x}
+\cos\theta_{\rm i}{\bm e}_{z}). 
\end{align}
From this and Eqs.~(\ref{S}), 
we have 
\begin{align}
T_{\parallel}\equiv\frac{ {\bm S}({\bm 0})\cdot{\bm e}_{z}}{{\bm S}_{\rm i}\cdot{\bm e}_{z}} 
=
\frac{\omega}{c}\frac{
{\rm Re}\left({\vare_{x}}/{q_{z}}\right)}{\cos\theta_{\rm i}}
\frac{|E_{x}|^2}{|E_{\rm i}|^2}. 
\label{Tpara1}
\end{align}
To calculate $|E_{x}|^2/|E_{\rm i}|^2$, 
we consider the following continuity conditions (Fig.~\ref{config-ap}(b)): 
\begin{align}
&({\bm E}_{\rm i}+{\bm E}_{\rm r})\cdot{\bm e}_{x}= {\bm E}\cdot{\bm e}_{x},\\
&({\bm B}_{\rm i} + {\bm B}_{\rm r})\cdot{\bm e}_{y} = {\bm B}\cdot{\bm e}_{y}, 
\label{conti-2}
\end{align}
which are expressed as 
\begin{align}
&(E_{\rm i}-E_{\rm r})\cos\theta_{\rm i} = E_{x}, \\
&E_{\rm i} + E_{\rm r} = \frac{c}{\omega}(q_{z}E_{x}-q_{x}E_{z}). 
\end{align}
By eliminating $E_{\rm r}$ and using Eqs.~(\ref{extra1}) and (\ref{dispersion-ew}), 
we have 
\begin{align}
\frac{E_{x}}{E_{\rm i}} =
 \dfrac{2\cos\theta_{\rm i}}
{1+\dfrac{\omega}{c}\dfrac{\vare_{x}}{q_{x}}\cos\theta_{\rm i}}
\end{align}
Substituting this into Eq.~(\ref{Tpara1}), 
we obtain 
\begin{align}
T_{\parallel}= 
\dfrac{4\dfrac{\omega}{c}{\rm Re}\left(\dfrac{\vare_{x}}{q_{z}}\right)\cos\theta_{\rm i}}{\left|
1+\dfrac{\omega}{c}\dfrac{\vare_{x}}{q_{z}}\cos\theta_{\rm i}
\right|^2}. 
\end{align}

\subsection{Calculation of transmittance, reflectance, and absorbance in a Rashba conductor slab} 
\begin{figure}[t]
\resizebox{8.4cm}{!}{\includegraphics{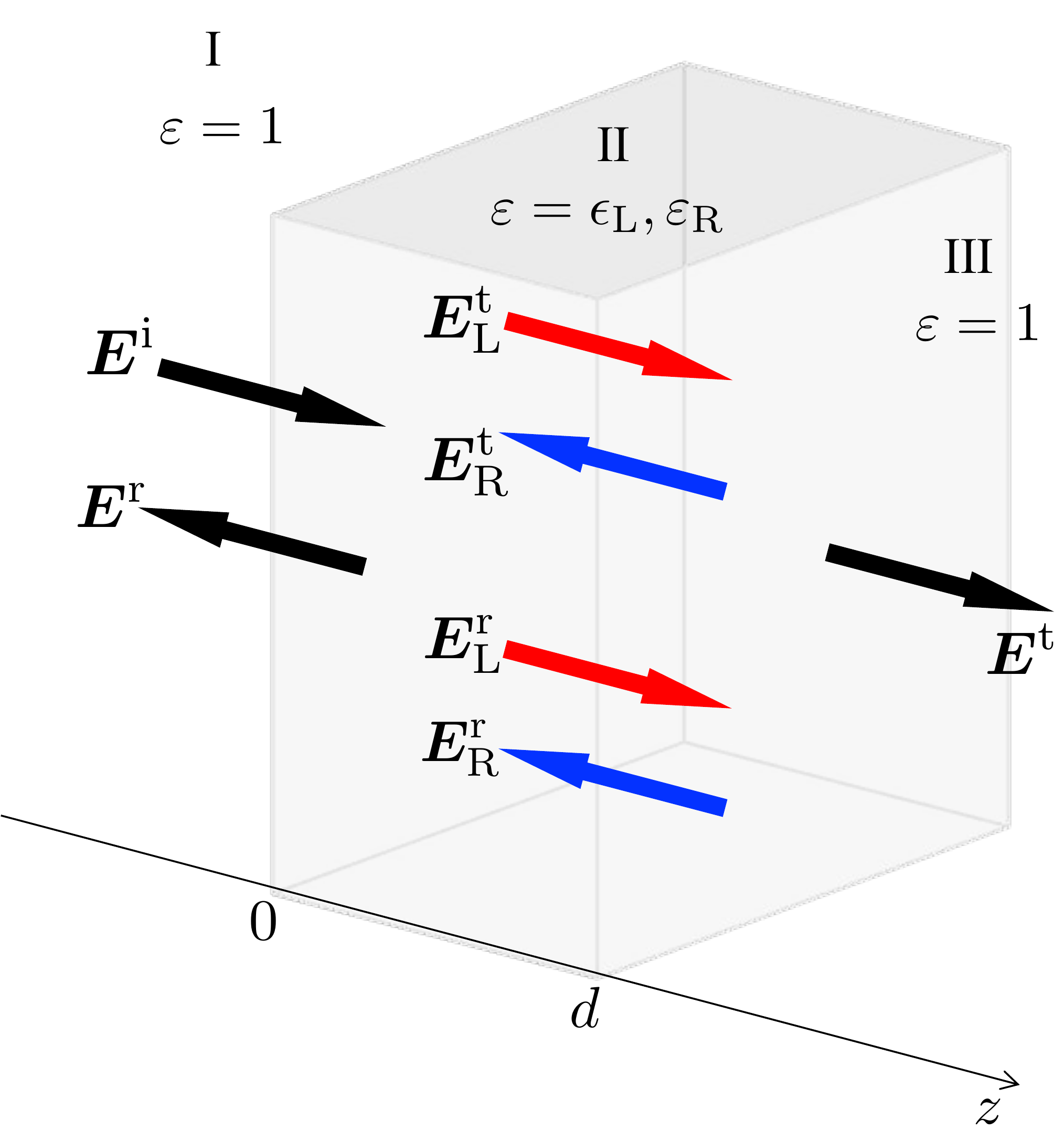}}
\caption{Schematic illustration of reflection and refraction 
of electromagnetic waves normally incident from vacuum on Rashba conductor slab 
with thickness $d$.  
}
\label{MCD-config2}
\end{figure}

Let us consider propagation of an electromagnetic wave through a Rashba conductor slab 
with thickness $d$ 
and evaluate the transmittance $T_{\rm L,R}$, reflectance $R_{\rm L,R}$, 
and absorbance $A_{\rm L,R}$ for the left- and right-handed circularly polarized waves. 
We assume that 
the wave propagates in the $z$-direction and 
is normally incident on the Rashba conductor interface at $z = 0$. 
We separate the wave propagation region into three parts: I, II and III. 
Regions I and III are vacuums for which the wave vector is denoted 
$q_{0} = \omega/c$ and region II covers the Rashba conductor, 
the wave vector of which is $q_{\rm R, L} = \omega \sqrt{\vare_{\pm}}/c$ 
in Eq.~(\ref{qLR})  
for the left- and the right-handed circularly polarized waves, respectively.  
In regions I and III, 
the electric waves with frequency $\omega$ 
can be expressed as 
\begin{align}
&{\bm E}_{\rm I} =
\begin{pmatrix}
 E^{\rm i}_{x} \\
E^{\rm i}_{y}
\end{pmatrix}
e^{i(q_{0}z-\omega t)} + 
\begin{pmatrix}
E^{\rm r}_{x} \\
E^{\rm r}_{y}
\end{pmatrix}
e^{-i(q_{0}z+\omega t)}, \\
&{\bm E}_{\rm III} = 
\begin{pmatrix}
E^{\rm t}_{x}\\
E^{\rm t}_{y}
\end{pmatrix}
e^{i(q_{0}z-\omega t)},
\end{align}
where the subscripts $i$, $r$, and $t$ on $E_{x,y}$ denote the incident, reflected, and transmitted components of the electric field in the vacuum, respectively, 
and we have used the following representation:
\begin{align}
\begin{pmatrix}
A\\
B
\end{pmatrix}
=A{\bm e}_{x} + B{\bm e}_{y}.
\end{align}
For region II, 
we assume that 
the electric wave consists of the transmitted and reflected 
left- and right-handed circularly polarized waves, 
the components of which are denoted by $E^{\rm t(r)}_{\rm L(R)}$, respectively. 
Thus, in regions II, we can write the electric waves, 
${\bm E}_{\rm II}$, as 
\begin{align}
{\bm E}_{\rm II} &=
\begin{pmatrix}
E^{\rm t}_{\rm L} \\
iE^{\rm t}_{\rm L}
\end{pmatrix}
e^{i(q_{\rm L}z-\omega t)} 
+\begin{pmatrix}
E^{\rm t}_{\rm R} \\
-iE^{\rm t}_{\rm R}
\end{pmatrix}
e^{i(q_{\rm R}z-\omega t)} 
+\begin{pmatrix}
E^{\rm r}_{\rm L} \\
iE^{\rm r}_{\rm L}
\end{pmatrix}
e^{-i(q_{\rm L}z+\omega t)} 
+ \begin{pmatrix}
E^{\rm r}_{\rm R} \\
-iE^{\rm r}_{\rm R}
\end{pmatrix}
e^{-i(q_{\rm R}z+\omega t)}. 
\end{align}
From Faraday's law, the corresponding magnetic fields are given by 
\begin{align}
&\omega{\bm B}_{\rm I}e^{i\omega t}=
q_{0}
\begin{pmatrix}
-E^{\rm i}_{y}\\
E^{\rm i}_{x}
\end{pmatrix}
e^{iq_{0}z} 
-q_{0}
\begin{pmatrix}
-E^{\rm r}_{y}\\
E^{\rm r}_{x}
\end{pmatrix}
e^{-iq_{0}z}, \\
&\omega{\bm B}_{\rm II}e^{i\omega t}=
q_{\rm L}
\begin{pmatrix}
-iE^{\rm t}_{\rm L} \\
E^{\rm t}_{\rm L}
\end{pmatrix}
e^{iq_{\rm L}z}
+q_{\rm R}
\begin{pmatrix}
iE^{\rm t}_{\rm R} \\
E^{\rm t}_{\rm R}
\end{pmatrix}
e^{iq_{\rm R}z} 
 -q_{\rm L}
\begin{pmatrix}
-iE^{\rm r}_{\rm L} \\
E^{\rm r}_{\rm L}
\end{pmatrix}
e^{-iq_{\rm L}z} 
-q_{\rm R}
\begin{pmatrix}
iE^{\rm r}_{\rm R} \\
E^{\rm r}_{\rm R}
\end{pmatrix}
e^{-iq_{\rm R}z},  \\
&\omega{\bm B}_{\rm III}e^{i\omega t} = 
q_{0}
\begin{pmatrix}
-E^{\rm t}_{y}\\
E^{\rm t}_{x}
\end{pmatrix}
e^{iq_{0}z}, 
\end{align}
At $z = 0$, the continuity condition gives 
\begin{subequations}
\begin{align}
E^{\rm i}_{x} + E^{\rm r}_{x} &= E^{\rm t}_{\rm L} + E^{\rm t}_{\rm R} 
+ E^{\rm r}_{\rm L} + E^{\rm r}_{\rm R},\\
E^{\rm i}_{y} +E^{\rm r}_{y}&= iE^{\rm t}_{\rm L} - iE^{\rm t}_{\rm R} 
+i E^{\rm r}_{\rm L} -i E^{\rm r}_{\rm R},\\
E^{\rm i}_{x} - E^{\rm r}_{x} &= \delta_{\rm L}E^{\rm t}_{\rm L} 
+ \delta_{\rm R}E^{\rm t}_{\rm R}-\delta_{\rm L}E^{\rm r}_{\rm L} -\delta_{\rm R} E^{\rm r}_{\rm R},\\
E^{\rm i}_{y} -E^{\rm r}_{y}&= i\delta_{\rm L}E^{\rm t}_{\rm L} 
- i\delta_{\rm R}E^{\rm t}_{\rm R} 
i\delta_{\rm L} E^{\rm r}_{\rm L} +i\delta_{\rm R} E^{\rm r}_{\rm R},
\end{align}
\end{subequations}
where $\delta_{\rm L,R} = q_{\rm L,R}/q_{0}$. 
These equations are expressed in matrix form as 
\begin{align}
\label{M1}
\begin{pmatrix}
E^{\rm i}_{+}\\
E^{\rm i}_{-}\\
E^{\rm r}_{+}\\
E^{\rm r}_{-}
\end{pmatrix}
=
\begin{pmatrix}
0&1+\delta_{\rm R}&0&1-\delta_{\rm R}\\
1+\delta_{\rm L}&0&1-\delta_{\rm L}&0\\
0&1-\delta_{\rm R}&0&1+\delta_{\rm R}\\
1-\delta_{\rm L}&0&1+\delta_{\rm L}&0
\end{pmatrix}
\begin{pmatrix}
E^{\rm t}_{\rm L}\\
E^{\rm t}_{\rm R}\\
E^{\rm r}_{\rm L}\\
E^{\rm r}_{\rm R}
\end{pmatrix},
\end{align}
where $E^{\rm i,r}_{\pm} = E^{\rm i,r}_{x}\pm i E^{\rm i,r}_{y}$ 
is a component representing the left- $(-)$ and right-handed $(+)$
circularly polarized incident and reflected waves, respectively.  
On the other hand, at $z=d$, the continuity condition gives 
\begin{subequations}
\begin{align}
&E^{\rm t}_{\rm L}e^{iq_{\rm L}d} + E^{\rm t}_{\rm R}e^{iq_{\rm R}d}
+E^{\rm r}_{\rm L}e^{-iq_{\rm L}d} + E^{\rm r}_{\rm R}e^{-iq_{\rm R}d}=E^{\rm t}_{x}e^{iq_{0}d},\\
&iE^{\rm t}_{\rm L}e^{iq_{\rm L}d} -iE^{\rm t}_{\rm R}e^{iq_{\rm R}d}
iE^{\rm r}_{\rm L}e^{-iq_{\rm L}d} -i E^{\rm r}_{\rm R}e^{-iq_{\rm R}d}=E^{\rm t}_{y}e^{iq_{0}d},\\
&-i\delta_{\rm L}E^{\rm t}_{\rm L}e^{iq_{\rm L}d} 
+ i\delta_{\rm R}E^{\rm t}_{\rm R}e^{iq_{\rm R}d}
+i\delta_{\rm L}E^{\rm r}_{\rm L}e^{-iq_{\rm L}d} 
-iq_{\rm R} E^{\rm r}_{\rm R}e^{-iq_{\rm R}d}=-E^{\rm t}_{y}e^{iq_{0}d},\\
&\delta_{\rm L}E^{\rm t}_{\rm L}e^{iq_{\rm L}d} 
+q_{\rm R}E^{\rm t}_{\rm R}e^{iq_{\rm R}d}
-q_{\rm L}E^{\rm r}_{\rm L}e^{-iq_{\rm L}d} 
-q_{\rm R}E^{\rm r}_{\rm R}e^{-iq_{\rm R}d}=E^{\rm t}_{x}e^{iq_{0}d}.
\end{align}
\end{subequations}
After some algebra, we have 
\begin{align}
\label{M2}
\begin{pmatrix}
E^{\rm t}_{\rm L}\\
E^{\rm t}_{\rm R}\\
E^{\rm r}_{\rm L}\\
E^{\rm r}_{\rm R}
\end{pmatrix}
=
\frac{e^{iq_{0}d}}{4}
\begin{pmatrix}
0&(1+\delta^{-1}_{\rm L})e^{-iq_{\rm L}d}&0&0\\
(1+\delta^{-1}_{\rm R})e^{-iq_{\rm R}d}&0&0&0\\
0&(1-\delta^{-1}_{\rm L})e^{iq_{\rm L}d}&0&0\\
(1-\delta^{-1}_{\rm R})e^{iq_{\rm R}d}&0&0&0
\end{pmatrix}
\begin{pmatrix}
E^{\rm t}_{+}\\
E^{\rm t}_{-}\\
0\\
0
\end{pmatrix}, 
\end{align}
where $E^{\rm t}_{\pm}=E^{\rm t}_{x}\pm iE^{\rm t}_{y}$ 
is the component representing the left- $(-)$ and the right-handed $(+)$ 
circularly polarized transmitted waves, respectively.  
Substituting Eq.~(\ref{M2}) into Eq.~(\ref{M1}), we 
have 
\begin{align}
\begin{pmatrix}
E^{\rm i}_{+}\\
E^{\rm i}_{-}\\
E^{\rm r}_{+}\\
E^{\rm r}_{-}
\end{pmatrix}
=
\begin{pmatrix}
{\cal T}_{\rm R}^{-1}&0&0&0\\
0&{\cal T}_{\rm L}^{-1}&0&0\\
{\cal R}_{\rm R}{\cal T}_{\rm R}^{-1}&0&0&0\\
0&{\cal R}_{\rm L}{\cal T}_{\rm L}^{-1}&0&0\\
\end{pmatrix}
\begin{pmatrix}
E^{\rm t}_{+}\\
E^{\rm t}_{-}\\
0\\
0
\end{pmatrix},
\end{align}
where 
\begin{align}
&{\cal T}_{\rm L,R}=\frac{(1-\chi_{\rm L,R})e^{i(q_{\rm L,R}-q_{0})d}}{1-\chi_{\rm L,R}^{2}e^{2iq_{\rm L,R}d}},\\
&{\cal R}_{\rm L,R}=\frac{\chi_{\rm L,R}(1-e^{2iq_{\rm L,R}d})}{1-\chi_{\rm L,R}^2e^{2iq_{\rm L,R}d}}
\end{align}
with 
\begin{align}
\chi_{\rm R, L} = \frac{1-\delta_{\rm L,R}}{1+\delta_{\rm L,R}}
= \frac{1-\sqrt{\vare_{\rm L,R}}}{1+\sqrt{\vare_{\rm L,R}}}.
\end{align}
Thus, the transmittance, $T_{\rm L,R}$, and reflectance, $R_{\rm L,R}$, 
are respectively given by  
\begin{align}
&T_{\rm L,R} = \left|\frac{E^{\rm t}_{\mp}}{E^{\rm i}_{\mp}}\right|^2 = |{\cal T}_{\rm L,R}|^2, \\
&R_{\rm L,R} = \left|\frac{E^{\rm r}_{\mp}}{E^{\rm i}_{\mp}}\right|^2 = |{\cal R}_{\rm L,R}|^2, \\
\end{align}
Therefore, we obtain the absorbance, $A_{\rm L,R}$, as 
\begin{align}
A_{\rm L,R} = 1 - T_{\rm L,R} - R_{\rm L,R}. 
\end{align}
\end{widetext}


\begin{thebibliography}{99}
\bibitem{perp}
In this paper, the symbols $\perp$ and  $\parallel$  
are used to represent the perpendicular and parallel directions with respect to the Rashba field ${\bm \alpha} $ direction, respectively. 
Note that we have used the opposite definition to that given in Refs. \onlinecite{{Tsutsui-Murakami12},{Maiti15},{Ye-Tokura15}}.
\bibitem{note1}
The equilibrium electron density, 
$n_{\rm e}$, for ${\bm M}=0$ has been calculated as 
$n_{\rm e} = -\sum_{s}\sum_{\kv}\delta(\epsilon^{s}_{\kv,{\bm M}={\bm 0}}-\epsilon_{\rm F})
= n_{\rm e}^{(0)}(1+ \Delta_{\tilde{\alpha}})$, 
where $n_{\rm e}^{(0)}= \dfrac{(k_{\rm F}^{\perp})^2k_{\rm F}^{\parallel}}{3\pi^2}$ 
is the electron density in the absence of the RSOI 
and 
$\Delta_{\tilde{\alpha}} = \dfrac{3}{2}\tilde{\alpha}^2\left(
1+ \dfrac{1+\tilde{\alpha}^2}{\tilde{\alpha}}\tan^{-1}\tilde{\alpha} 
\right)
$ 
is the correction. 
Here, $\tilde{\alpha} = \dfrac{m_{\perp} \alpha}{\hbar^2 k^{\perp}_{\rm F}}$ 
is a non-dimensional Rashba interaction strength.  
Note that the electron density, $n_{\rm e}$, 
increases monotonically with increasing $\tilde{\alpha}$.
\end{thebibliography}
\end{document}